\documentclass[useAMS,usenatbib]{mnras}
\usepackage{times}
\usepackage{graphicx}
\usepackage[none]{hyphenat}
\usepackage{comment} 
\usepackage[normalem]{ulem}
\usepackage{booktabs,caption,fixltx2e}
\usepackage{amssymb}
\usepackage[flushleft]{threeparttable}
\usepackage{blindtext}
\usepackage{subfig}
\usepackage{multirow}
\usepackage{adjustbox}
\usepackage{algorithm}
\usepackage[noend]{algpseudocode}

\title[Clump formation in stellar wind collisions]{3D simulations of clump formation in stellar wind collisions}
\author[D. Calder\'on et al.]{D. Calder\'on$^{1,2,3}$\thanks{E-mail:diego.calderon@utf.mff.cuni.cz},  J. Cuadra$^1$, M. Schartmann$^{4,5,2}$, A. Burkert$^{5,4,2}$, J. Prieto$^6$, and \newauthor C. M. P. Russell$^1$\\
$^1$Instituto de Astrof\'isica, Facultad de F\'isica, Pontificia Universidad Cat\'olica de Chile, 782-0436 Santiago, Chile\\
$^2$Max Planck Institute for Extraterrestrial Physics, P.O. Box 1312, Giessenbachstr. 1, D-85741 Garching, Germany\\
$^3$Institute of Theoretical Physics, Faculty of Mathematics and Physics, Charles University, 180 00 Prague, Czech Republic\\
$^4$Excellence Cluster ORIGINS, Ludwig-Maximilians-Universit\"at M\"unchen, Boltzmannstr. 2, D-85748 Garching, Germany\\
$^5$Universit\"atssternwarte der Ludwig-Maximilians-Universit\"at, Scheinerstr. 1, D-81679 M\"unchen, Germany \\ 
$^6$Departamento de Astronom\'ia, Universidad de Chile, Casilla 36-D, Santiago, Chile
}

\begin{document}

\label{firstpage}

\date{Draft \today}

\pagerange{\pageref{firstpage}--\pageref{lastpage}} \pubyear{2020}
\maketitle

\begin{abstract}
	The inner parsec of our Galaxy contains tens of Wolf-Rayet stars whose powerful outflows are constantly interacting while filling the region with hot, diffuse plasma.
	Theoretical models have shown that, in some cases, the collision of stellar winds can generate cold, dense material in the form of clumps. 
	However, their formation process and properties are not well understood yet.
	In this work we present, for the first time, a statistical study of the clump formation process in unstable wind collisions. 
	We study systems with dense outflows (\hbox{${\sim}10^{-5}\rm\ M_{\odot}\ yr^{-1}$}), wind speeds of \hbox{$500$--$1500\rm\ km\ s^{-1}$}, and stellar separations of \hbox{${\sim}20$--$200\rm\ au$}. 
	We develop 3D high resolution hydrodynamical simulations of stellar wind collisions with the adaptive-mesh refinement grid-based code \textsc{Ramses}. 
	We aim to characterise the initial properties of clumps that form through hydrodynamic instabilities, mostly via the non-linear thin shell instability (NTSI). 
	Our results confirm that more massive clumps are formed in systems whose winds are close to the transition between the radiative and adiabatic regimes. 
	Increasing either the wind speed or the degree of asymmetry increases the dispersion of the clump mass and ejection speed distributions. 
	Nevertheless, the most massive clumps are very light (\hbox{${\sim}10^{-3}$--$10^{-2}\rm\ M_{\oplus}$}), about three orders of magnitude less massive than theoretical upper limits. 
	Applying these results to the Galactic Centre we find that clumps formed through the NTSI should not be heavy enough either to affect the thermodynamic state of the region or to survive for long enough to fall onto the central super-massive black hole. 
\end{abstract}


\section{Introduction}

	Massive stars experience strong mass-loss episodes during their lives. 
	These powerful outflows have mass-loss rates that can reach up to \hbox{${\sim}10^{-4}\rm\ M_{\odot}\ yr^{-1}$} and velocities that can exceed \hbox{$2000\rm\ km\ s^{-1}$} \citep{P08}. 
	This kind of activity occurs mainly during the Wolf-Rayet (WR) and Luminous Blue Variable (LBV) stages.  
	It has been observed that in binary systems such winds collide producing very energetic signatures such as particle acceleration, X-ray and $\gamma$-ray emission \citep{A10,Ha16,H18,P18}. 
	In this case the material launched at supersonic speeds on opposite directions collides generating dense shells of compressed shocked material at temperatures typically in the range \hbox{${\sim}10^6$--$10^7\rm\ K$}. 
	However, binaries are not the only environments where stellar winds interact leaving energetic observational signatures. 
	Crowded stellar systems like the nuclear star clusters in the centre of the Milky Way are clear examples.
	The immediate vicinity of the central super-massive black hole (SMBH), Sgr~A*, is populated by hundreds of massive O, B, and evolved stars \citep[see][for a review]{G10}. 
	Out of them 30 have been spectroscopically identified as WR stars. 
	They have significant mass-loss rates (\hbox{$\gtrsim10^{-5}\rm\ M_{\odot}\ yr^{-1}$}) in the form of stellar winds at very high speeds \citep[\hbox{$500$--$2500\rm\ km\ s^{-1}$;}][]{M07}. 
	It is thought that these outflows are responsible for filling the region with hot (\hbox{${\sim}10^7\rm\ K$}), diffuse plasma ($n\approx10\rm \ cm^{-3}$) that irradiates in X-ray  \citep[\hbox{e.g.}][]{B03,W13,R17}.  
	On top of this, it is reasonable to argue that a fraction of this material might fall onto the SMBH. 
	However, the amount and way the gas flows towards Sgr~A* or escape from the environment outwards is not well understood yet \citep{W13}. 
		
	In the last decades several groups have been monitoring the stars located within the inner parsec \citep[e.g.][]{P06,Y14,G17}. 
	This has made it possible to infer the physical properties of the stars and their winds, and, in some cases, determine their orbital motion around the SMBH with high precision. 
	Therefore, modelling the hydrodynamics of this environment is a unique opportunity to study the gas dynamics at small distances from a SMBH, and also the multiple stellar wind interactions that are constantly taking place in this region. 
	 \citet{C05,C06,C08,C15} developed hydrodynamics simulations of the complete system of WR stars moving on the observed orbits around Sgr~A*. 
	Simultaneously, the stars were feeding their environment via stellar winds. 
	
	\citet{L16} conducted a similar work but studied another stellar component, the so-called S-stars.  
	These objects correspond to B-type stars which are orbiting around the SMBH more closely compared to the WR stars. 
	The wind properties of the S-stars have not been constrained accurately yet, but their winds are certainly not as dense as the WR stars. 
	Specifically, their mass-loss rates are about two orders of magnitude lower.
	Although both works have managed to model very complex systems they have not been able to obtain reliable estimates of the amount of cold, low-angular momentum material, i.e. material that is more likely to be accreted by Sgr~A*. 
	This is a direct consequence of the use of the traditional Smoothed-Particle Hydrodynamics (SPH) approach, which presents problems when simulating strong shocks, discontinuities, and two-phase medium. 
	In some cases, this fact can produce artificial clumping instead of describing the expected filamentary structure properly \citep{H13}. 
	For instance, the simulations of \citet{C08}  show that stellar wind collisions constantly generate dense, cold, clumpy material. 
	However, our analytical study showed that clump formation should not be as frequent as seen in such simulations \citep{C16}. 
	Therefore, there is a need for describing these processes in detail with more appropriate computational tools, especially in this environment. 
	
	In a recent study, \citet{R18} modelled the WR outflows in the Galactic Centre with a grid-based hydrodynamical code. 
	This approach is significantly better suited to simulate shocks and a two-phase medium. 
	As a result, they did not observe cold, dense clumps as originating from the stellar wind collisions. 
	Nonetheless, this was not the main focus of their study.
	Instead their model was optimised to resolve the inner region as accurate as possible, and not necessarily for modelling the wind collisions in detail. 
	Thus, despite the efforts of several works dedicated to studying how stellar winds feed Sgr~A*, the wind interactions themselves have not been the focus of any previous study yet. 
	
	In this context, it is important to remark that the thermodynamic state of the gas in the inner parsec can have a significant impact on the accretion rate onto the central black hole. 	
	For example if cold material can be formed and survive long enough to fall onto the SMBH it could cause changes in the accretion activity. 
	\citet{C08} showed how the accretion of gas clumps can cause variability episodes on timescales of hundreds of years on the activity of Sgr~A*. 
	Theoretically, the accretion of a single clump of large enough mass could be responsible for the more active past of the SMBH inferred from observations of X-ray echoes \citep[e.g.][]{S93,S98,MB07,P10}. 
	In addition, the question arises whether the cold, small gas cloud G2 on a tight orbit around the central SMBH could be a result of wind interactions \citep{G12,B12,C16,C18}.
	Therefore, in order to understand the current and past activity of the Galactic Centre, it is necessary to describe its stellar-wind collisions, more specifically the potential formation and evolution of gas clumps.

	In general, studies of stellar wind collisions have focused on binary systems, a.k.a. colliding wind binaries. 
	Early work was done by \citet{S92} studying unstable wind collisions through numerical 2D simulations. 
	\citet{P09} developed sophisticated 3D simulations of binary systems aiming to describe the hydrodynamics of wind interactions. 
	These models included many physical ingredients such as gravity, wind acceleration, radiative cooling, and orbital motion. 
	Within their findings they showed that clumps could be formed and, in some cases, they could live for long enough to escape from the system. 
	\citet{L11} revisited 2D models with the aid of the adaptive-mesh refinement (AMR) technique in order to simulate unstable wind collisions with high resolution.   
	They could formally identified that the so-called non-linear thin shell instability (NTSI) dominates the shape of unstable slabs over long timescales. 
	Also, they warned about the tremendous computational challenge that one faces when modelling these systems.
	\citet{V11} presented 3D simulations of colliding wind binaries to study the shape and structure of the slabs formed in wind collisions. 
	Moreover, they also explored the structure of shells formed from the interaction of WR winds with material previously ejected by the same star during earlier stages of its life \citep{V12}. 
	They found that such interactions can create very complex structures through the development of different thin-shell instabilities depending on the radiative properties of the WR wind shock. 
	If the material swept-up was dense enough, it could lose its thermal support becoming unstable very easily and creating very distinctive small scale patterns. 
	\citet{K14} studied the effects of the NTSI on X-ray emission through numerical 2D simulations. 
	\citet{H16} carried out 3D simulations with extremely high-resolution and dust formation, aiming to reproduce infrared observations of the spiral patterns created by the interaction of the winds combined with binary orbital motion. 
	
	Up to now there has not been any detailed quantitative study of the cold gas, sometimes in the form of clumps, produced in stellar wind collisions. 
	In this work we present, for the first time, a statistical analysis of the clump formation process in such systems. 
	Motivated by the WR stars in the Galactic Centre, we study systems with dense outflows (\hbox{${\sim}10^{-5}\rm\ M_{\odot}\ yr^{-1}$}), speeds of \hbox{$500$--$1500\rm\ km\ s^{-1}$}, and stellar separations of \hbox{${\sim}20$--$200\rm\ au$}. 
	In general, these systems are wider than typical colliding wind binaries studied in the literature. 
	We use the code \textsc{Ramses} \citep{T02} to run 3D hydrodynamic simulations of unstable wind collisions. 
	The code includes an AMR module for reaching higher resolution without increasing the computational cost significantly.
	We do not focus on modelling specific colliding wind binary systems. 
	Instead, we explore a rather specific set of parameters, suited for the WR stellar system within the innermost parsec in order to: 
	$i)$ understand the process of clump formation,  
	$ii)$ determine the initial clump physical properties and characterise their dependence on system parameters, 
	and $iii)$ compare the results with previous theoretical estimates of clump formation in the Galactic Centre.
	
	This work is structured as follows. 
	In Section~\ref{sec:physics}, we describe the physics involved in stellar wind collisions. 
	Also, we briefly review the physical mechanisms responsible for clump formation and discuss relevant parameters. 
	Then, we present the numerical setup of our simulations in Section~\ref{sec:setup}. 
	Here, we include a description of the models explored. 
	The results of our study are shown in Section~\ref{sec:results}, where we describe the hydrodynamics of each model, as well as the characteristics of the clumps that form. 
	Then, we compare our results with previous analytical work, study the impact of resolution, and discuss implications on the hydrodynamics state of the Galactic Centre in Section~\ref{sec:discussion}. 
	Finally, we present our conclusions and future work guidelines in Section~\ref{sec:conclusions}.  

\section{Stellar wind collisions}
\label{sec:physics}

	\subsection{Structure of the interaction zone}
	\label{sec:slab}
				
		In general, stellar winds of massive stars propagate with supersonic speeds developing shock waves that compress material in shells behind their fronts.  
		The shape and physical properties of such shells depend strongly on the nature of the shocks, in particular the ability of the shocked material to radiate away its thermal energy.  
		Accordingly, there are two regimes into which shocks fall: radiative and adiabatic. 
		In the former, the compressed material radiates its thermal energy rapidly and forms thin shells of cold, dense material just behind the shock front. 
		For the latter, the compressed material predominantly loses energy through adiabatic expansion, leaving the layer hot and thick. 
		Naturally, there is also an intermediate case where the energy losses through radiation and adiabatic expansion are complimentary, and in a binary system the two shocks can have different properties.
		
		In their seminal work, \citet{S92} introduced the parameter $\chi$ to characterise the radiative nature of a stellar wind such that we can interpret in advance if it is associated to a radiative or an adiabatic shock.
		The \textit{cooling parameter} $\chi$ is defined as the ratio of the cooling timescale $t_{\rm cool}$ to the adiabatic expansion timescale $t_{\rm ad}$ for the shocked material:
		
			\begin{equation}
				\chi = \frac{t_{\rm cool}}{t_{\rm ad}}\approx\frac{V_{\rm w,8}^4d_{12}}{\dot{M}_{-7}},
				\label{eq:chi}
			\end{equation}
		
		\noindent where $V_{\rm w, 8}$ is the wind speed in units of \hbox{$1000\rm\ km\ s^{-1}$}, 
		$d_{12}$ is the distance from the star to the contact discontinuity in units of \hbox{$10^{12}\rm\ cm$}, $\dot{M}_{-7}$ is the wind mass-loss rate in units of \hbox{$10^{-7}\rm\ M_{\odot}\ yr^{-1}$}.
		Originally, this expression was obtained under the assumption of solar abundances; however we modified it in order to account for different values of metallicity by introducing a factor $(Z/Z_{\odot})^{-1}$, where $Z$ is the mass fraction in metals (see Appendix~\ref{app:cool}, for a discussion). 
		Based on this definition, if \hbox{$\chi<1$} the wind is radiative while if \hbox{$\chi>1$} the wind is adiabatic. 
		
\begin{figure}
	 	\centering
		\includegraphics[width=0.475\textwidth]{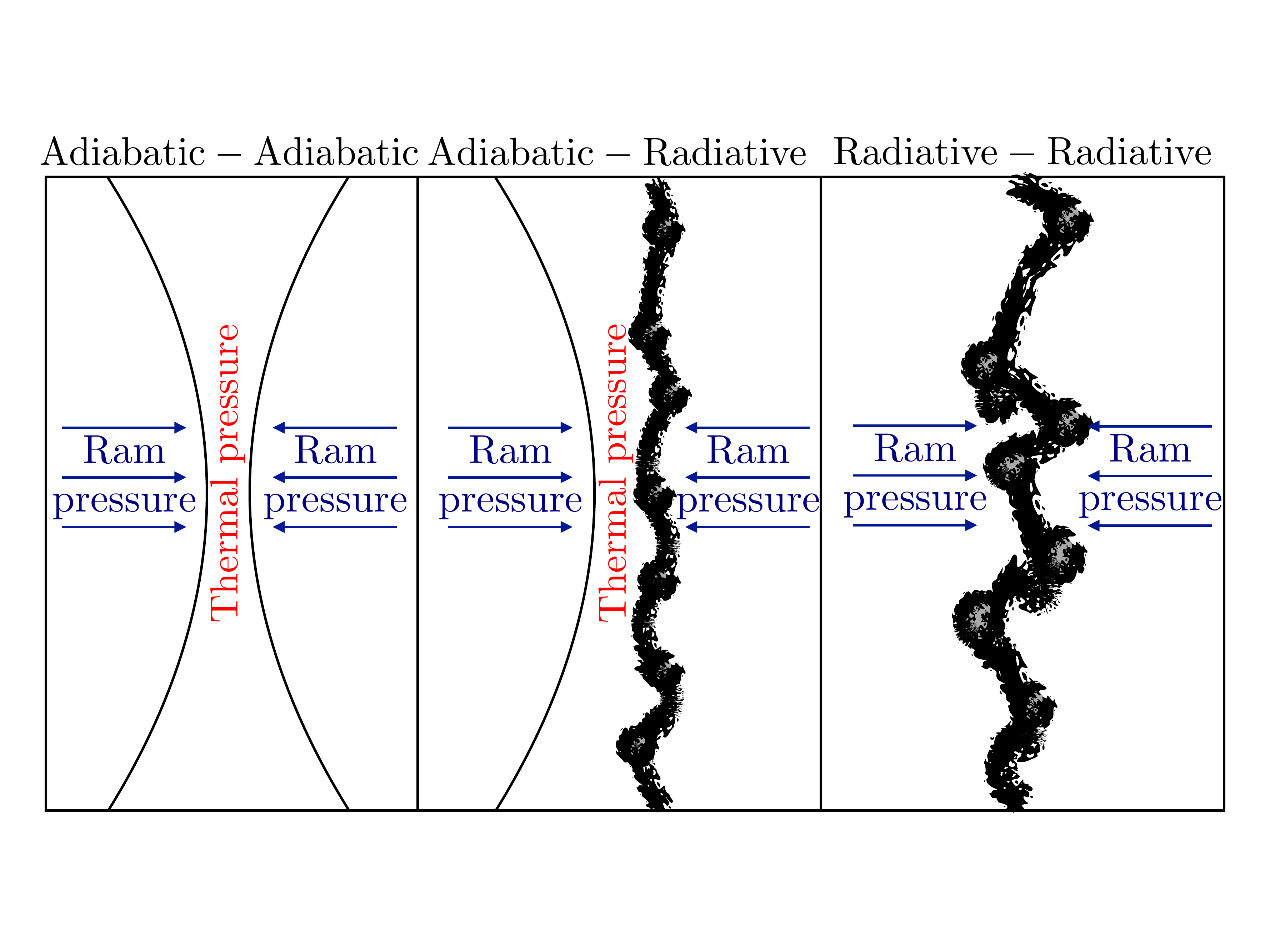}
		\caption{Schematic representation of different types of stellar wind collisions according to radiative properties. 
		Left panel shows the result of a collision of two adiabatic winds: a thick, hot slab of shocked material. 
		The middle panel contains the outcome of a radiative wind colliding with an adiabatic wind: a dense, thin shell of cold material subject to the Vishniac's instability.  
		Right panel illustrates the result of a collision of two radiative winds: a dense, thin shell of cold material subject to the NTSI. 
		It is important to remark that in case the winds have different speeds the KHI can be excited in any case.}
		\label{fig:sketch}
	\end{figure}

		Once the winds collide they generate a slab of shocked material whose properties depend strongly on the wind radiative nature. 
		In the case where both winds are adiabatic, the resulting slab is smooth, thick, hot, and held up by thermal pressure on both sides (see left panel of Figure~\ref{fig:sketch}). 
		If one wind is adiabatic while the other is radiative, the radiative shock is supported by its ram pressure and a thin shell of cold material forms (see central panel of Figure~\ref{fig:sketch}). 
		Such a thin shell can become unstable very easily; however, the thermal pressure of the adiabatic shock acts as a dampener of such instabilities. 
		This is the so-called Vishniac instability \citep{V83}. 
		Finally, if both winds are radiative, a dense and cold slab is formed, confined by ram pressure on both sides. 
		If it is perturbed, the non-linear thin shell instability \citep[NTSI;][]{V94} can be excited. 
		This mechanism is caused by the misbalance of the thermal pressure inside the cold slab with respect to the ram pressure of the winds. 
		Consequently, material tends to accumulate on the knots of the perturbation of the slab (gray regions in right panel of Figure~\ref{fig:sketch}). 
		It is important to remark that in the case where the winds have different speeds, regardless of the radiative nature of the winds, the Kelvin-Helmholtz instability (KHI) can be excited. 
		However, high-resolution numerical models of unstable wind collisions have shown that, if excited, the NTSI tends to dominate over other instabilities. 
		Specifically, it is the main shaper of the slab structure due to its large-scale perturbations \citep{L11}.
	
	\subsection{Clump formation}
	\label{sec:form}
	
		Theoretically, only a limited range of wavelengths can excite the NTSI. 
		\citet{V94} showed that the unstable wavelengths should be at least of the width of the slab. 
		Otherwise, such shells could not be effectively corrugated. 
		On the other side, the upper limit is given by the sound crossing length\footnote{Defined as the distance a sound wave travels in a given timescale, i.e. \hbox{$l_{s}=c_s\Delta t$}}, so
		
		\begin{equation}
			l_{\rm slab} \lesssim \lambda_{\rm NTSI}\lesssim l_{s}.
		\end{equation}
		
		\noindent This analytical description assumes an isothermal equation of state, i.e. infinitely efficient cooling. 
		Nonetheless, in reality we expect cooling to occur on a finite amount of time. 
		Within the ideal gas assumption the sound speed is proportional to the square root of the temperature of the gas, i.e. \hbox{$c_{\rm s}\propto\sqrt{T}$}. 
		Thus, a longer cooling timescale implies also a longer unstable wavelength upper limit, which potentially means that larger clumps could be formed. 
		Nevertheless, let us bear in mind that still it is a necessary condition of having a thin, cold slab. 
		Hence the largest clumps will form whenever radiative cooling is efficient but takes place as slow as possible. 
		Therefore, winds whose $\chi$ approaches unity should generate the largest and most massive clumps \citep{C16}.
		Simply by assuming a geometry and using the value of the density in the slab we can estimate the approximate clump masses. 
				
\section{Numerical simulations}
\label{sec:setup}

	\subsection{Equations}
	
		Our numerical simulations are carried out with the AMR hydrodynamics code \textsc{Ramses} \citep{T02}. 
		The code solves the Euler equations in their conservative form, i.e. 
	
		\begin{eqnarray}
			\frac{\partial\rho}{\partial t}+\nabla\cdot(\rho\mathbf{u}) &= & 0,\\
			\frac{\partial}{\partial t}(\rho\mathbf{u})+\nabla\cdot(\rho\mathbf{u}\otimes\mathbf{u}) & = & \rho\mathbf{f}(\mathbf{x}) - \nabla P,\\
			\frac{\partial}{\partial t}(\rho e)+\nabla\cdot\left[\rho\mathbf{u}\left(e+\frac{P}{\rho}\right)\right] & = & -\frac{\rho^2}{\left(\mu m_{\rm H}\right)^2}\Lambda(T)
			\label{eq:energy}
		\end{eqnarray}
	
		\noindent where $\rho$, $\mathbf{u}$ and $P$ are the mass density, velocity, and pressure of the fluid, respectively; 
		$\mathbf{f}$ is the gravitational force per mass unit and $e$ the total specific energy density which is given by
	
		\begin{equation}
			e=\frac{1}{2}\mathbf{u}\cdot\mathbf{u}+\frac{P}{(\gamma - 1)\rho},
		\end{equation}
	
		\noindent where $\gamma$ is the adiabatic index that is set to $5/3$ for adiabatic gases. 
		Furthermore, $\mu$ is the mean molecular weight, $m_{\rm H}$ is the proton mass, $T$ is the temperature of the gas, and $\Lambda(T)$ represents the energy losses due to optically-thin radiative cooling (see Appendix~\ref{app:cool}, for further details). 
		Additionally, we include a prescription for the stellar wind generation which is presented in Section~\ref{sec:winds}. 
		
	\subsection{Numerical setup}
		\label{sec:ns}
		
		We run 3D simulations on a Cartesian grid making use of the AMR technique, such that the resolution is enhanced in regions of the domain where specified physical criteria are met. 
		The domain is a cube of side length $2a$ where $a$ is the stellar separation of the system. 
		Every side of the domain has an outflow boundary condition. 
	
\begin{figure}
	 		\centering
			\includegraphics[width=0.4\textwidth]{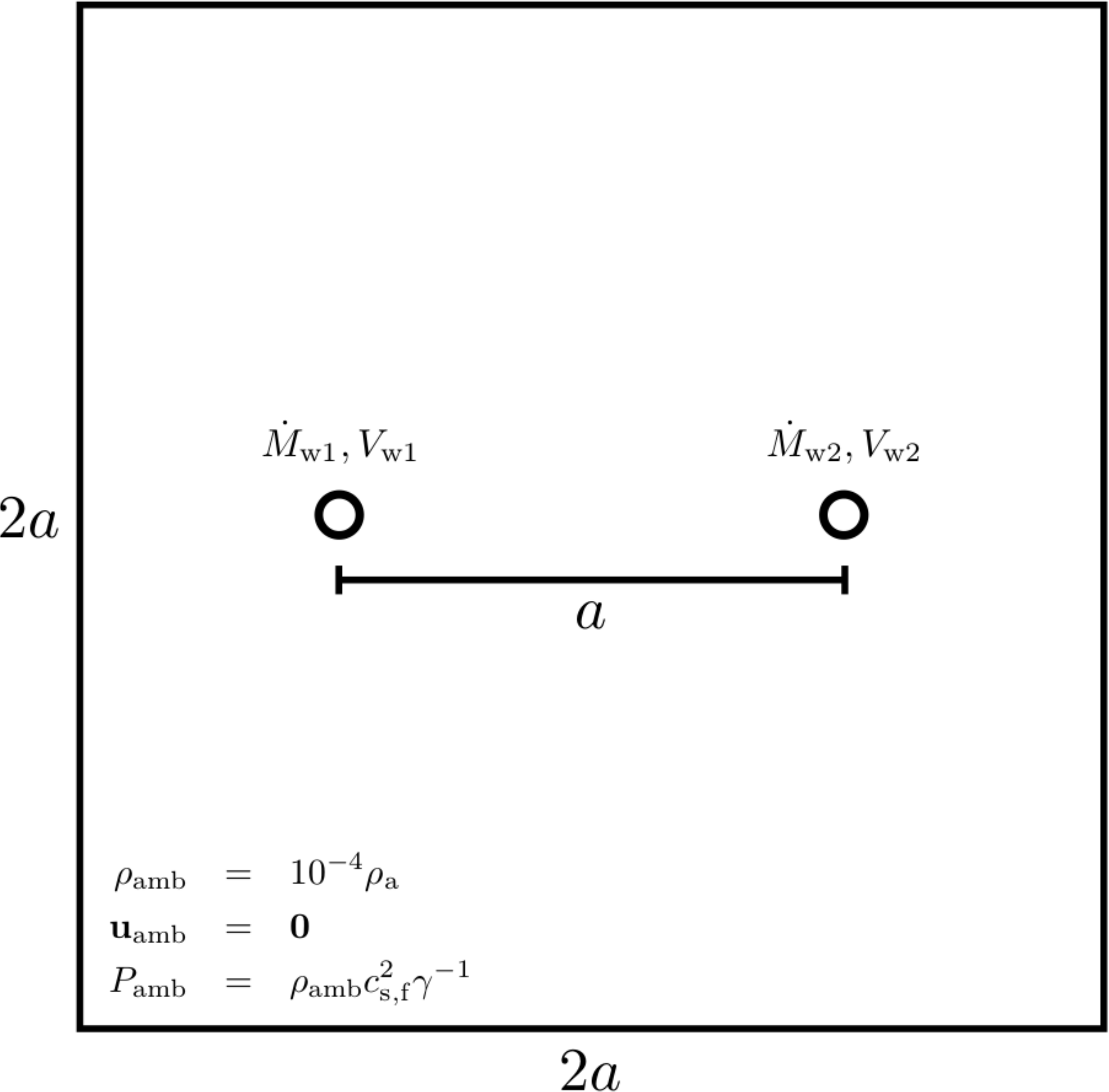}\hspace{0.5cm}
			\caption{Schematic representation of the initial conditions of simulations at proportional scale. 
			The domain is a cubic box with a side length of $2a$, hence volume $8a^3$. 
			The hydrodynamic variables of the ambient medium initially are set to \hbox{$\rho_{\rm amb}=10^{-4}\rho_a$}, \hbox{$\mathbf{u}_{\rm amb}=\mathbf{0}$}, \hbox{$P_{\rm amb}=\rho_{\rm amb}c_{\rm s,f}^2\gamma^{-1}$}. 
			Two stars blowing stellar winds are fixed at $\mathbf{r}_{\rm w,1}=(-0.5a,0,0)$ and $\mathbf{r}_{\rm w,2}=(+0.5a,0,0)$, i.e. their separation is $a$. 
			The stellar winds are generated in spherical regions of radius \hbox{$a_{\rm w}=0.04a$} centred at those locations. 
			Each stellar wind is characterised by its mass-loss rate $M_{\rm w,i}$ and terminal velocity $V_{\rm w,i}$. }
			\label{fig:setup}
		\end{figure}

		The setup was chosen in order to capture the development of instabilities in the wind interactions as accurate as possible. 
		We follow the guidelines provided by \citet{L11} in their extensive 2D study.
		We used an exact Riemann solver with a MinMod flux limiter. 
		These choices avoid the quenching of instabilities by numerical diffusion.  
		The refinement strategy is set to be based on density gradients, thus the resolution increases mostly in shocks and discontinuities.
		The coarse grid resolution of our simulations is $64^3$ cells, and there are four levels of refinement (standard resolution), creating an effective resolution of $1024^3$ cells. 
		Therefore, each resolution element reaches a length of \hbox{$a/512\approx0.002a$}. 
		Each simulation in this work consists of the hydrodynamics evolution of two stellar winds that are being blown from stars fixed in space from positions \hbox{$\mathbf{r}_{\rm w,1}=(-0.5a,0,0)$} and \hbox{$\mathbf{r}_{\rm w,2}=(+0.5a,0,0)$} in a cubic volume of length $2a$. 		
		The domain size is chosen in order to maximise the resolution in the region where clumps are formed, which is the main scope of this work.  
		The environment is initialised at low density, specifically four orders of magnitude smaller than the wind density at the distance that it is blown \hbox{$\rho_{\rm amb}=10^{-4}\rho_{\rm a}$}. 
		The medium is set at rest \hbox{$\mathbf{u}=\mathbf{0}$}, and at the lowest temperature allowed \hbox{$P_{\rm amb}=\rho_{\rm amb}c^2_{\rm s,f}\gamma^{-1}$}, where $c_{\rm s,f}$ is the sound speed at the temperature floor. 
		These specifications allow the stellar winds to flow freely, filling the domain without difficulties until they collide. 
		Figure~\ref{fig:setup} shows a 2D schematic representation at $z=0$ of the setup and initial conditions.  
		We ran each simulation for at least five wind crossing timescales, defined as the time the slowest wind takes to cross the domain, i.e. \hbox{$t_{\rm cross}=2a/V_{\rm w,1}$}. 
		The simulation time of each model corresponds to a small fraction of the orbital period in case the stars were in a binary system. 
		However, the orbital speed of such a system would be $\lesssim10$~per~cent of the stellar wind velocity, which justifies our choice of considering motionless stars. 
		
		Although other physical effects, such as thermal conduction as well as the presence of magnetic fields, could play a role in the evolution of the wind-confined slab, our models do not take them into account. 
		In principle, these physical mechanisms could suppress the growth of the shorter modes of instabilities in the slab \citep{B00,H07}. 
		Furthermore, it has been showed that the presence of tangled magnetic fields in gas clouds help them to survive for longer when moving in the presence of a strong wind \citep{M15}. 
		However, their inclusion adds more free parameters into our modelling and, at the same time, can increase significantly the computational cost of the simulations, especially when aiming to perform a systematic parameter study of high-resolution 3D models. 
		Thus, we chose to focus on studying the evolution of purely hydrodynamical systems (plus radiative cooling) in 3D and focus our computational resources on maximising their spatial resolution and simulation time. 
		
	\subsection{Stellar wind generation}
	\label{sec:winds}
		The setup includes a module to generate stellar winds, largely inspired by the approach of \citet{L07}. 
		Specifically, it consists of resetting the hydrodynamic variables to the 1D free-wind solutions inside a spherical region of the domain (hereafter ``masked region") after every time step. 
		Here we assumed that the winds are instantaneously accelerated up to their terminal speeds, which is justified as we study models whose stellar separations are significantly larger than the radius of the stars ($a\gg R_*$).
		In order to capture the spherical symmetry of the wind model, we forced the masked region to be refined up to the maximum level. 
		Each stellar wind in our simulations is determined by three parameters: mass-loss rate $\dot{M}_{\rm w,i}$, terminal velocity $V_{\rm w,i}$, and mask radius $a_{\rm w}$. 
		Therefore, the (primitive) hydrodynamic variables within the mask are kept fixed in time with the following values:
	
		\begin{eqnarray}
			\rho(|\mathbf{x}-\mathbf{r}_{\rm w,i}|^2 < a_{\rm w}^2)	&	=	&	\rho_{a}\left(\frac{a_{\rm w}}{|\mathbf{x}-\mathbf{r}_{\rm w,i}|}\right)^2,\\
			\mathbf{u}(|\mathbf{x}-\mathbf{r}_{\rm w,i}|^2 < a_{\rm w}^2)	&	=	&	V_{\rm w, i}\left(\frac{\mathbf{x}-\mathbf{r}_{\rm w,i}}{|\mathbf{x}-\mathbf{r}_{\rm w,i}|}\right),\\
			P(|\mathbf{x}-\mathbf{r}_{\rm w,i}|^2 < a_{\rm w}^2)	&	=	&	P_{a}\left(\frac{a_{\rm w}}{|\mathbf{x}-\mathbf{r}_{\rm w,i}|}\right)^{10/3},
		\end{eqnarray}
		
		\begin{eqnarray}
			\rho_a	&	=	&	\frac{1}{4\pi a_{\rm w}^2}\frac{\dot{M}_{\rm w,i}}{V_{\rm w,i}},\\
			P_a		&	=	&	\frac{1}{\gamma}\rho_ac_{\rm s,f}^2,
		\end{eqnarray}
		
		\noindent where the subscript $\rm i=1,2$ is the label of each wind in the simulation. 
		The sound speed $c_{\rm s,f}$ is obtained by choosing the temperature of the wind. 
		We set \hbox{$T_{\rm w}=10^4\rm\ K$} for the winds at \hbox{$|\mathbf{x}-\mathbf{r}_{\rm w,i}|=a_{\rm w}$}, which implies \hbox{$c_{\rm s,f}\approx10\rm\ km\ s^{-1}$}.  
		This temperature also corresponds to the floor temperature set in the simulations.  
		In reality, we expect the strong UV radiation field of the massive stars to be the responsible for setting this floor. 
		To check the validity of this assumption we also confirmed that the optical depth is lower than unity in every simulation run (see Appendix~\ref{app:cool}).
		The size of the computational region where winds are generated was chosen to be of radius \hbox{$a_{\rm w}=0.04a$}. 
		With this choice, the free wind region profile agrees to $1$~per~cent with the analytical density profile. 
		The size of this masked region is the same for both stars in every run. 
		
		We consider that the metallicity of the stellar winds is $Z=3Z_{\odot}$.  
		This choice is inspired by our motivation to apply the results to the Galactic Centre environment. 
		Although metallicity is not strongly constrained for such stars, this is the value typically assumed in the latest theoretical studies of the region \citep{C08,C16,R18}.	
	
	\subsection{Models}
	
\begin{figure}
	 		\centering
			\includegraphics[width=0.45\textwidth]{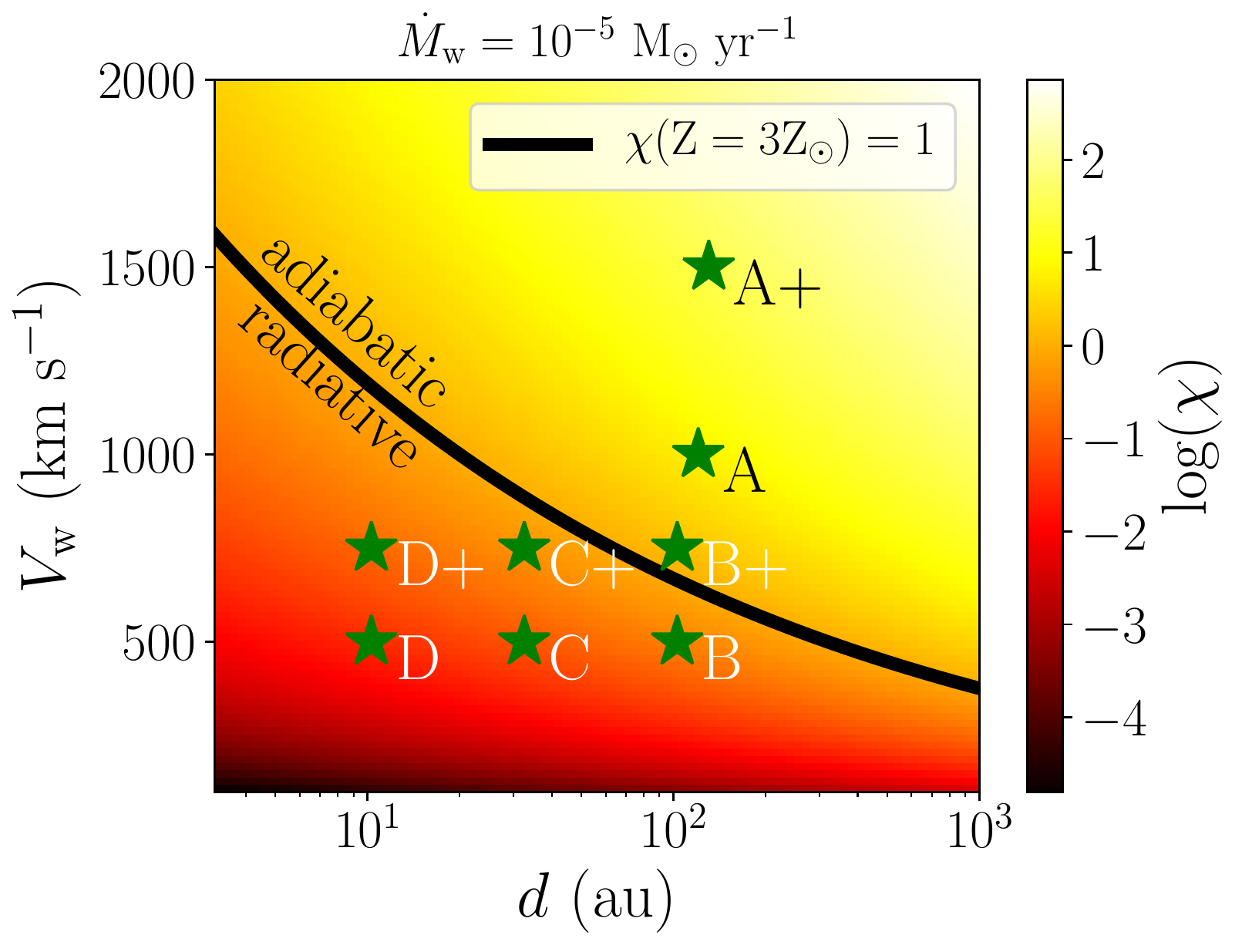}
			\caption{
			The cooling parameter $\chi$ computed from Equation~\ref{eq:chi} for a fixed mass-loss rate as a function of wind speed and distance to the contact discontinuity. 
			The solid black line stands for \hbox{$\chi=1$} which divides the radiative and adiabatic regime.  
			Green stars represent the wind of each star from each model studied.}
			\label{fig:chi}
		\end{figure}

		Figure~\ref{fig:chi} presents the cooling parameter $\chi$ of a single stellar wind computed from Equation~\ref{eq:chi} as a function of its wind speed and distance to the contact discontinuity for the metallicity chosen in this work. 
		The solid black line corresponds to $\chi=1$, i.e. the transition from the radiative (\hbox{$\chi<1$}) to the adiabatic (\hbox{$\chi>1$}) regime.  
		In this diagram the mass-loss rate is \hbox{$\dot{M}_{\rm w}=10^{-5}\rm\ M_{\odot}\ yr^{-1}$}, which is a typical value for WR stars. 
		However, their wind terminal velocities span a very wide range \citep[\hbox{$500$--$2000\rm\ km\ s^{-1}$;}][]{M07}. 
		Based on this, winds can be radiatively efficient if their speeds are slow, and/or if the position of the wind interaction region is very close to one of the stars (see Figure~\ref{fig:chi}). 
		The former condition can be satisfied for some types of WR stars, for example by the Ofpe/WN9 class whose winds are relatively slow of about \hbox{$400$--$600\rm\ km~s^{-1}$} \citep{C94,M07,V17}. 
		The latter can occur in close encounters of single stars, including the case when stars with very different momentum flux in their winds are involved. 
		In such a case, the stronger wind ``pushes" the weaker one to remain closer to its star and, in some cases can force it to be radiative. 	
		
		In this work we set the mass-loss rate of the stars to \hbox{$\dot{M}_{\rm w}=10^{-5}\rm\ M_{\odot}\ yr^{-1}$}, therefore the stellar separation $a$ and the wind terminal speed of each star $V_{\rm w,i}$ are the free parameters of each model. 
		We refer to models of two identical stars, i.e. with identical winds, as \textit{symmetric systems}. 
		On the other hand, models whose stars have winds with different terminal velocities are referred to as \textit{asymmetric systems}. 
		In order to quantify the degree of asymmetry in the wind interaction we use the ratio of wind momentum fluxes $\eta$, which can be calculated by

		\begin{equation}
			\eta=\frac{\dot{M}_{\rm w,1}V_{\rm w,1}}{\dot{M}_{\rm w,2}V_{\rm w,2}},		
		\end{equation}
		
		\noindent where subscripts $1$ and $2$ represent the weaker and the stronger wind, respectively \citep{L90}. 
		Thus, it is satisfied that \hbox{$\eta\leq1$}. 
		Moreover as the mass-loss rates are fixed, the momentum flux ratio is determined by the terminal velocity ratio, i.e. \hbox{$\eta=V_{\rm w,1}/V_{\rm w,2}$}. 
		
		\citet{C16} conducted an analytic study in order to predict the mass of clumps formed in unstable stellar wind collisions. 
		That study focused on models whose parameters were motivated by the WR stars present in the Galactic Centre.  
		Here, we study a subsample of those models in order to make a direct comparisons between the simulations and that work. 
		Table~\ref{tab:models} presents the parameters of every model explored in this work. 
		We simulate six symmetric models with three different stellar separations \hbox{$a=210,66,21\rm\ au$}, and two wind speeds \hbox{$V_{\rm w}=500,750\rm\ km\ s^{-1}$}.  
		Additionally, we study two asymmetric models by fixing the stellar separation and the weaker wind speed to \hbox{$a=210\rm\ au$} and \hbox{$V_{\rm w,1}=500\rm\ km\ s^{-1}$}, and using \hbox{$V_{\rm w,2}=1000,1500\rm\ km\ s^{-1}$} for the stronger wind (\hbox{$\eta=0.5,0.33$}). 
		Finally, we run two more simulations, one with one level less and another with an extra level of refinement (up to three and five levels, respectively), for analysing the impact of resolution and convergence. 
		Figure~\ref{fig:chi} highlights the position of every stellar wind in the parameter space to determine in advance its radiative nature. 
		Notice that most of them are below the $\chi=1$ line, which means they correspond to radiative winds. 
		The fastest winds, A and A+, are located well above the transition line in the adiabatic regime.  
		Although the $B+$ wind is in the adiabatic wind region, the transition is not sharply defined, thus it is more accurate to refer to this region as a transition zone. 

		\begin{table}
		\begin{adjustbox}{width=\columnwidth,center}
		\begin{threeparttable}
            		\caption{Parameters of each model.}
            		\begin{tabular}{|l|l|l|r|l|r|c|}
			\hline
			Name	&	$\eta$	& $V_{\rm w,1},V_{\rm w,2}$	&	$a$	&	$t_{\rm cross}$	&	$\chi$	&	Max. Res.\\
					&			& $(\rm km\ s^{-1})$			&	(au)	&	(yr)			&			&	(cells)\\
			\hline	
			B10		&	1.0		& 500,500			&	210	&	4.0			&	0.321	&	$1024^3$\\
			B+10	&	1.0		& 750,750			&	210	&	2.67			&	1.627	&	$1024^3$\\
			C10		&	1.0		& 500,500			&	66	&	0.6			&	0.032	&	$1024^3$\\
			C+10	&	1.0		& 750,750			&	66	&	0.4			&	0.163	&	$1024^3$\\
			D10		&	1.0		& 500,500			&	21	&	0.2			&	0.003	&	$1024^3$\\
			D+10	&	1.0		& 750,750			&	21	&	0.13			&	0.016	&	$1024^3$\\
			\hline
			B9		&	1.0		& 500,500			&	210	&	4.0			&	0.321	&	$512^3$\\
			B11		&	1.0		& 500,500			&	210	&	4.0			&	0.321	&	$2048^3$\\
			\hline
			BA10	&	0.5		& 500,1000		&	210	&	2.0			&	6.069	&	$1024^3$\\
			BA+10	&	0.33		& 500,1500		&	210	&	1.33			&	32.805	&	$1024^3$\\
			\hline
		        	\end{tabular}
			\label{tab:models}
			\begin{tablenotes}
				\item \textit{Notes.} 
				The mass-loss rate of the stars is set to \hbox{$\dot{M}_{\rm w}=10^{-5}\rm\ M_{\odot}\ yr^{-1}$} in every model. 
				Column~1: ID of a single simulation run.
				Column~2: ratio of the momentum fluxes of the winds. 
				Column~3: stellar wind speed of each star in $\rm\ km\ s^{-1}$. 
				Column~4: stellar separation in astronomical units. 
				Column~5: wind crossing timescale defined as \hbox{$t_{\rm cross}=2a/V_{\rm w,1}$} in years. 
				Column~6: cooling parameter calculated from Equation~\ref{eq:chi}. 
				Column~7: maximum effective resolution of the simulation in number of cells.
			\end{tablenotes}
		\end{threeparttable}
		\end{adjustbox}
        		\end{table}	
		
		\begin{figure*}
	 		\centering
			\subfloat[][Model B10: density at $z=0$ and $t=1.2\rm\ yr$]{\includegraphics[width=0.4\textwidth]{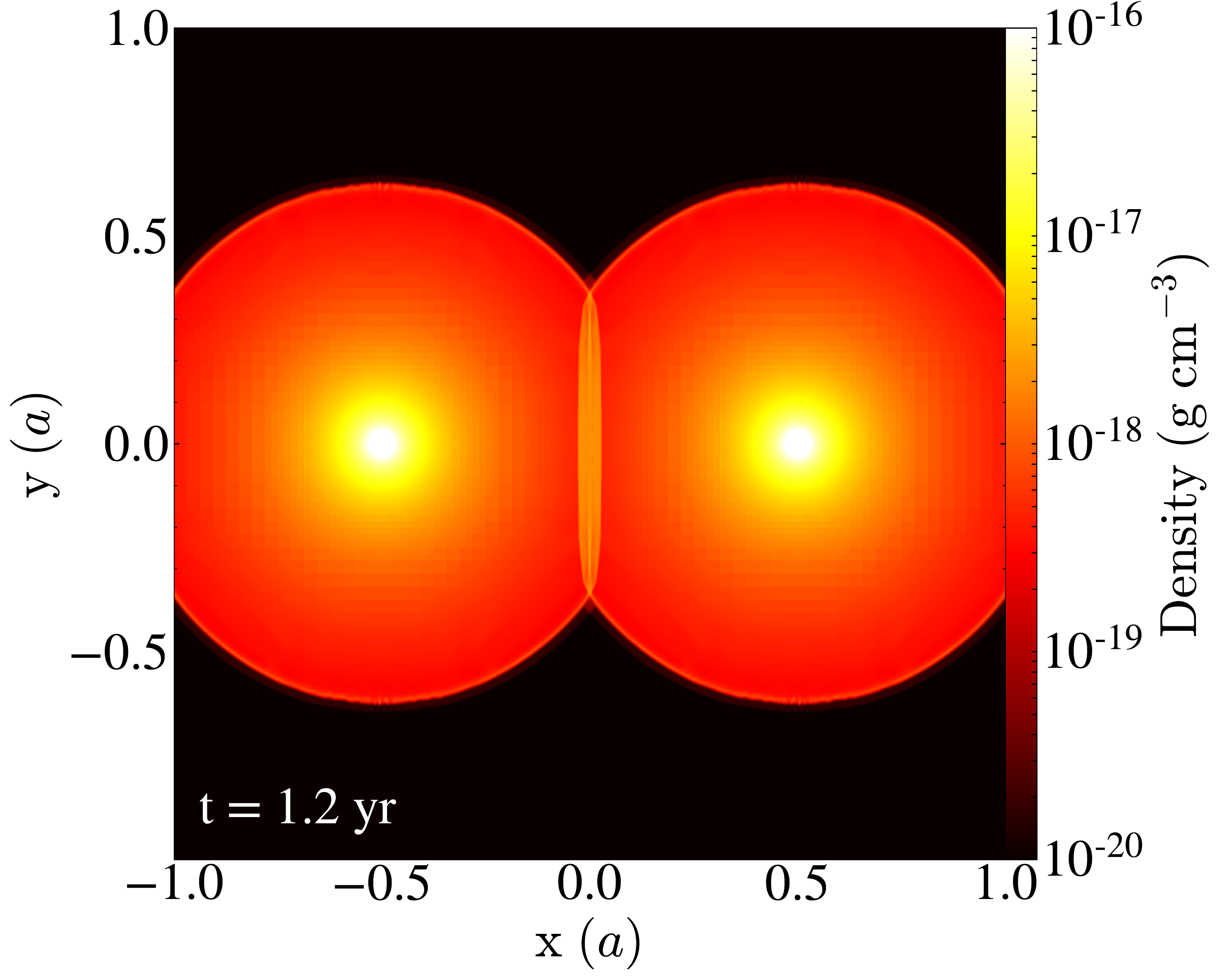}\label{fig:B10:1}}
			\hspace{0.25cm}
			\subfloat[][Model B10: density at $z=0$ and $t=2.6\rm\ yr$]{\includegraphics[width=0.4\textwidth]{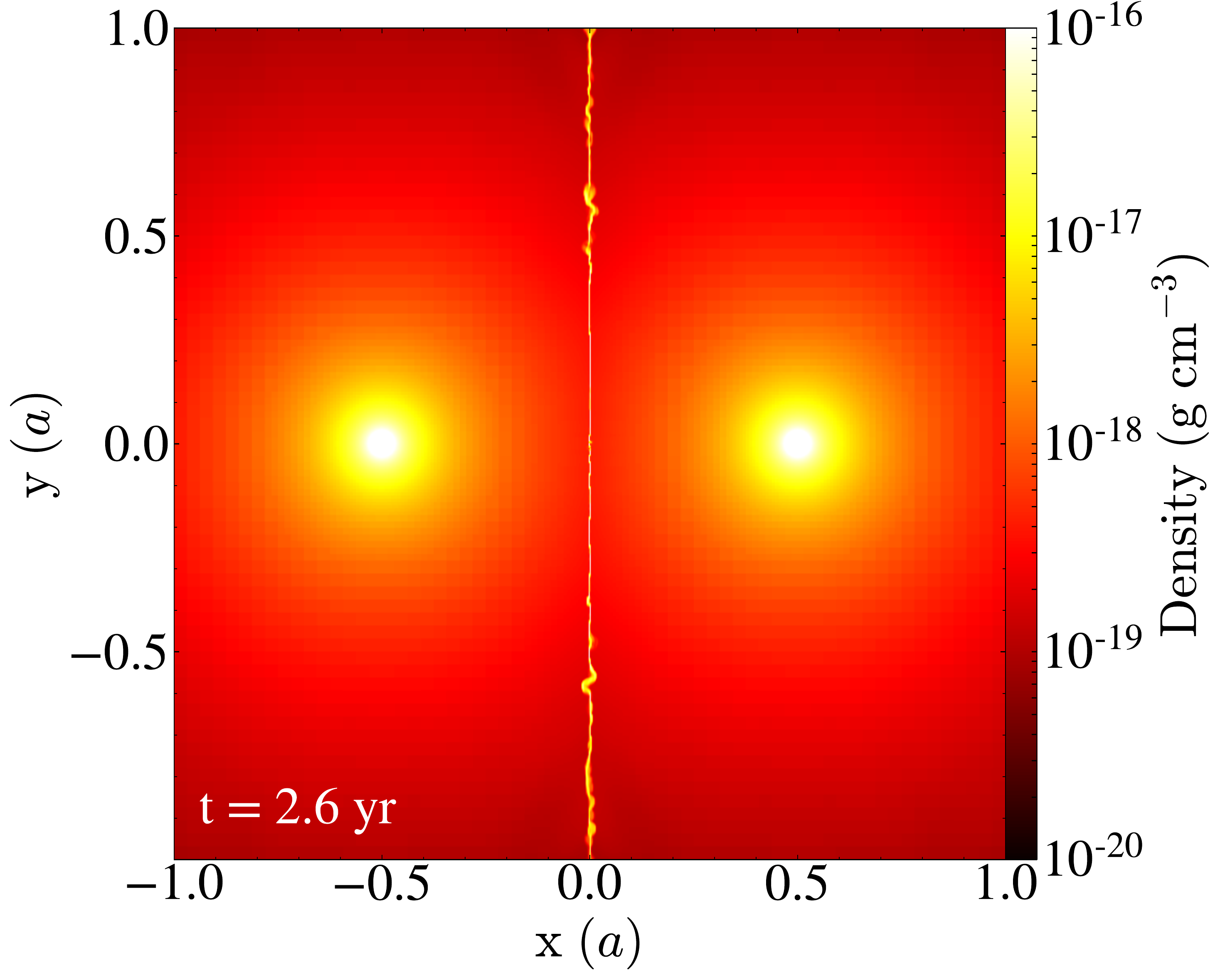}\label{fig:B10:2}}
			
			\subfloat[][Model B10: density at $z=0$ and $t=3.3\rm\ yr$]{\includegraphics[width=0.4\textwidth]{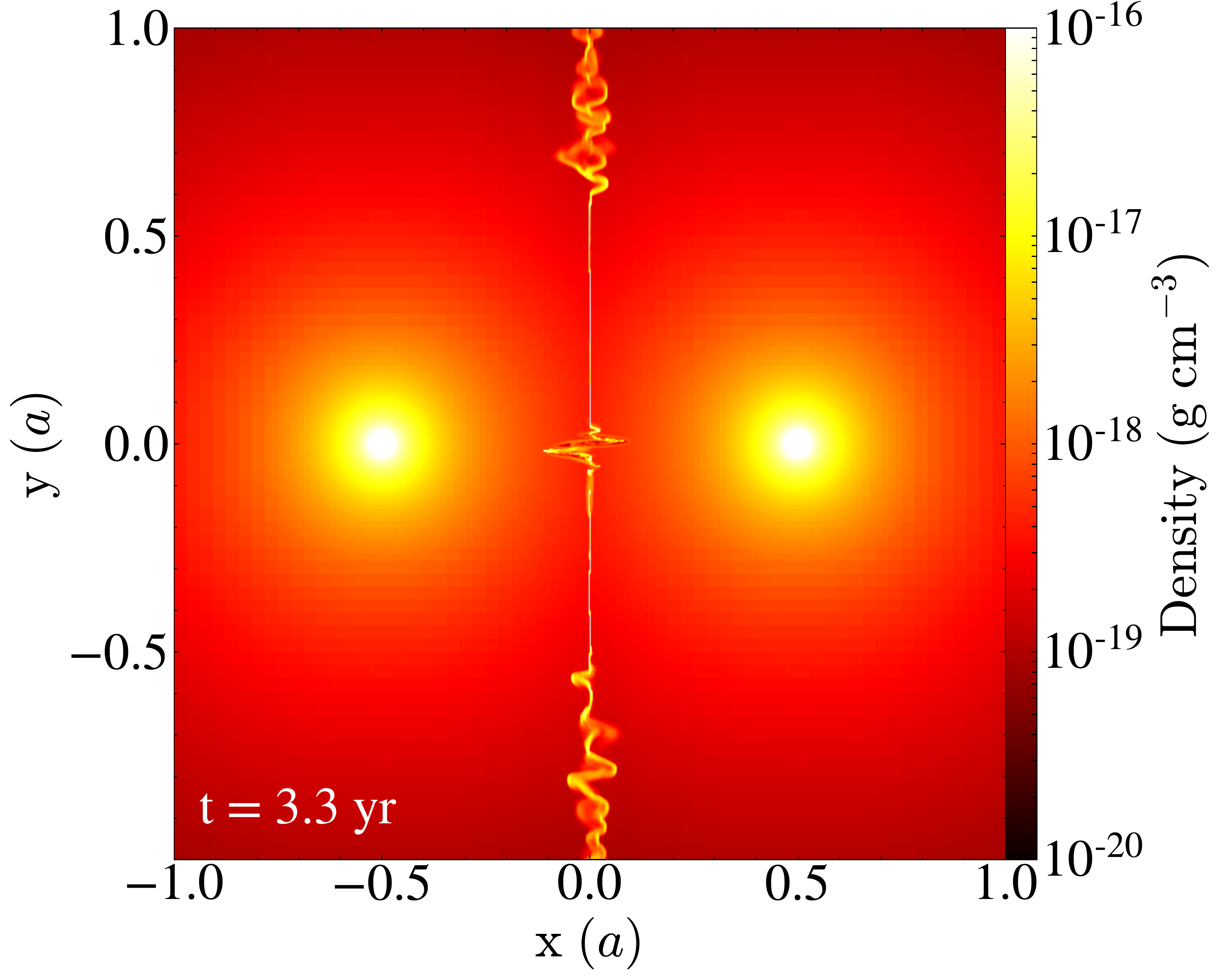}\label{fig:B10:3}}
			\hspace{0.25cm}
			\subfloat[][Model B10: density at $z=0$ and $t=11.2\rm\ yr$]{\includegraphics[width=0.4\textwidth]{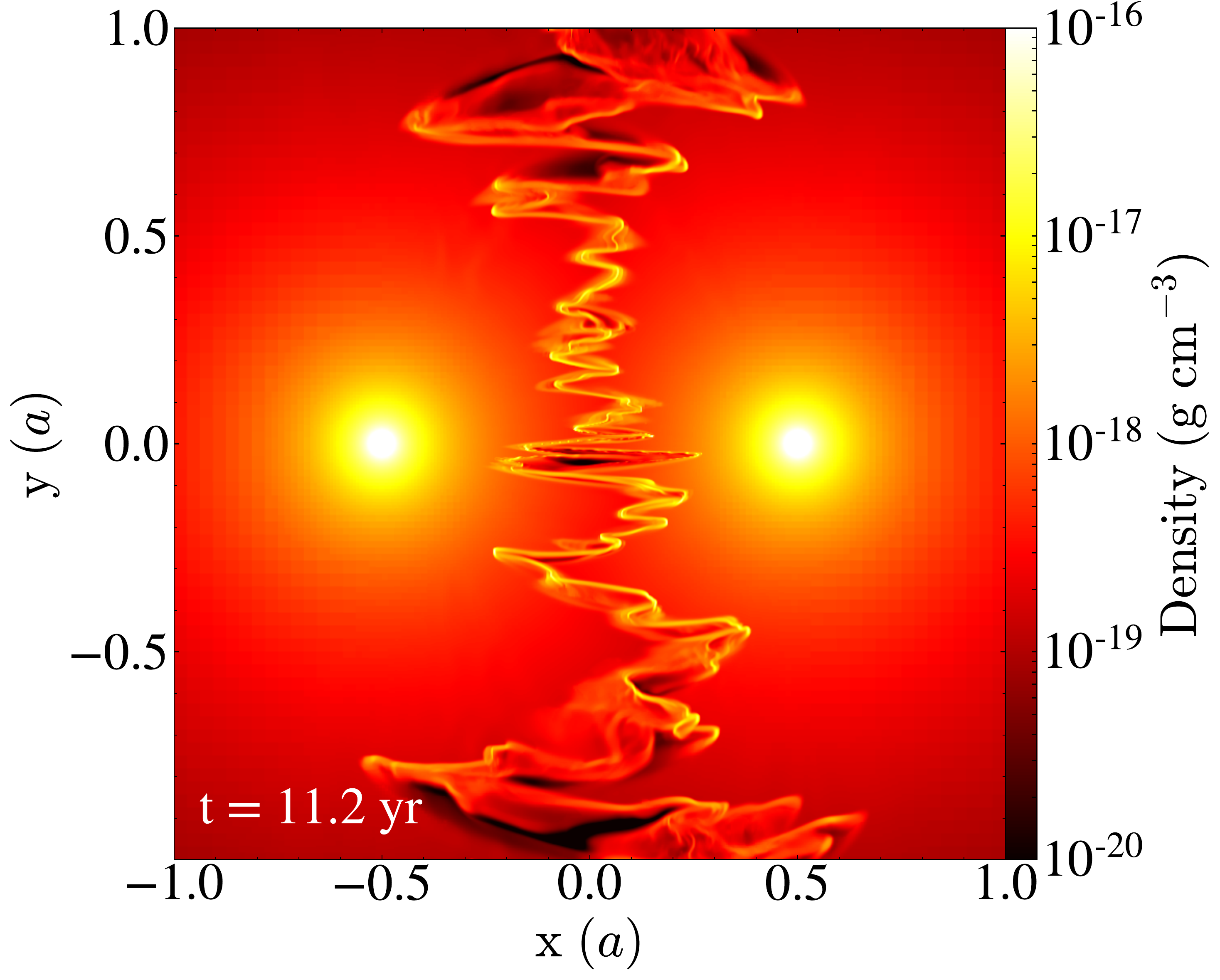}\label{fig:B10:4}}
			\caption{
			Density maps of cuts along $z=0$ plane of model B10. 
			Each panel shows different stages in the simulation. 
			Panel (a) shows the model at \hbox{$t=1.2\rm\ yr$} (\hbox{$0.3t_{\rm cross}$}) when the initial wind collision creates a thick dense, hot slab. 
			Panel (b) contains the system at \hbox{$t=2.6\rm\ yr$} (\hbox{$0.65t_{\rm cross}$}) highlighting the dense thin-shell formed after the slab cooled down. Here it is also possible to spot the first wiggles observed off-axis. 
			Panel (c) shows the simulation at \hbox{$t=3.3\rm\ yr$} (\hbox{$0.83t_{\rm cross}$}) when a roughly sinusoidal perturbation appears near the apex. This perturbation starts to shape the entire slab. 
			Panel (d) illustrates the system at \hbox{$t=11.2\rm\ yr$} (\hbox{$2.8t_{\rm cross}$}). At this point, the system is already in stationary state. Notice that the slab is completely unstable. 
			This Figure has an associated animation attached (Figure4\_B10\_density\_slice\_z.mov).}
			\label{fig:B10}
		\end{figure*}
		
		\begin{figure*}
	 			\centering
				\subfloat[][B10: density integrated along $z$-axis at $t=11.2\rm\ yr$]{\includegraphics[width=0.4\textwidth]{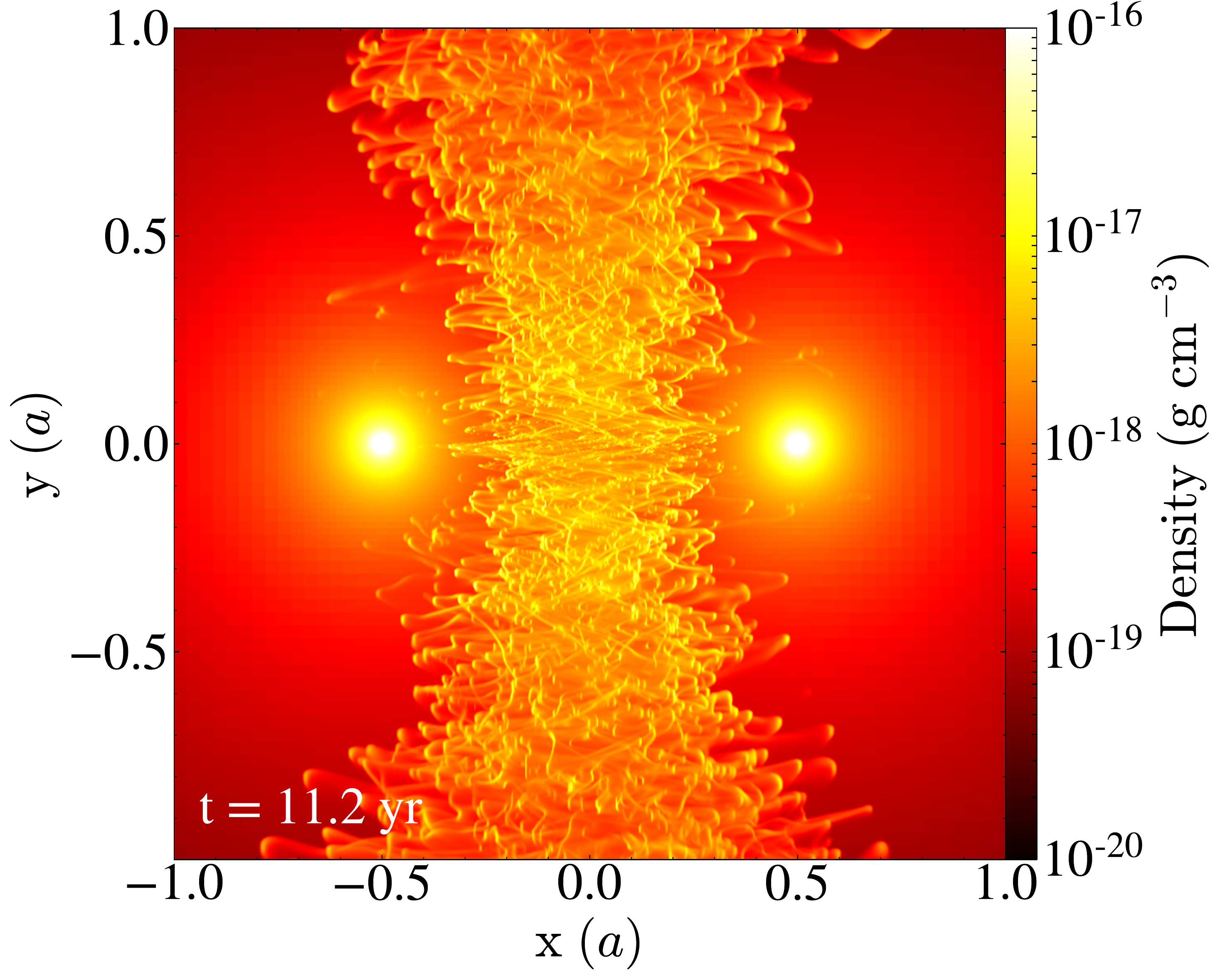}\label{fig:B10_3d:1}}
				\hspace{0.25cm}
				\subfloat[][B10: density integrated along $x$-axis at $t=11.2\rm\ yr$]{\includegraphics[width=0.4\textwidth]{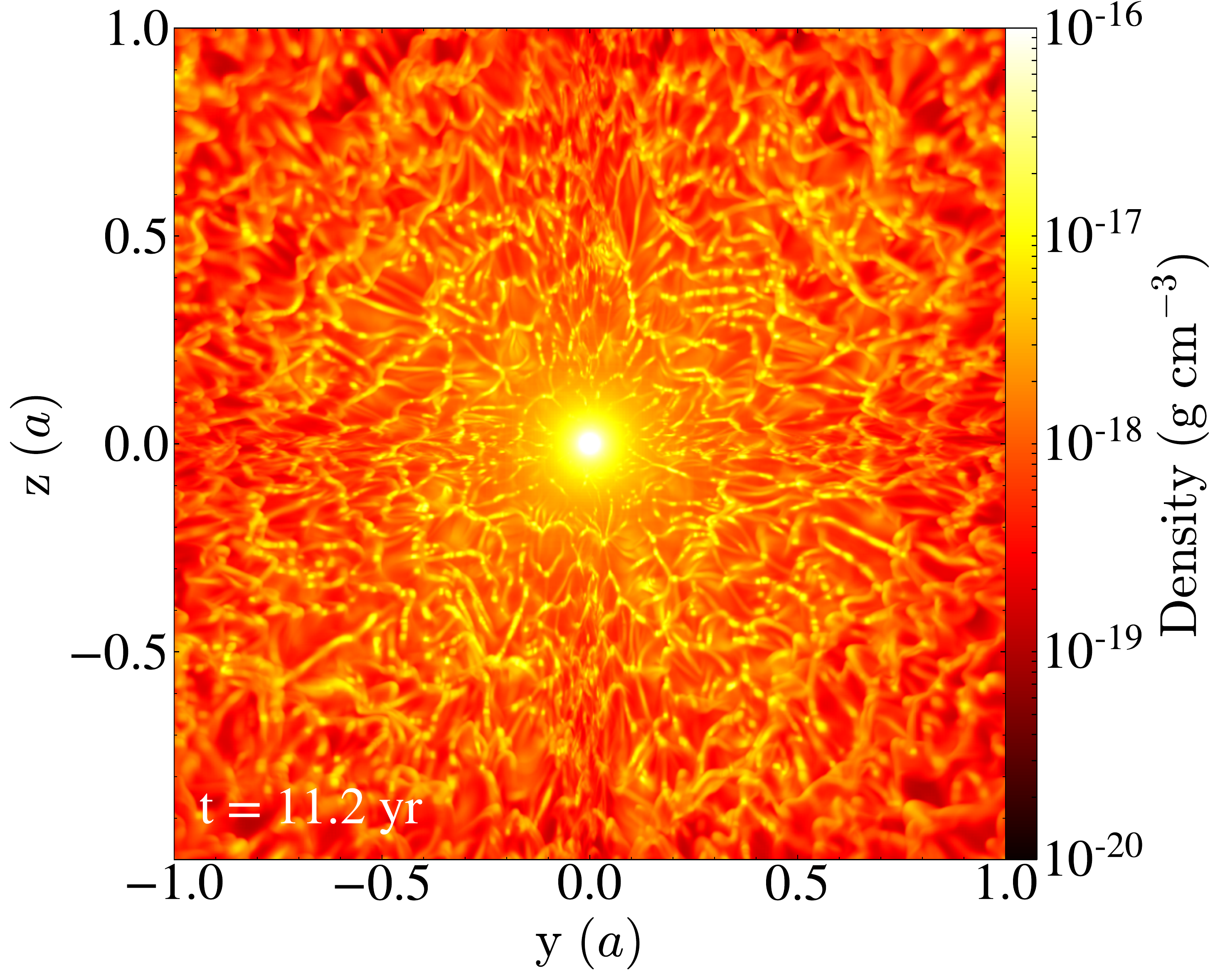}\label{fig:B10_3d:2}}
				\caption{Projected density maps of model B10 at \hbox{$t=11.2\rm\ yr$} (\hbox{$2.8t_{\rm cross}$}). 
				Panel (a) and (b) correspond to projections along the $z$- and $x$-axis, respectively. 
				The integrated density was calculated using the density field as weight so volumetric density values can be shown. 
				This quantity helps to highlight the dense gas, which corresponds to the cold material and to the most refined regions of the domain.  
				Notice that the simulation time is the same as in Figure~\ref{fig:B10:4}. 
				Panel (b) has an associated animation attached (Figure5b\_B10\_density\_projection\_x.mov).}
				\label{fig:B10_3d}
		\end{figure*}
		
		\begin{figure}
	 			\centering
				\subfloat[][B+10: $t=8\rm\ yr$]{\includegraphics[width=0.24\textwidth]{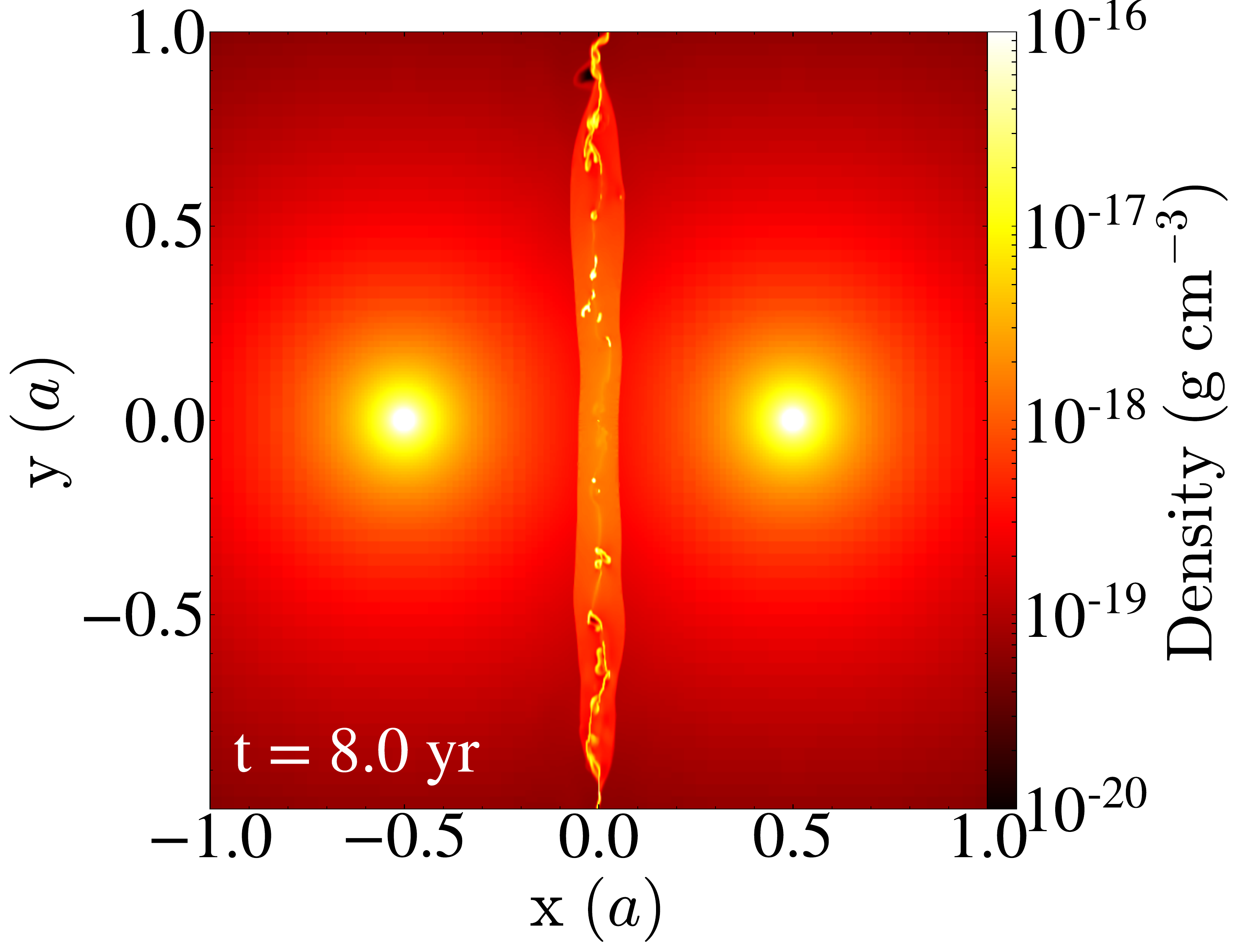}\label{fig:B+10:1}}
				\subfloat[][B+10: $t=34.8\rm\ yr$]{\includegraphics[width=0.24\textwidth]{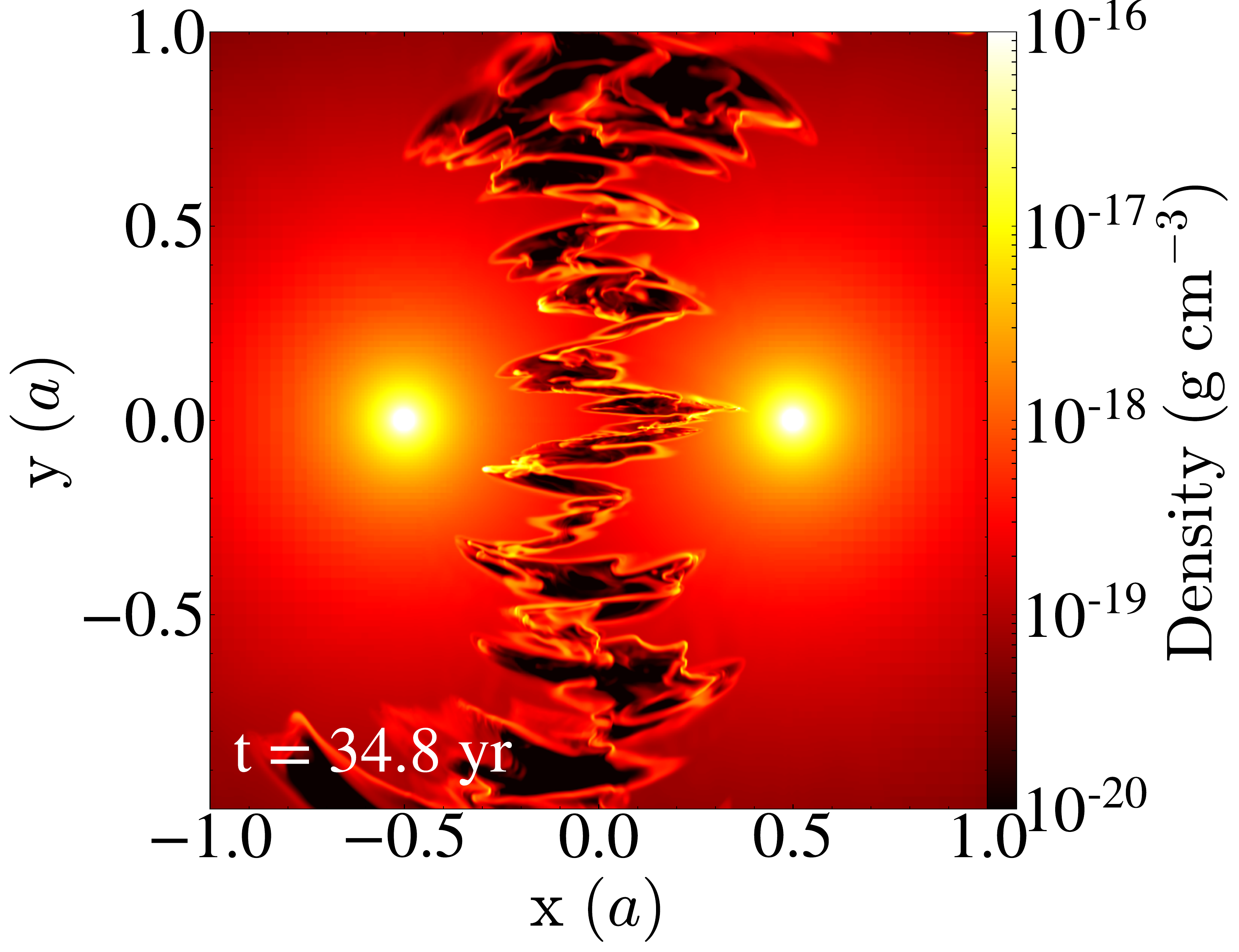}\label{fig:B+10:2}}
				\caption{
				Density maps on the $z=0$ plane of model B+10. 
				Notice that the left panel shows the evolution of the system at similar time compared to Figure~\ref{fig:B10:4} (\hbox{$t\approx3t_{\rm cross}$}).
				However, given that in this case the winds are faster the slab has not cooled down yet. 
				The right panel presents the evolution of the system at \hbox{$t=34.8\rm\ yr$} (\hbox{$13t_{\rm cross}$}) 
				Notice that in this case the instability looks more violent compared to the model B10. 
				In particular, the density contrast in this model is starker. 
				This Figure has an associated animation attached (Figure6\_B+10\_density\_slice\_z.mov).
				}
				\label{fig:B+10}
		\end{figure}
		
		\begin{figure}
	 			\centering
				\subfloat[][C10: $t=2.2\rm\ yr$]{\includegraphics[width=0.24\textwidth]{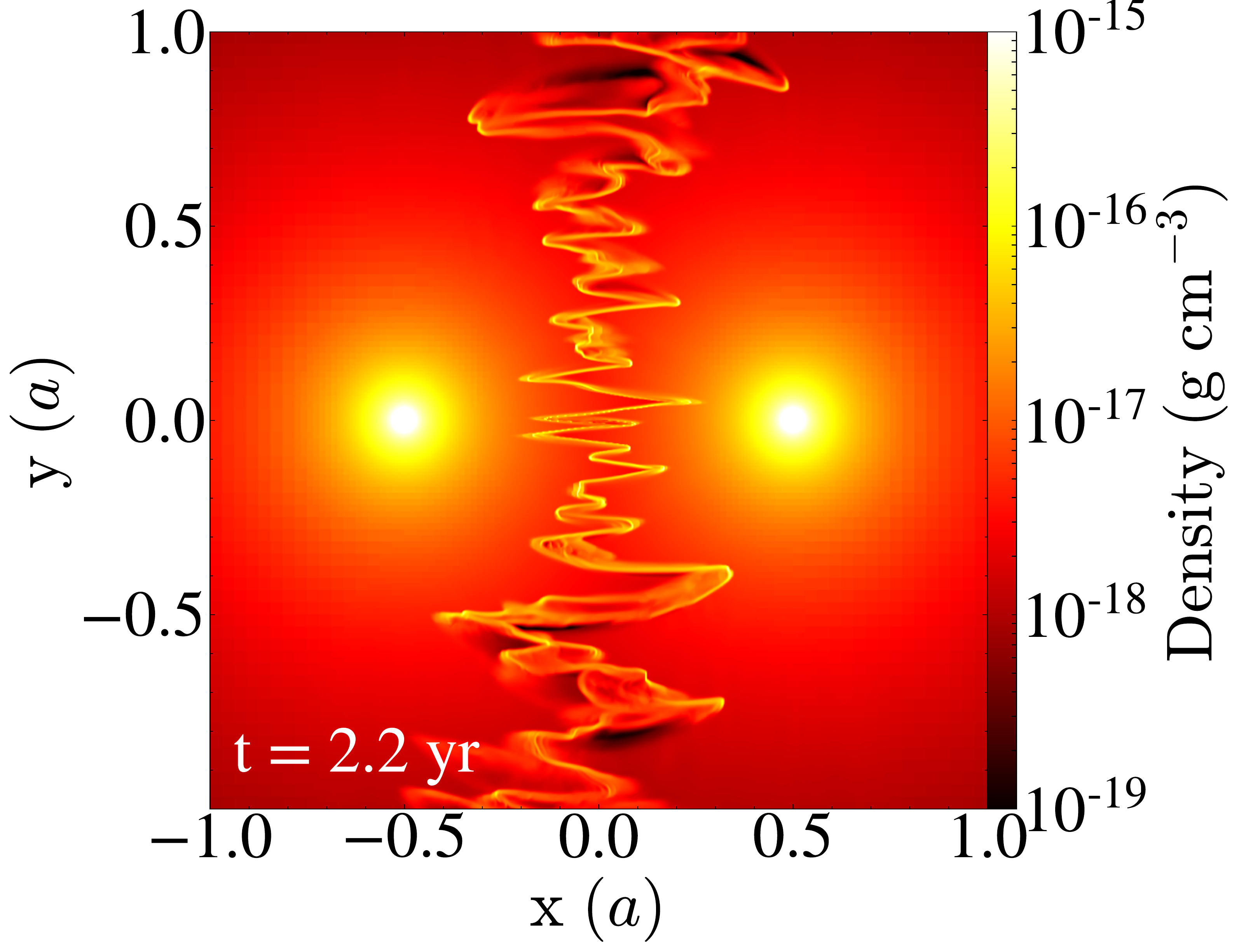}\label{fig:CC+:1}}
				\subfloat[][C+10: $t=2.2\rm\ yr$]{\includegraphics[width=0.24\textwidth]{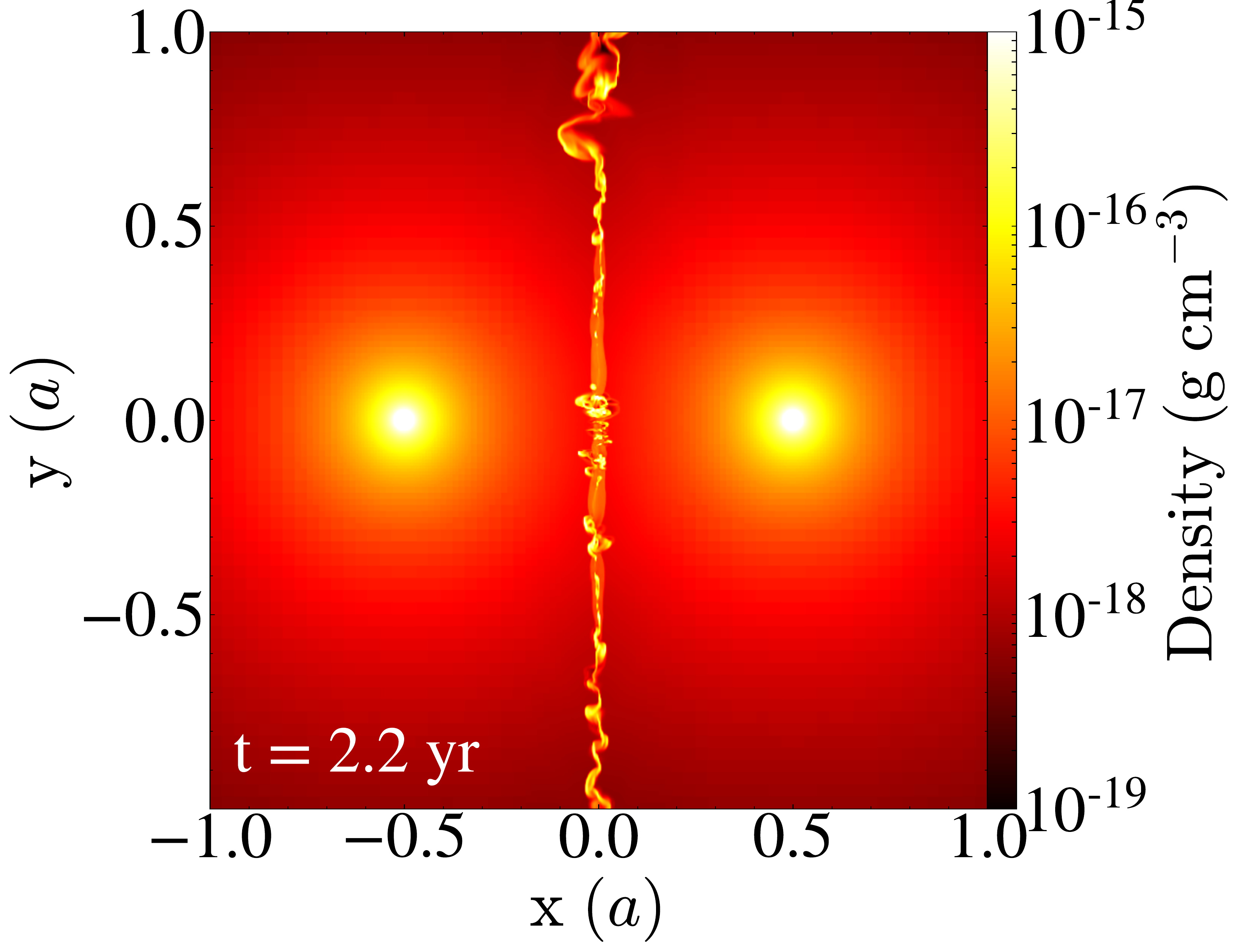}\label{fig:CC+:2}}
				\caption{
				Density maps at $z=0$ plane of models C10 and C+10 in panels (a) and (b), respectively. 
				Both show the state of the system at exactly the same simulation time \hbox{$t=2.2\rm\ yr$}. 
				Notice that in the model C10 the instability has already developed in the entire slab. 
				Meanwhile, in the model C+10 the instability is developing but has not reached stationary state yet.
				}
				\label{fig:CC+}
		\end{figure}
		
			\begin{figure*}
				\subfloat[][B10: $t=1.2\rm\ yr$]{\includegraphics[width=0.25\textwidth]{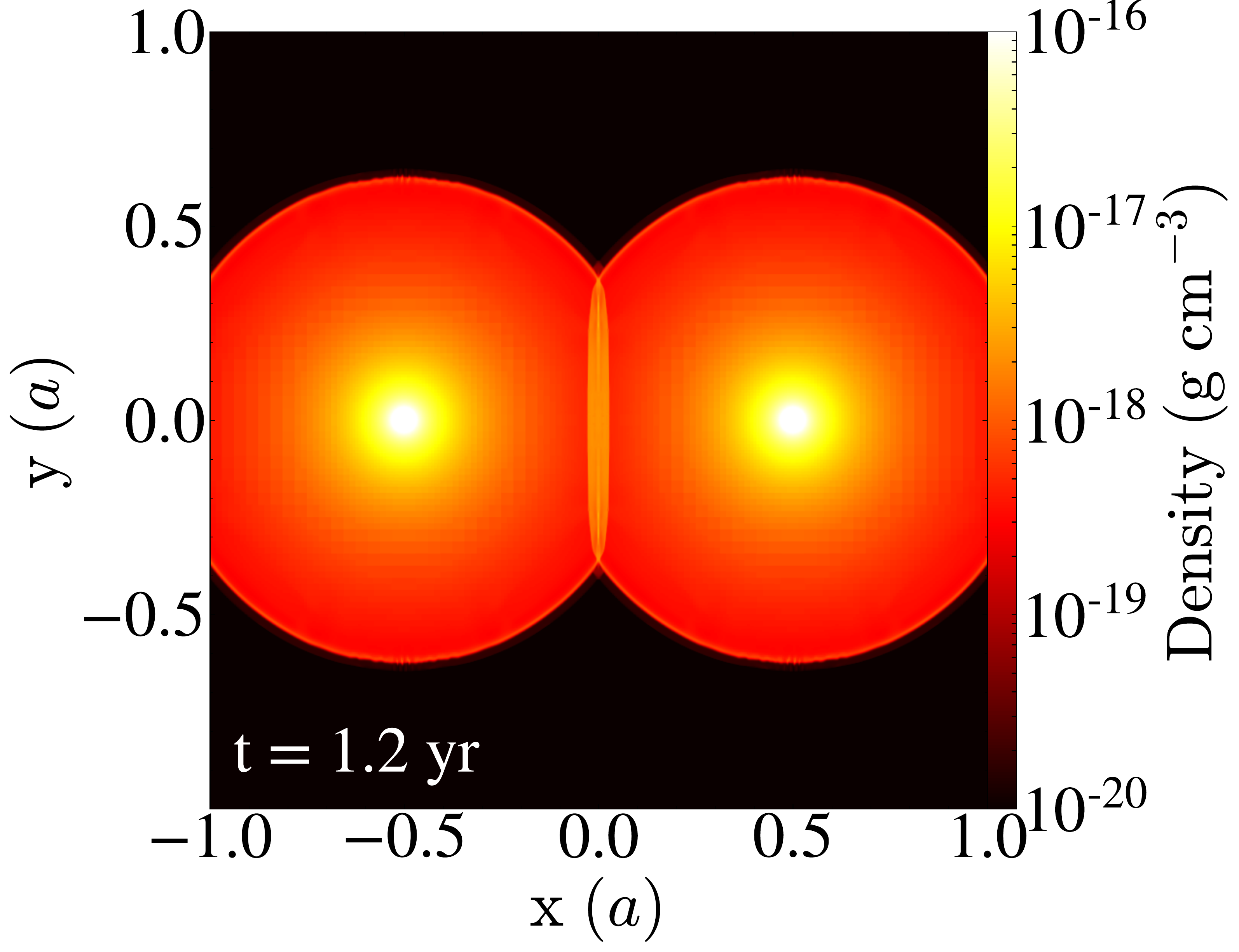}\label{fig:evol:B10-1}}
				\subfloat[][B10: $t=2.6\rm\ yr$]{\includegraphics[width=0.25\textwidth]{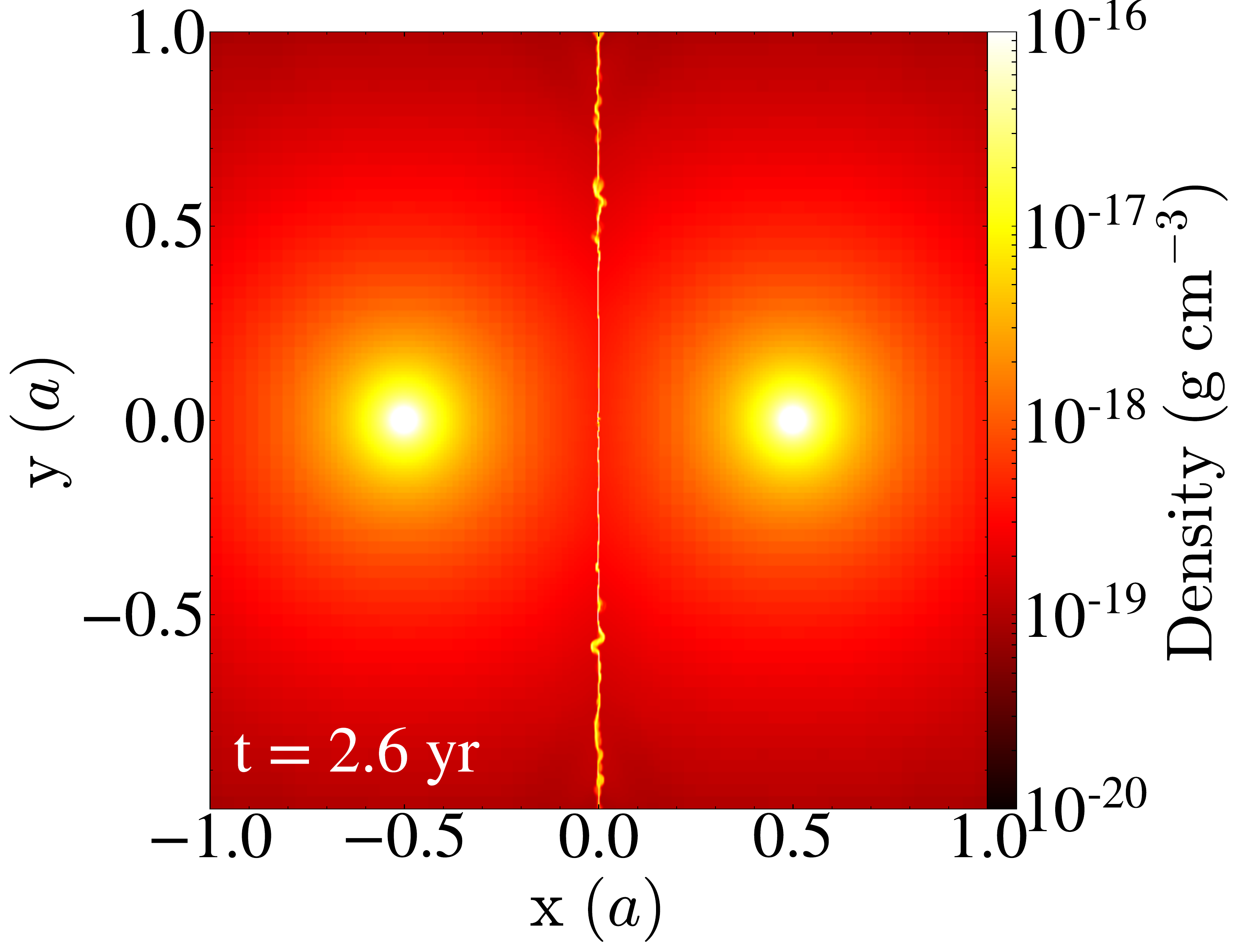}\label{fig:evol:B10-2}}
				\subfloat[][B10: $t=3.3\rm\ yr$]{\includegraphics[width=0.25\textwidth]{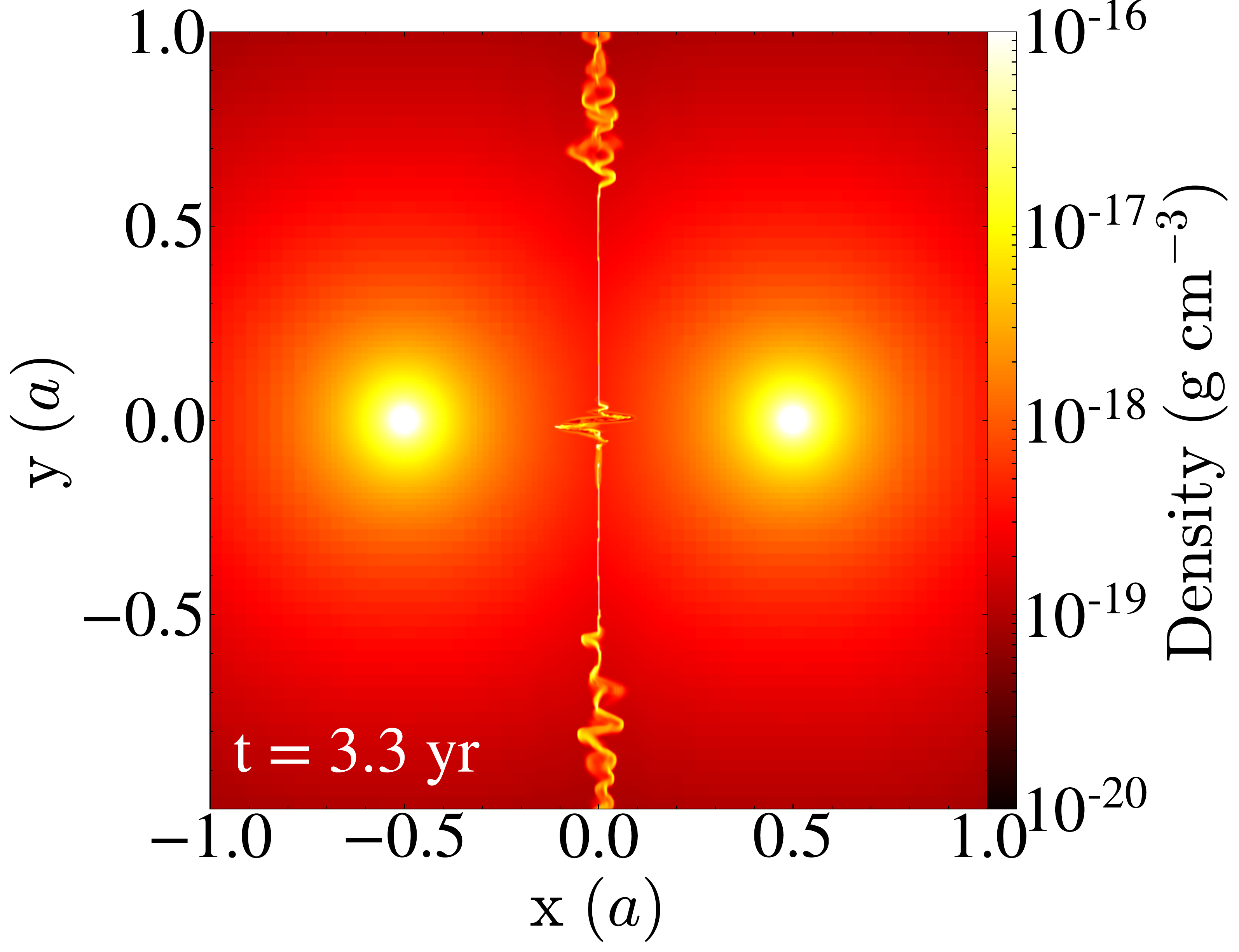}\label{fig:evol:B10-3}}
				\subfloat[][B10: $t=11.2\rm\ yr$]{\includegraphics[width=0.25\textwidth]{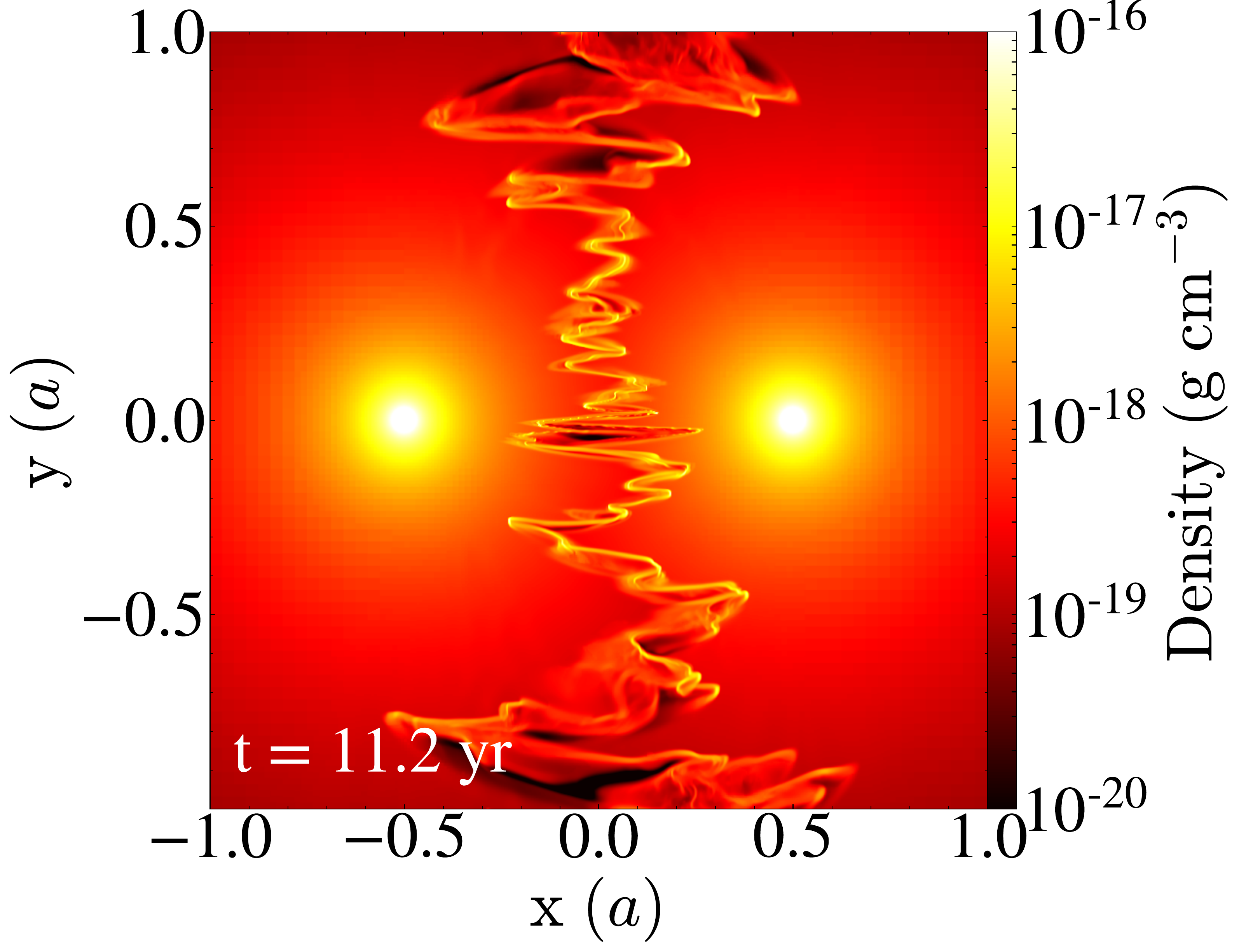}\label{fig:evol:B10-4}}
				\vspace{-0.25cm}
				\subfloat[][BA10: $t=1.2\rm\ yr$]{\includegraphics[width=0.25\textwidth]{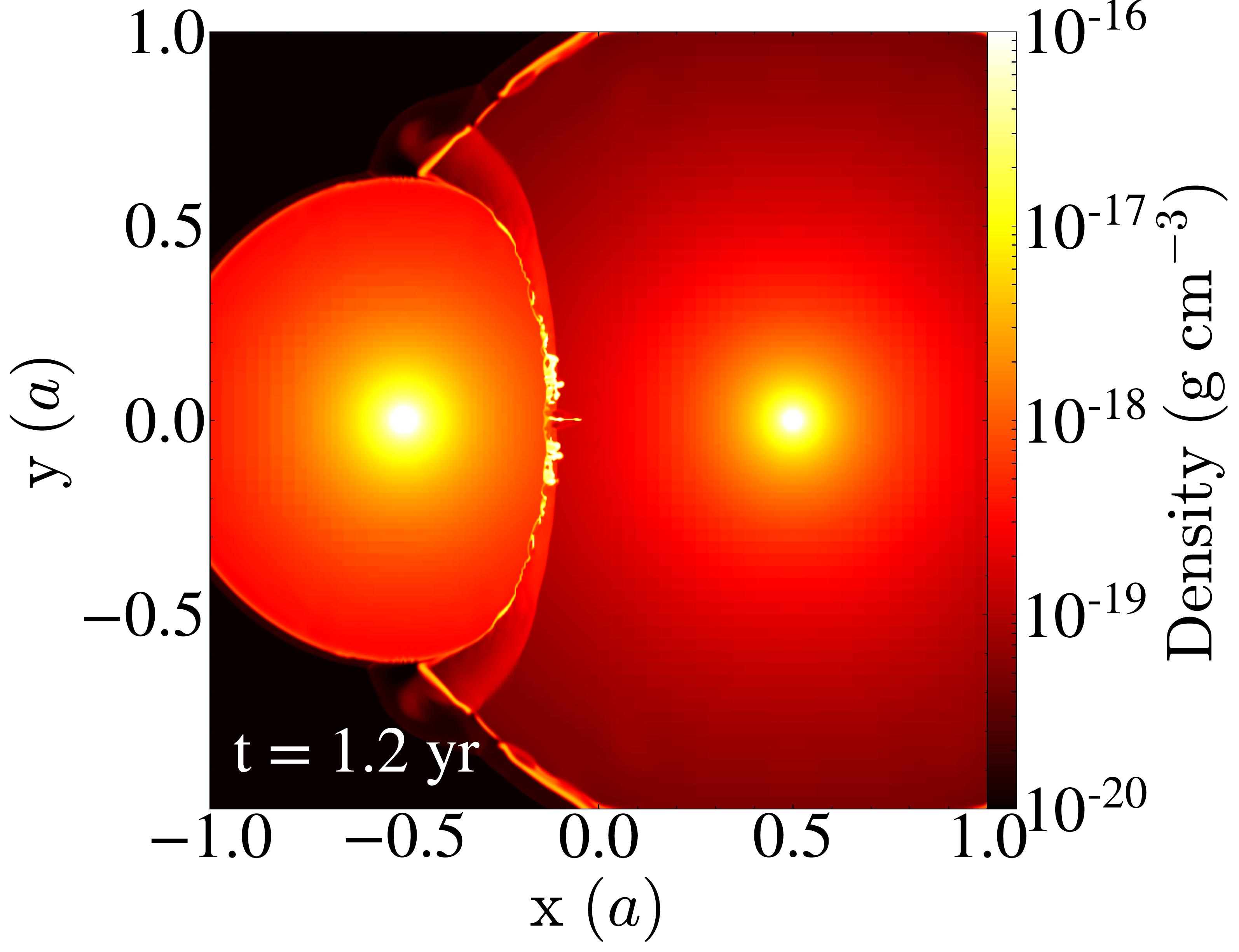}\label{fig:evol:BA10-1}}
				\subfloat[][BA10: $t=2.6\rm\ yr$]{\includegraphics[width=0.25\textwidth]{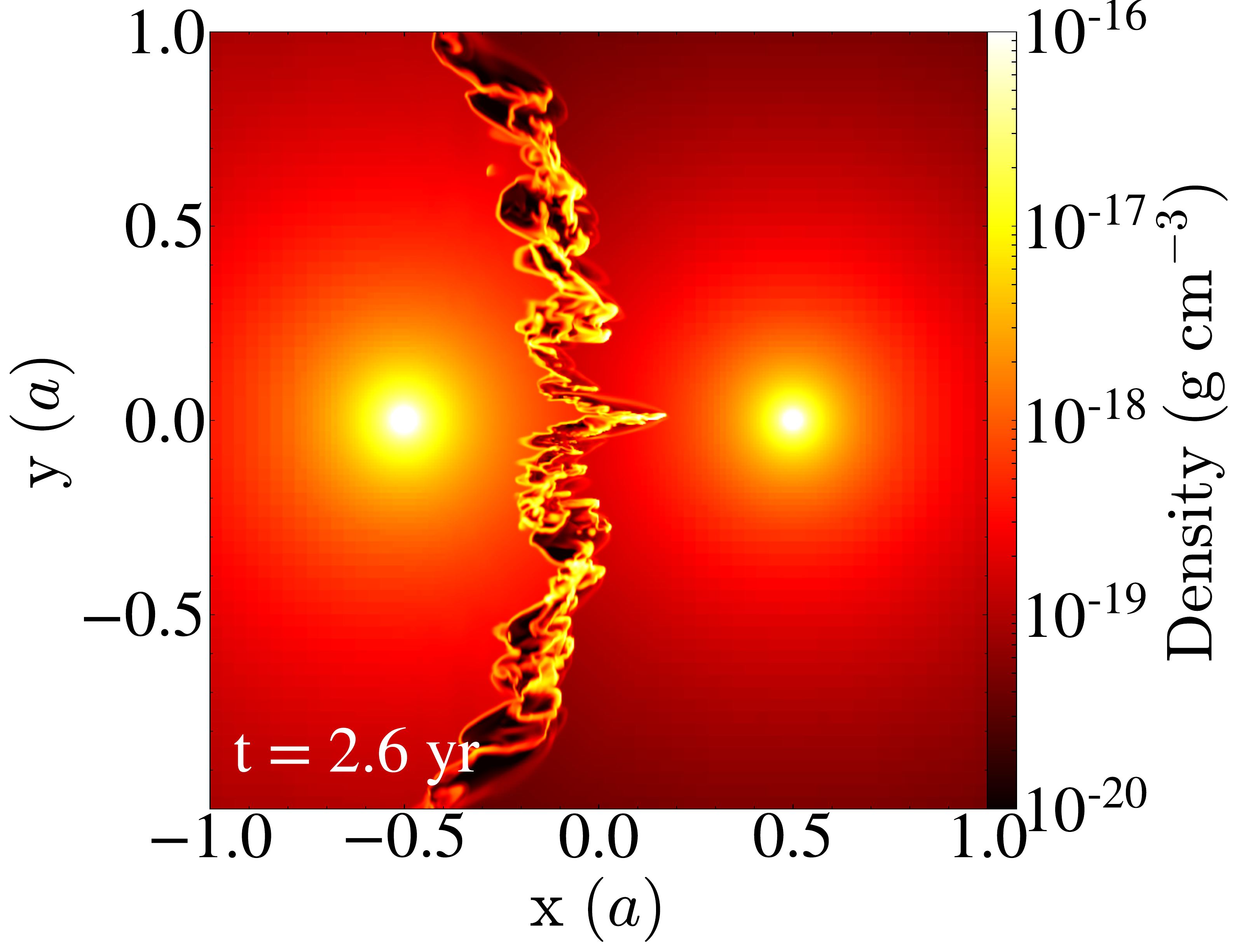}\label{fig:evol:BA10-2}}
				\subfloat[][BA10: $t=3.3\rm\ yr$]{\includegraphics[width=0.25\textwidth]{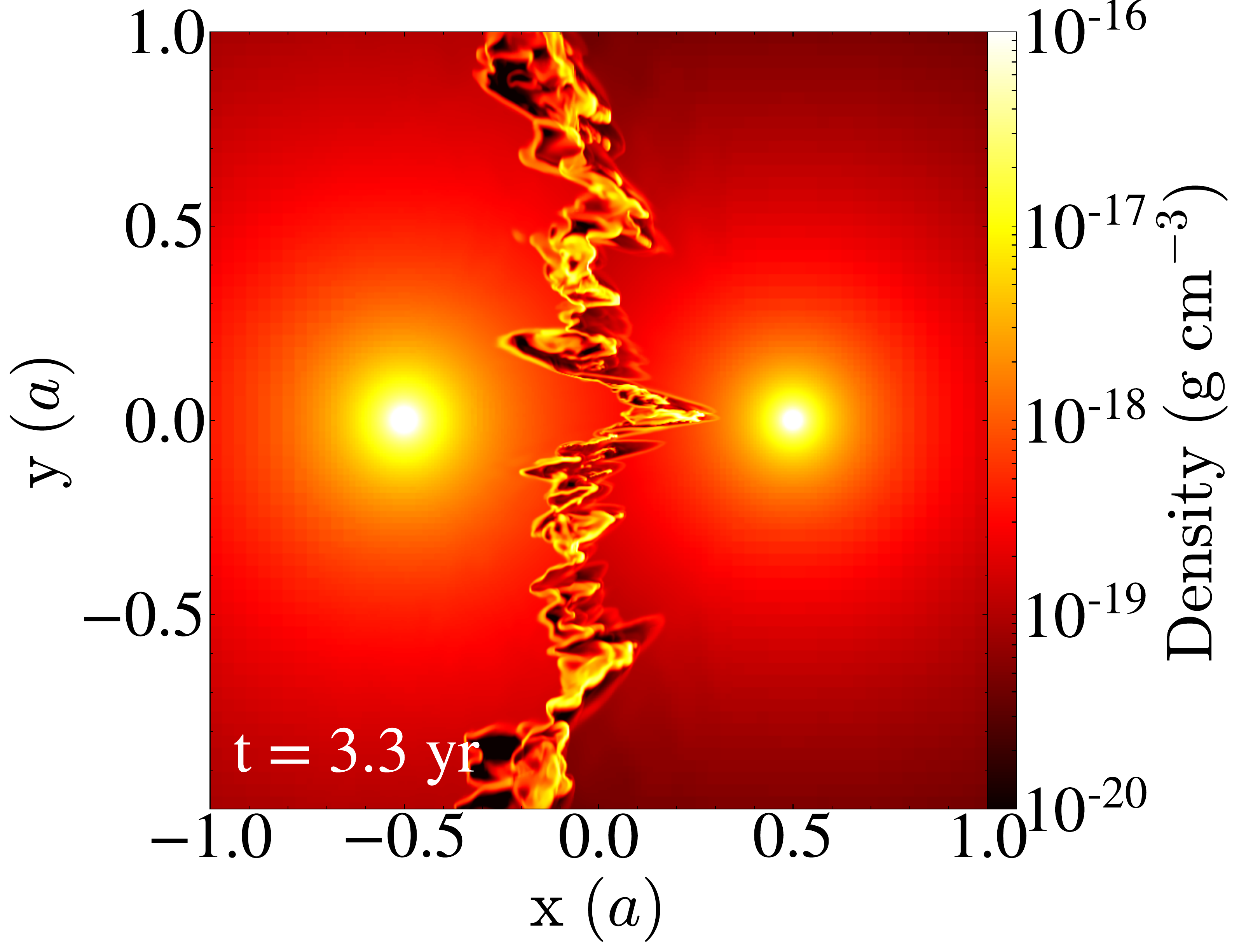}\label{fig:evol:BA10-3}}
				\subfloat[][BA10: $t=11.2\rm\ yr$]{\includegraphics[width=0.25\textwidth]{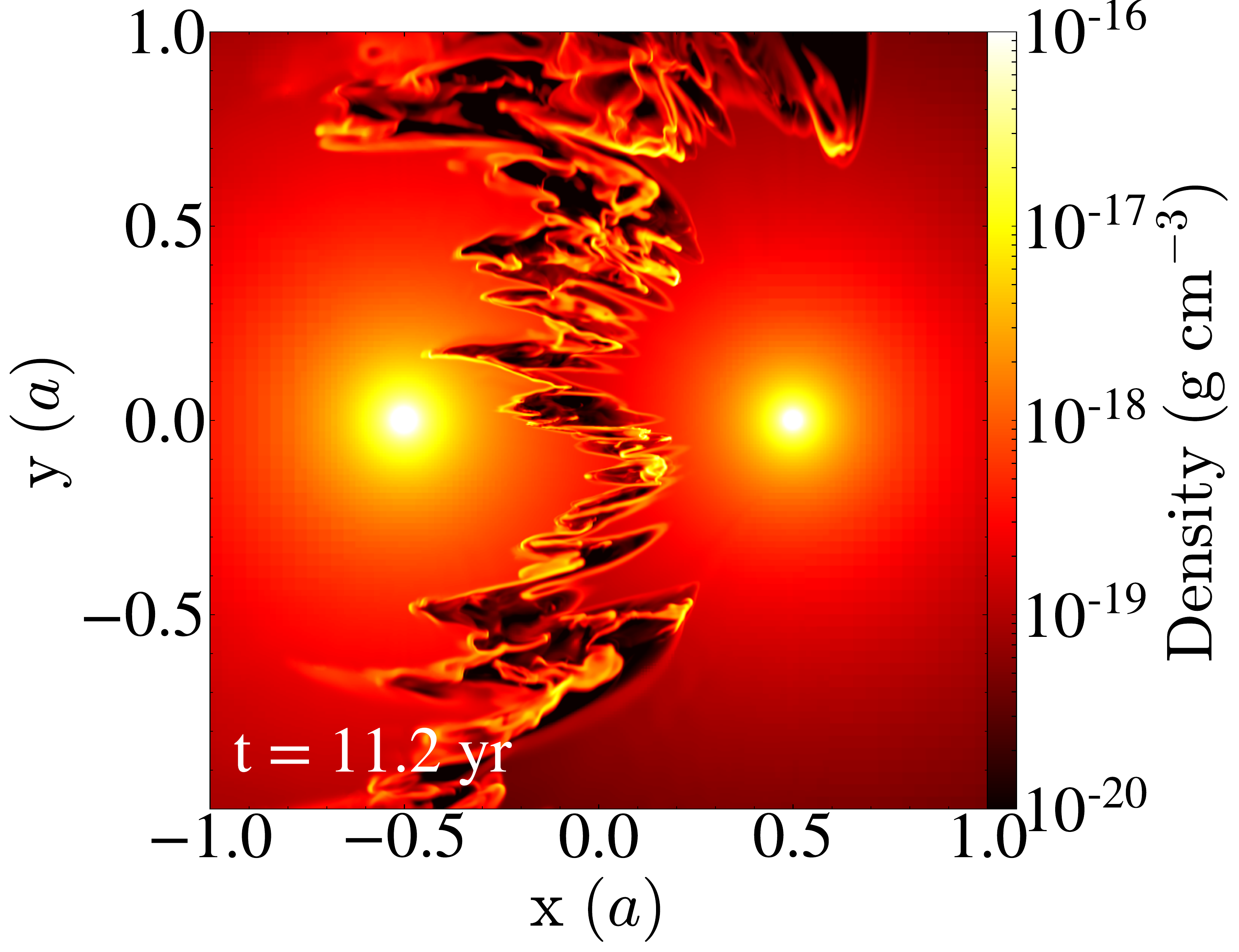}\label{fig:evol:BA10-4}}
				\vspace{-0.25cm}
				\subfloat[][BA+10: $t=1.2\rm\ yr$]{\includegraphics[width=0.25\textwidth]{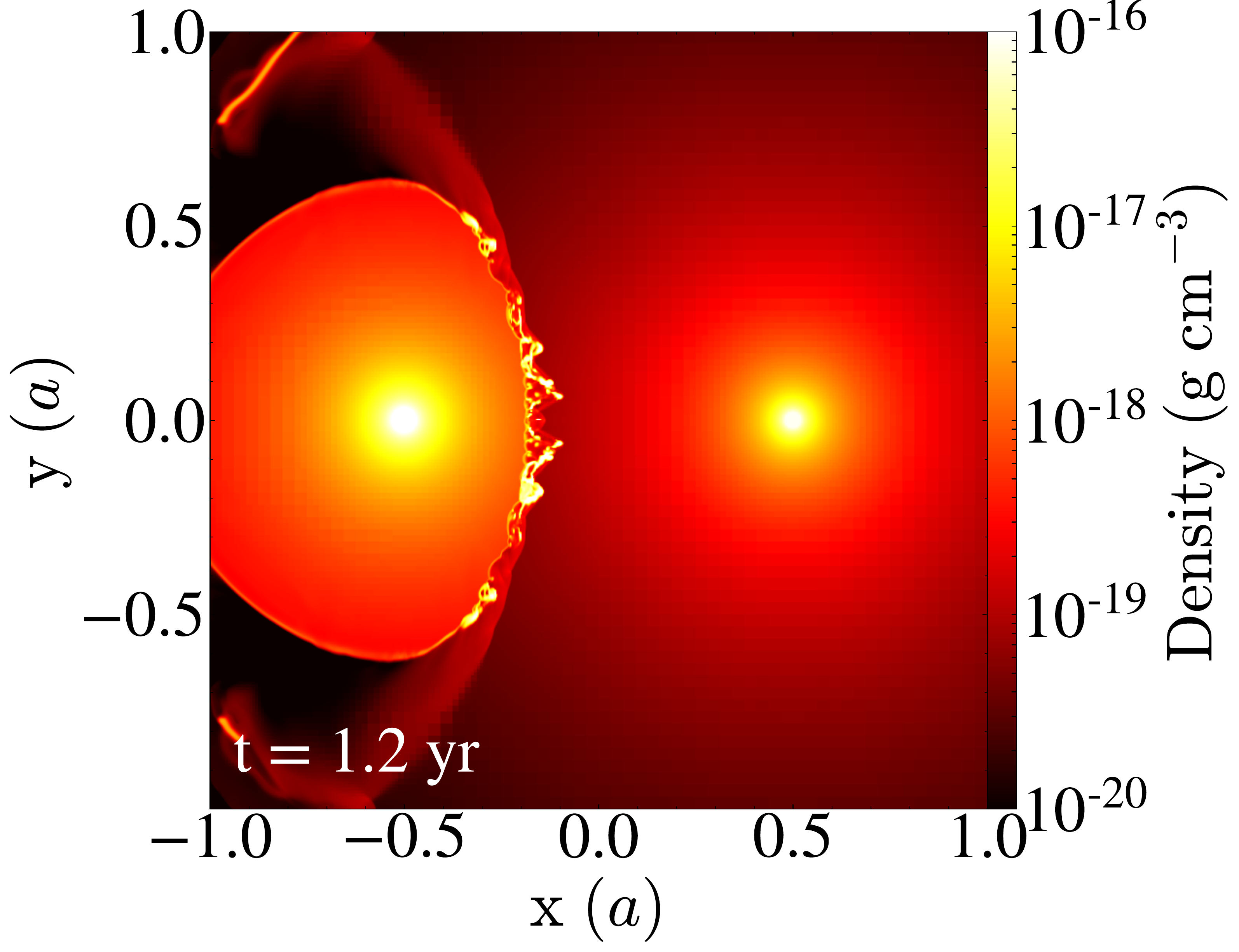}\label{fig:evol:BA+10-1}}
				\subfloat[][BA+10: $t=2.6\rm\ yr$]{\includegraphics[width=0.25\textwidth]{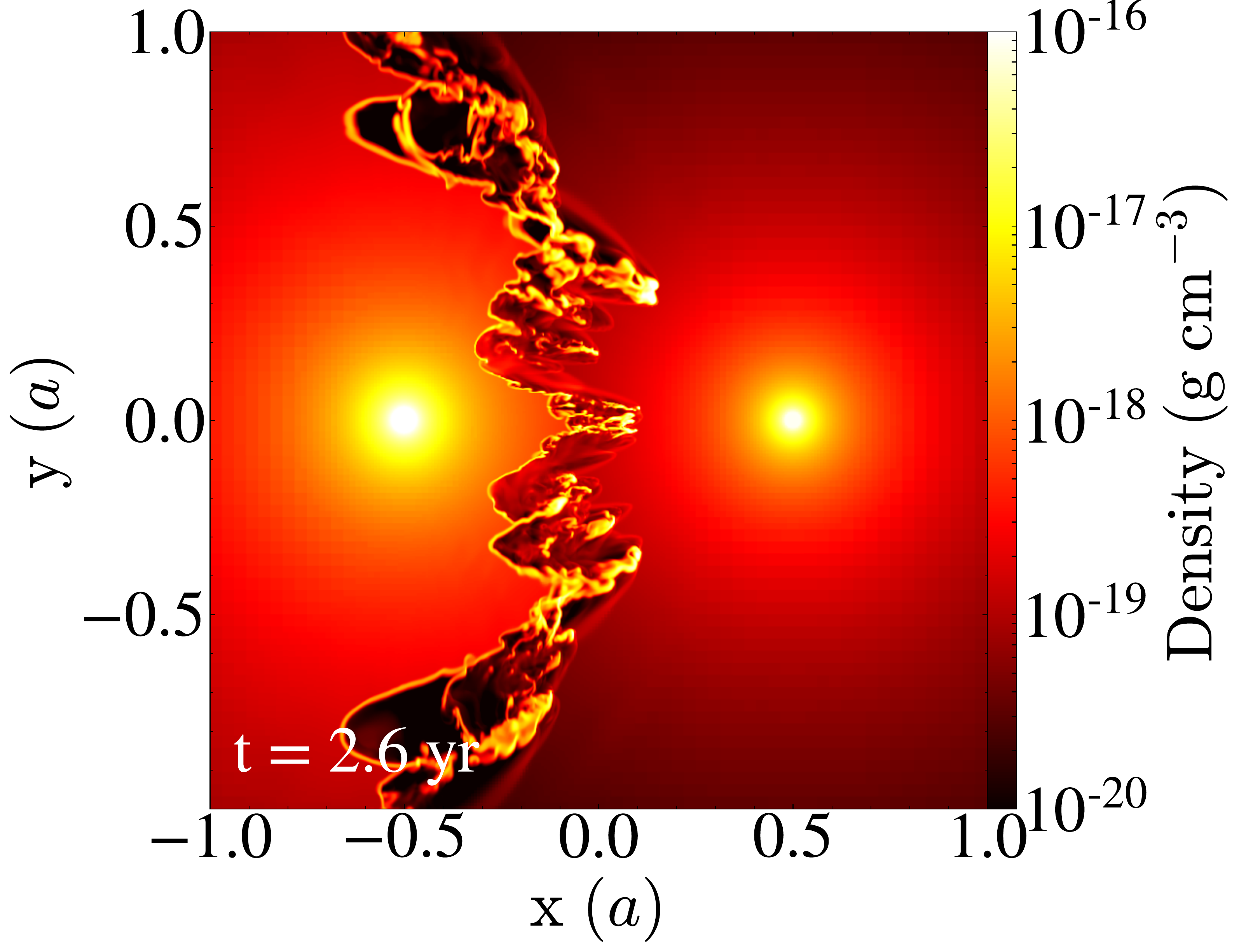}\label{fig:evol:BA+10-2}}
				\subfloat[][BA+10: $t=3.3\rm\ yr$]{\includegraphics[width=0.25\textwidth]{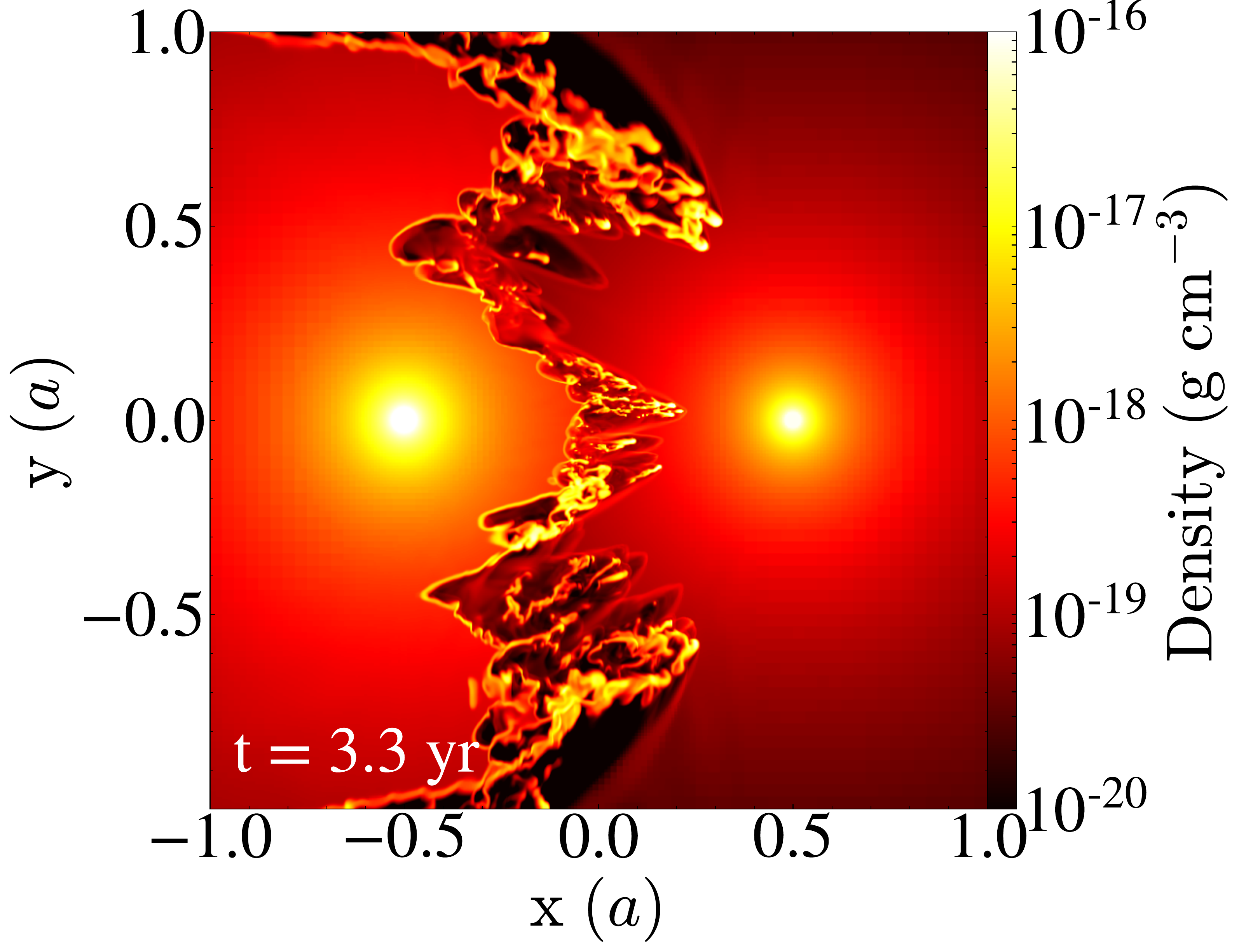}\label{fig:evol:BA+10-3}}
				\subfloat[][BA+10: $t=11.2\rm\ yr$]{\includegraphics[width=0.25\textwidth]{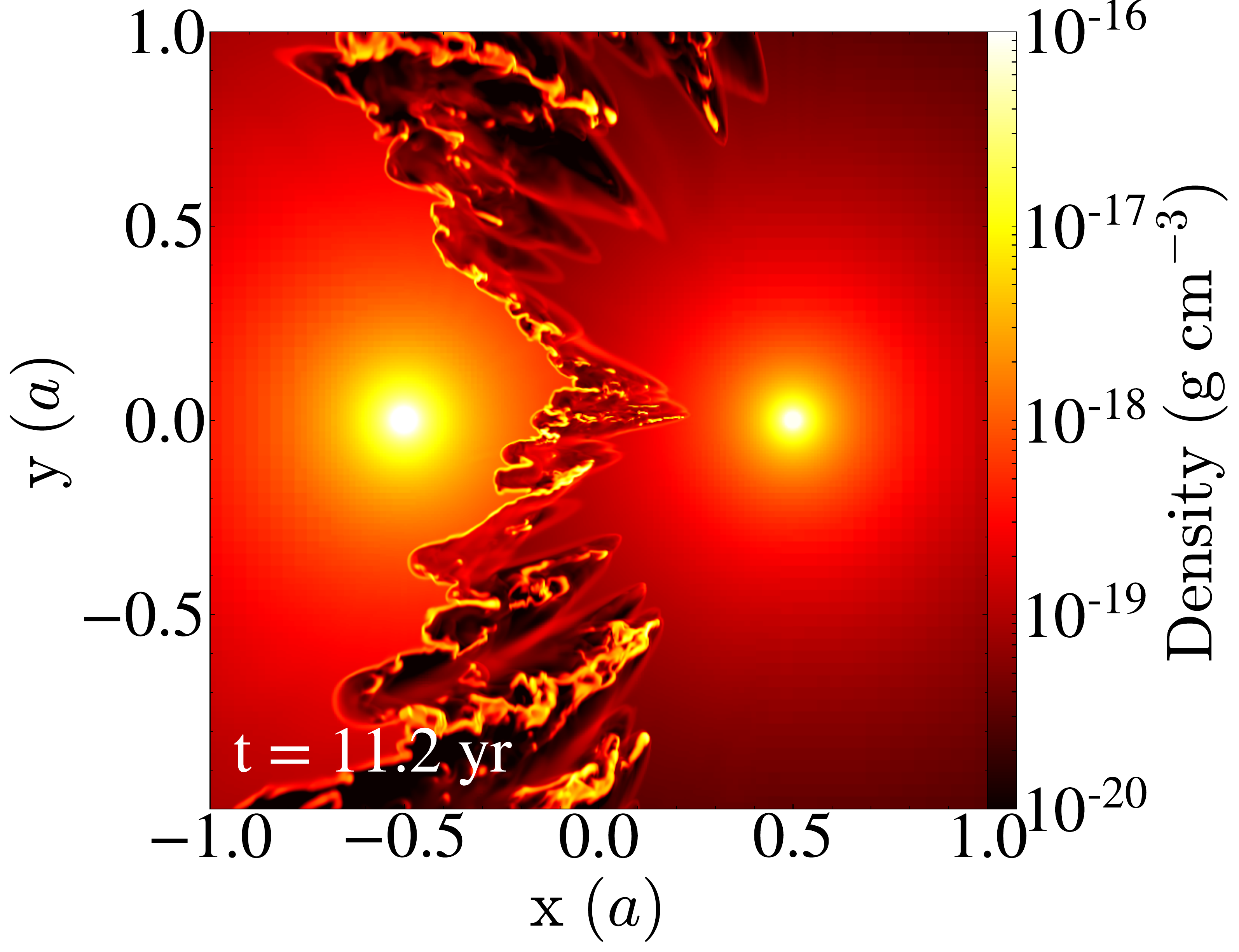}\label{fig:evol:BA+10-4}}
				\caption{
				Density maps at $z=0$ of different models at the same simulation time. 
				Rows stand for models B10 (upper), BA10 (central), and BA+10 (lower). 
				It is important to remark that these models consider the same domain size and resolution.
				From left to right, columns represent simulation times \hbox{$t=1.2\rm\ yr$}, \hbox{$t=2.6\rm\ yr$}, \hbox{$t=3.3\rm\ yr$}, and \hbox{$t=11.2\rm\ yr$}. 
				Notice that the more different the wind speed is the faster the slab becomes unstable. 
				Furthermore, a large velocity difference produces more violent instabilities excited in the slab. 
				Models B10 (upper row), BA10 (central row), and BA+10 (lower row) have associated animations attached (Figure6\_B10\_density\_slice\_z.mov, Figure8efgh\_BA10\_density\_slice\_z.mov, and Figure8ijkl\_BA+10\_density\_slice\_z.mov, respectively).}
				\label{fig:asym}
			\end{figure*}
			
			\begin{figure*}
				\centering
				\subfloat[][BA10: $\rho(x,y,z=0)$]{\includegraphics[width=0.4\textwidth]{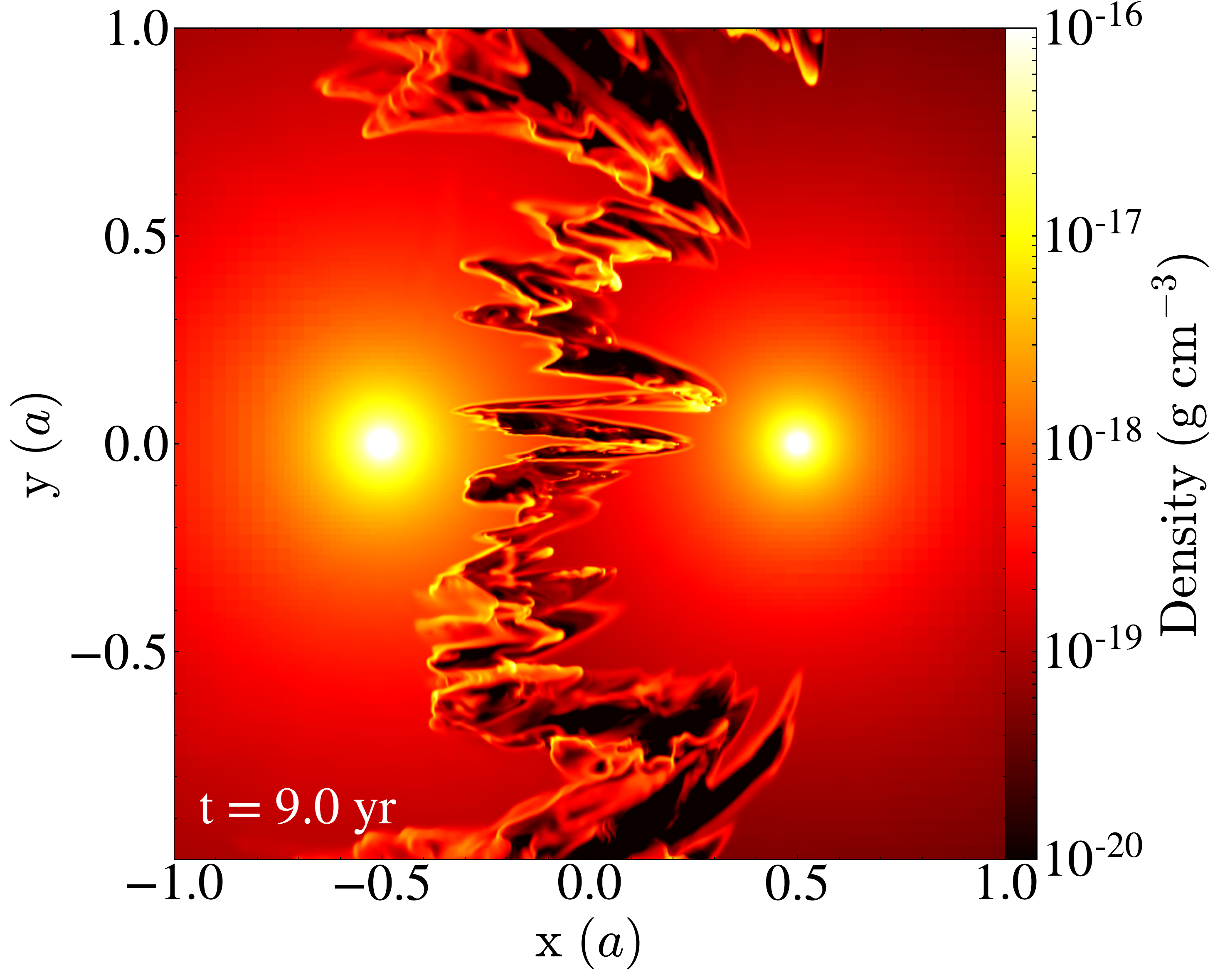}\label{fig:BA10-xz:1}}
				\hspace{0.25cm}
				\subfloat[][BA10: $\rho(x={-}0.09a,y,z)$]{\includegraphics[width=0.4\textwidth]{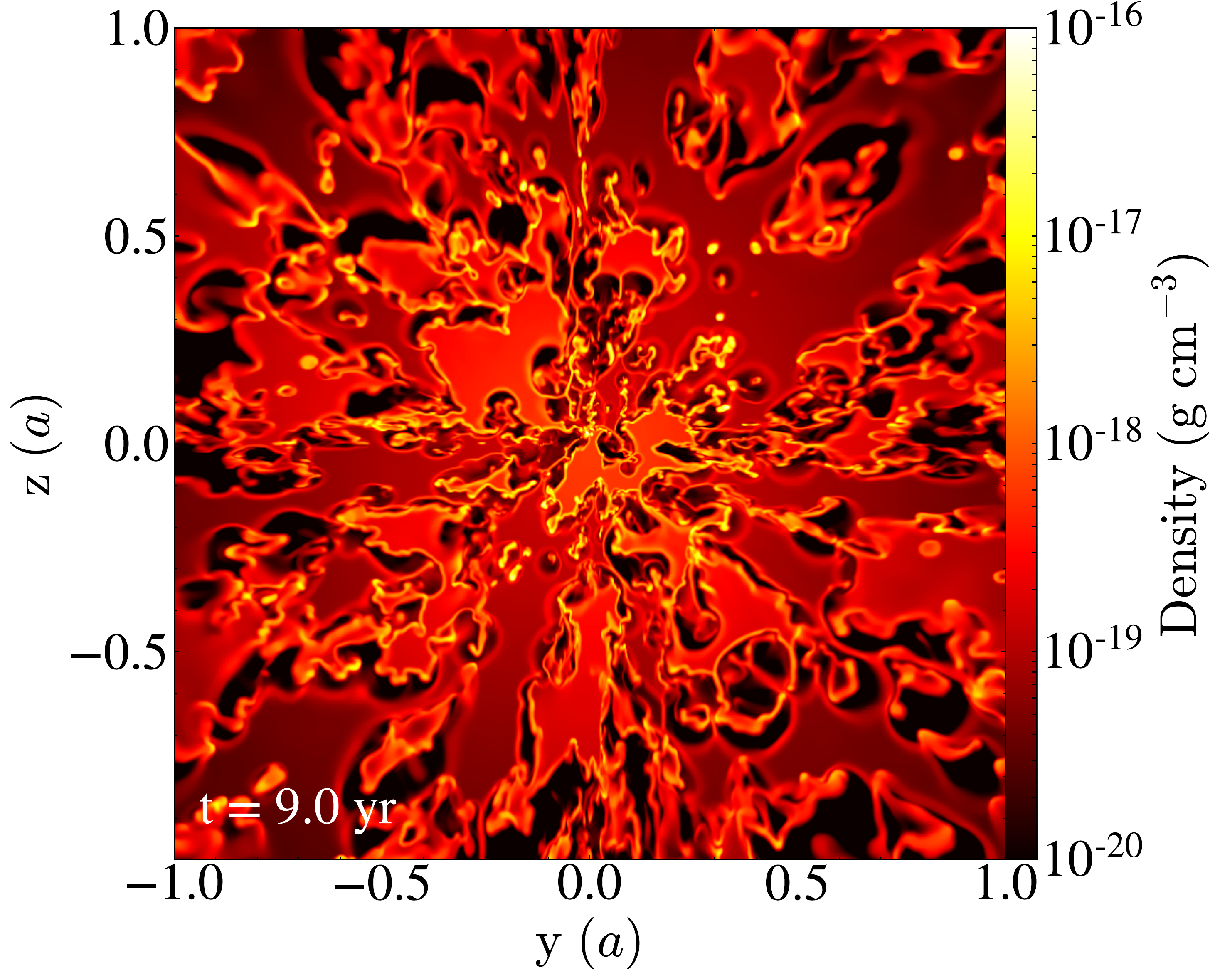}\label{fig:BA10-xz:2}}
				\vspace{-0.25cm}
				\subfloat[][BA10:  $T(x,y,z=0)$]{\includegraphics[width=0.4\textwidth]{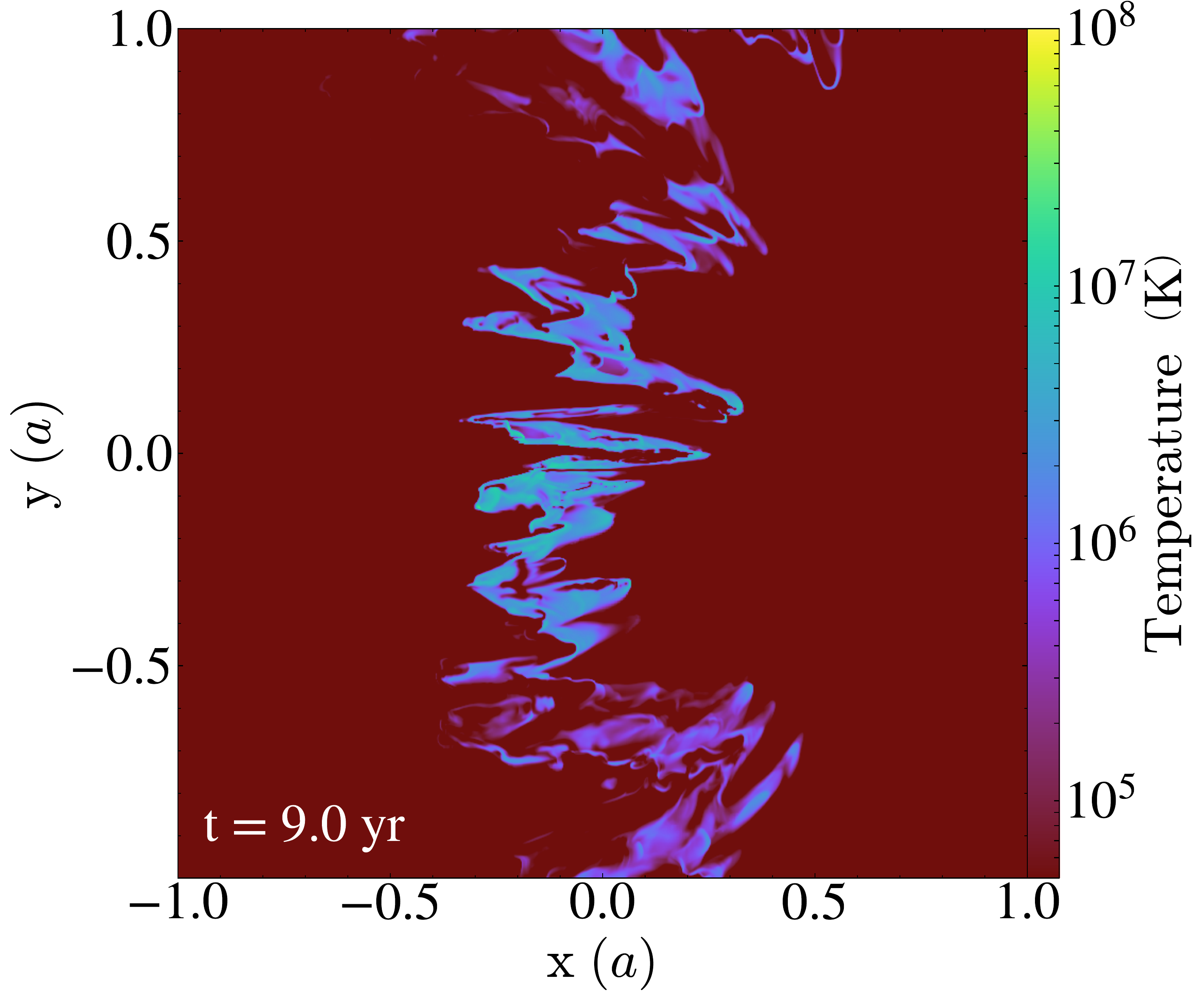}\label{fig:BA10-xz:3}}
				\hspace{0.25cm}
				\subfloat[][BA10:  $T(x={-}0.09a,y,z)$]{\includegraphics[width=0.4\textwidth]{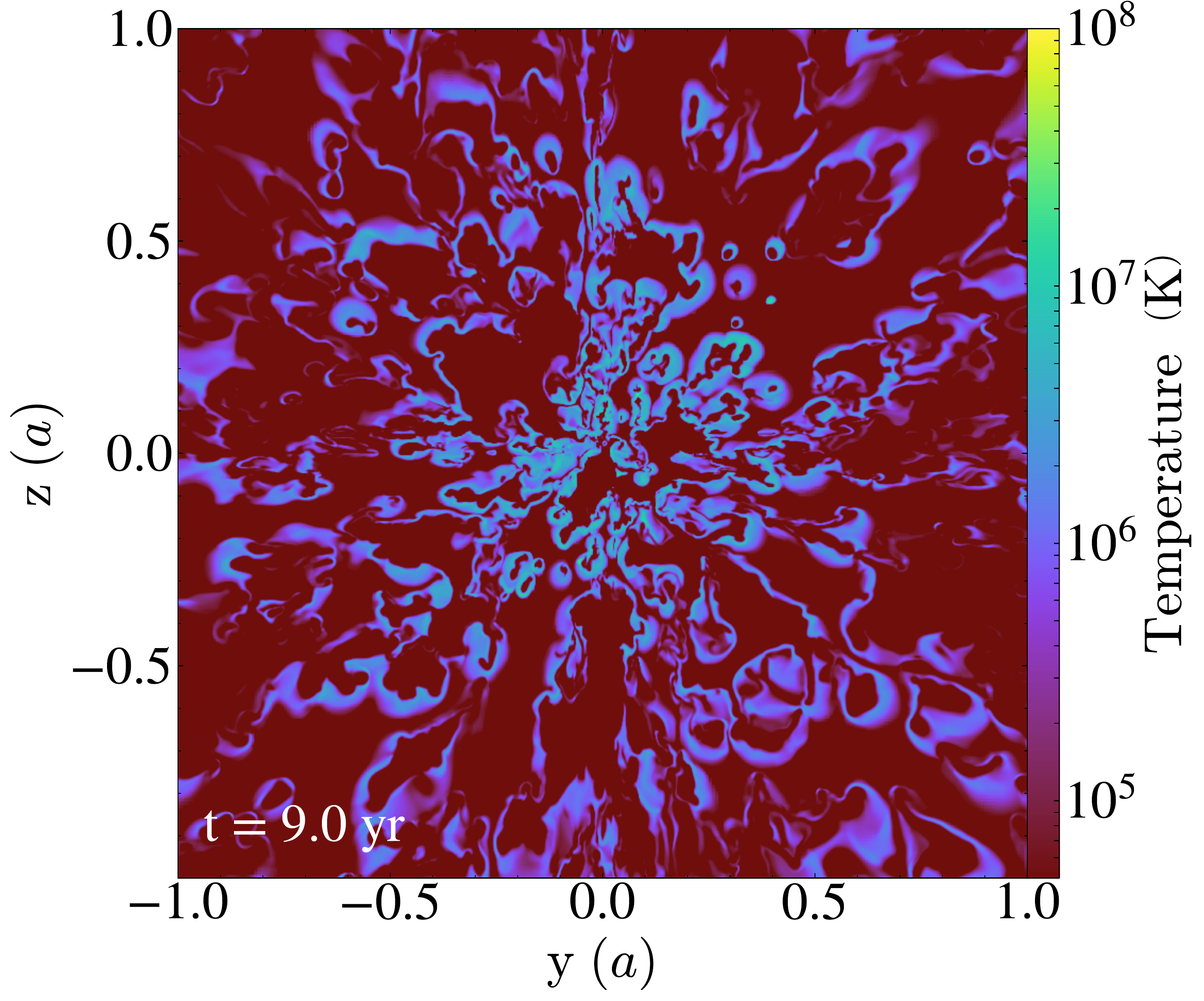}\label{fig:BA10-xz:4}}
				\caption{
				Density (upper row) and temperature maps (lower row) of model BA10. 
				The left and right columns contain maps at \hbox{$z=0$} and \hbox{$x={-}0.09a$} (slab equilibrium position), respectively. 
				All panels show the state of the system at exactly the same simulation time $t=9.0\rm\ yr$. 
				High temperatures in the slab are due to the compression and inefficient cooling of the fast wind material (blown by the star on the right). 
				Notice that the densest regions in the unstable slab are at low temperatures (\hbox{$T\lesssim10^5\rm\ K$}). 
				Meanwhile lower density regions in the slab are kept at very high temperatures (\hbox{$T\approx10^6$--$10^7\rm\ K$}). 
				Figure~\ref{fig:BA10-xz:3} has an associated animation attached (Figure9c\_BA10\_temperature\_slice\_z.mov).
				}
				\label{fig:BA10-xz}
			\end{figure*}
			
			\begin{figure*}
				\subfloat[][BA+10: $\rho(x,y,z=0)$]{\includegraphics[width=0.4\textwidth]{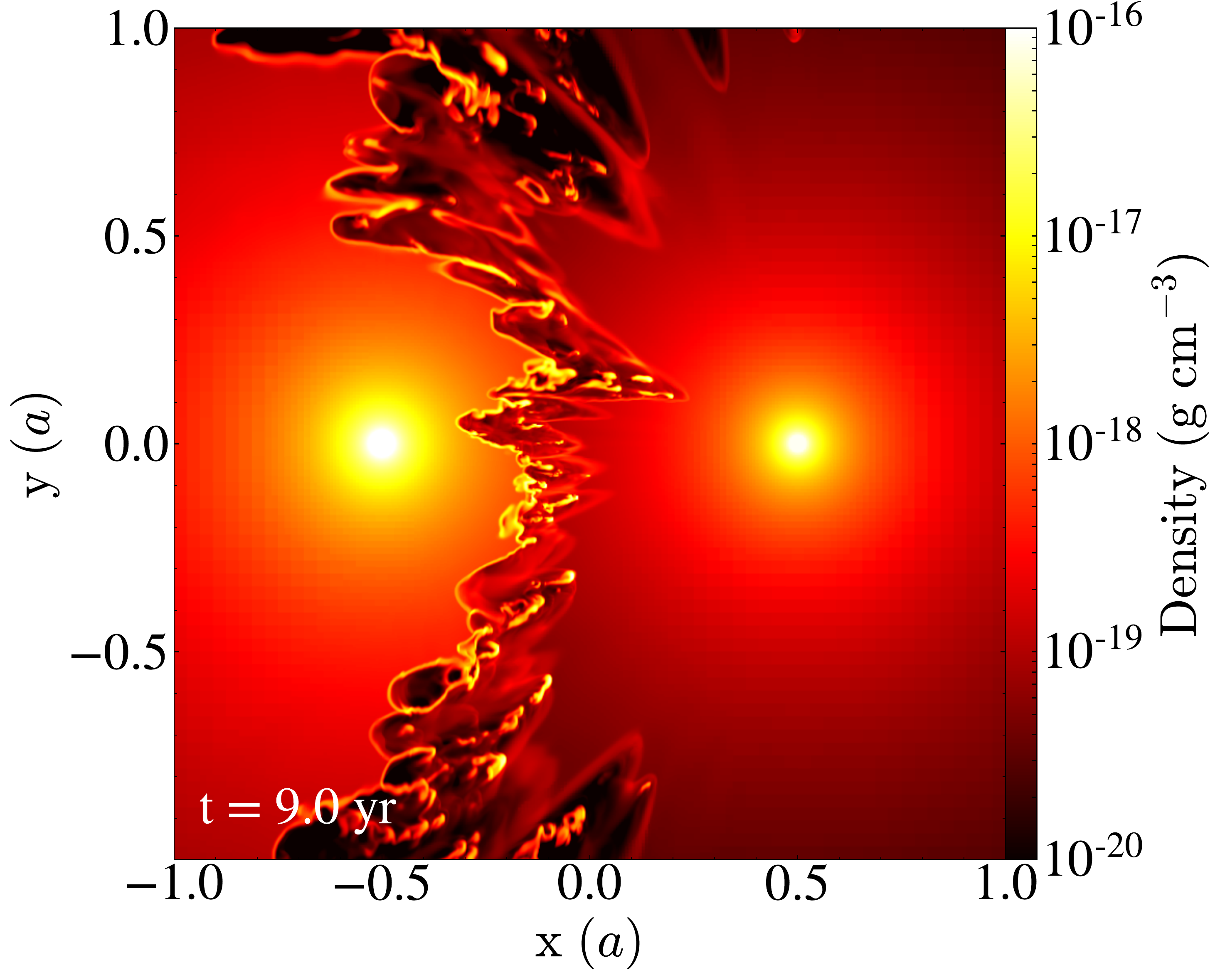}\label{fig:BA+10-xz:1}}
				\hspace{0.25cm}
				\subfloat[][BA+10: $\rho(x={-}0.13a,y,z)$]{\includegraphics[width=0.4\textwidth]{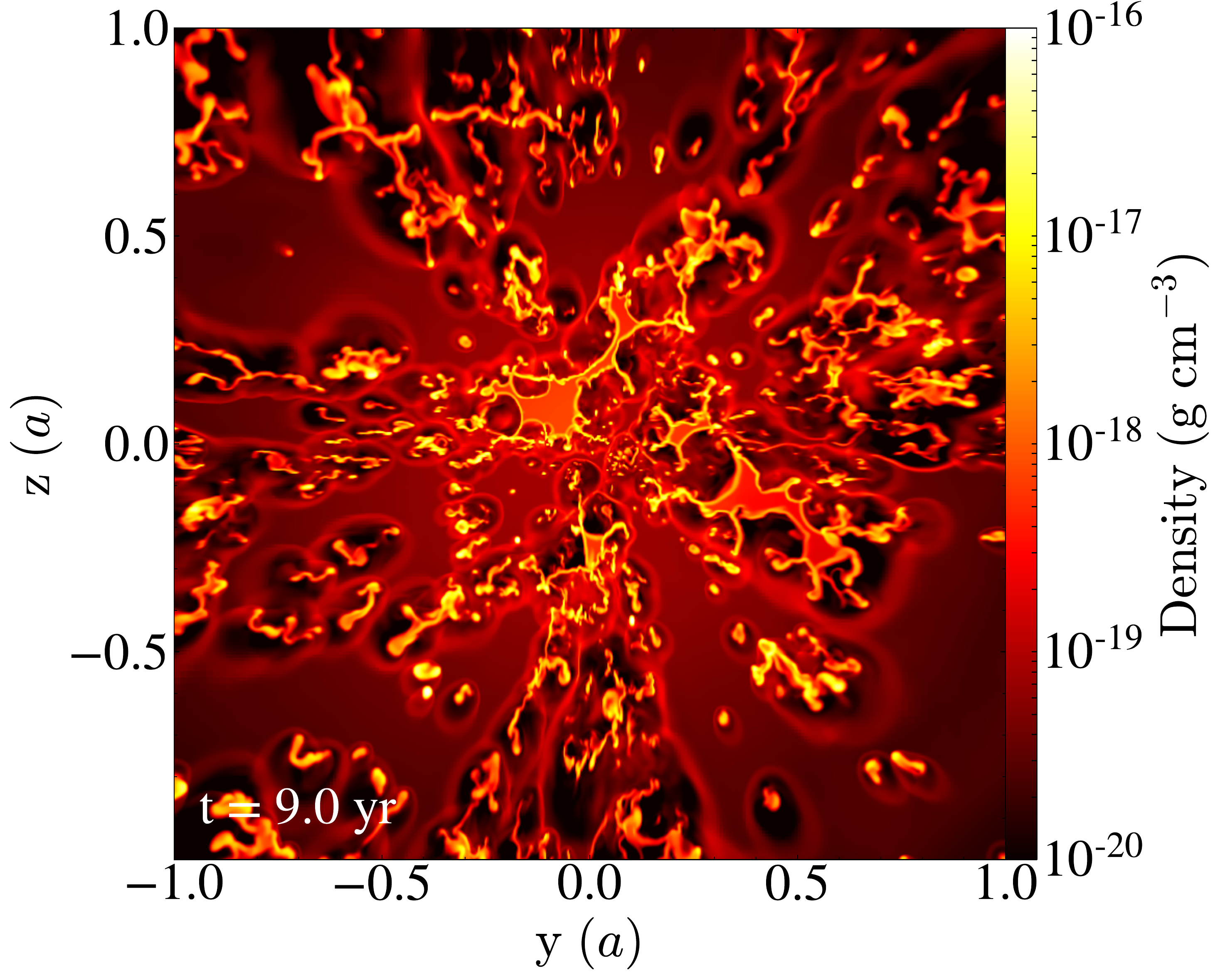}\label{fig:BA+10-xz:2}}
				\vspace{-0.25cm}
				\subfloat[][BA+10: $T(x,y,z=0)$]{\includegraphics[width=0.4\textwidth]{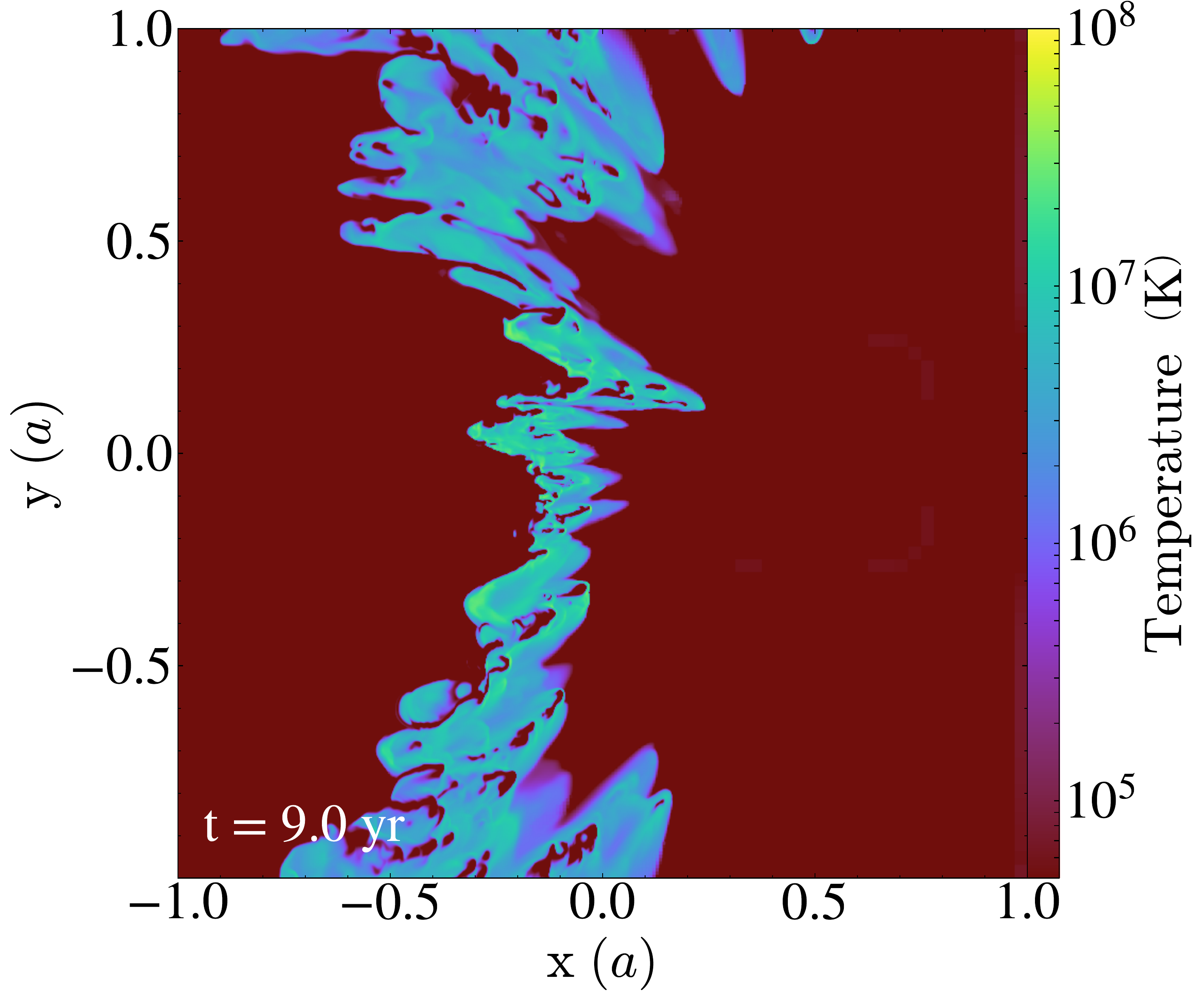}\label{fig:BA+10-xz:3}}
				\hspace{0.25cm}
				\subfloat[][BA+10: $T(x={-}0.13a,y,z)$]{\includegraphics[width=0.4\textwidth]{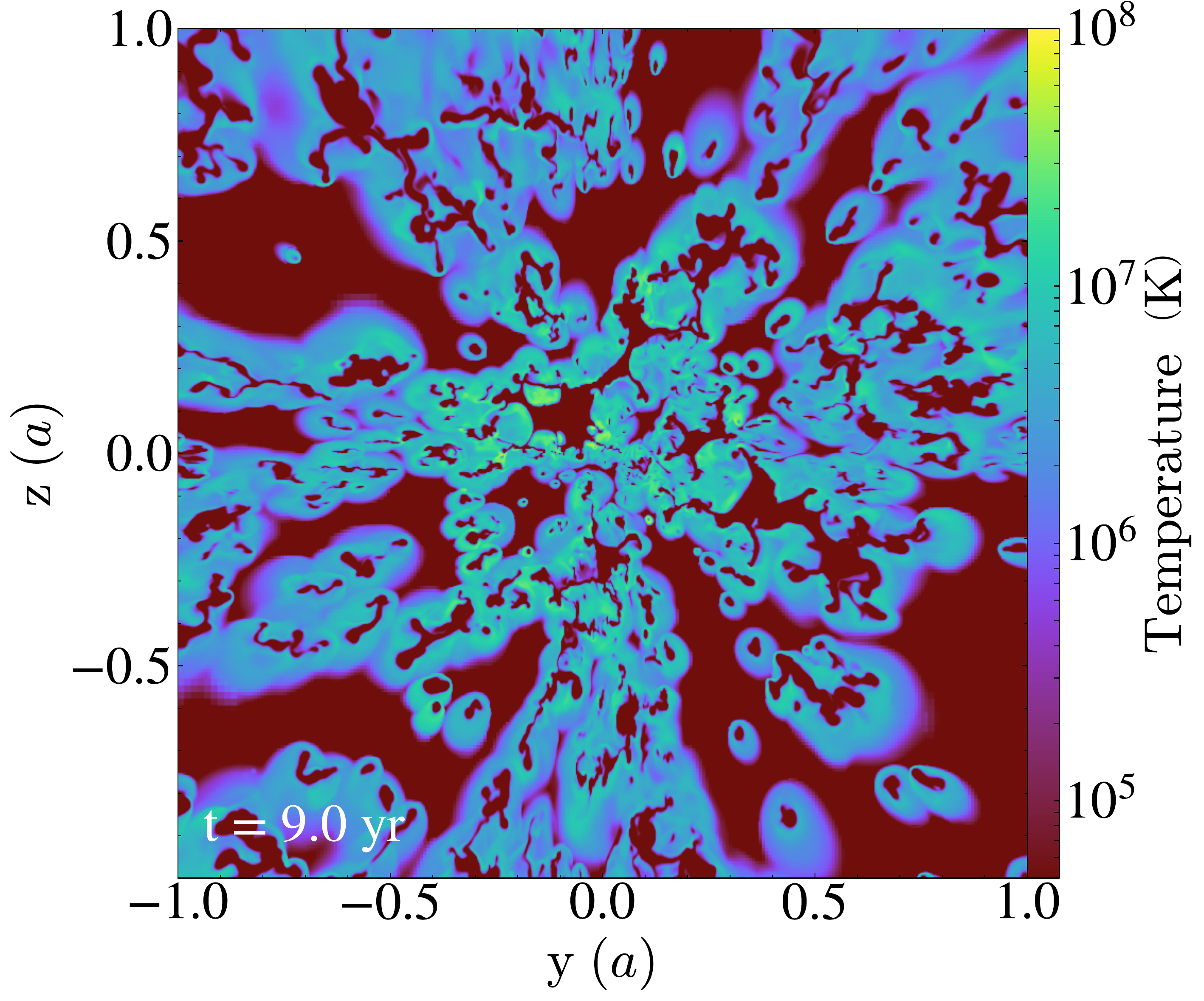}\label{fig:BA+10-xz:4}}
				\caption{
				Density and temperature maps of model BA+10. 
				This figure is analogous to Figure~\ref{fig:BA10-xz} but for model BA+10. 
				The left and right columns contain maps at \hbox{$z=0$} and \hbox{$x={-}0.13a$} (slab equilibrium position), respectively.
				Notice that density and temperature reached are larger compared to model BA10. 
				Figure~\ref{fig:BA+10-xz:3} has an associated animation attached (Figure10c\_BA+10\_temperature\_slice\_z.mov).}
				\label{fig:BA+10-xz}
			\end{figure*}
			
\section{Results}
\label{sec:results}

	In this section we present and analyse each model. 
	Firstly, we give a description of each run and highlight differences among them. 
	Then, we present a quantitative analysis and characterisation of the structures formed in wind collisions. 
	
	\subsection{Hydrodynamics}
	\label{sec:hydro}
		
		As we described previously, our simulations are divided in two groups: symmetric (\hbox{$\eta=1$}); and asymmetric (\hbox{$\eta<1$}) wind collisions. 
		The overall structure of the slabs formed and mechanisms acting can differ significantly between them (see Section~\ref{sec:slab}), thus we present the results of the symmetric and asymmetric models separately in the following sections.
		 
		\subsubsection{Symmetric models}
		
			In order to describe the evolution of these systems, we use as reference the model B10, which resembles very well the general behaviour of the symmetric models. 
			Immediately after the winds collide a thick slab of compressed and hot material is formed (see Figure~\ref{fig:B10:1}). 
			Given that the winds are radiative (by construction) the material cools down very rapidly. 
			As a consequence, the slab loses its thermal pressure support, so it becomes thinner and denser as shown in Figure~\ref{fig:B10:2}. 
			In general, such a thin slab can be easily perturbed from its rest position. 
			In this case numerical noise is enough to seed instabilities in the shell. 
			Thus, wiggles appear on the slab and are located away from the apex\footnote{In the context of stellar wind collisions, the apex is defined as the intersection point between the slab and the line connecting both stars.}. 
			They can be clearly observed in Figures~\ref{fig:B10:2} and~\ref{fig:B10:3}, though they are quickly advected out of the domain. 
			Meanwhile, close to the apex a roughly sinusoidal displacement of the slab starts to grow in amplitude (see Figure~\ref{fig:B10:3}).  
			Now, in this region winds are no longer colliding completely perpendicular to the slab anymore. 
			As a result, the slab becomes slightly wider (and less dense) while material seems to concentrate on the extremities of the perturbation. 
			The density enhancement can be observed already in Figure~\ref{fig:B10:3}. 
			Then, the central displacement of the slab starts to be advected away from the apex, which moves the sinusoidal perturbation to the rest of the slab. 
			Simultaneously, more modes are excited, especially after the perturbation propagated across the whole slab. 
			Approximately after two wind crossing timescales, the slab is completely shaped by the instability (see Figure~\ref{fig:B10:4}). 
			Here, it is even clearer that the density enhancements (clumps) occur on the most displaced regions of slab. 
			Beyond this point the system is in an approximately stationary state, at least until the end of the simulation, which is a minimum of four wind crossing timescales for the standard resolution runs. 
			
			The general behaviour of the symmetric models described so far resembles very well the evolution observed in the 2D isothermal models by \citet{L11}. 
			Since these simulations are 3D, we can move beyond just analysing the simulations in the \hbox{$z=0$} plane.
			However, in 3D space the evolution of the system and the shape of the slab is harder to study due to its complexity. 
			Figure~\ref{fig:B10_3d} contains projected density weighted by density\footnote{This quantity helps to highlight the dense gas, which corresponds to the cold material and to the most refined regions of the domain.} maps along the $z$- and $x$-axes across the entire computational domain in the left and right panels, respectively. 
			Figure~\ref{fig:B10_3d:1} shows once again that the densest regions of the interaction are the most displaced extremities of the slab, while Figure~\ref{fig:B10_3d:2} demonstrates how complex the structure can be at small scale as the structure is dense and filamentary-like. 
			Section~\ref{sec:clumps} presents the study of the physical properties of such overdensities.
				
			The most relevant consequence of considering models with faster winds is the increase of the cooling timescale of the shocked gas. 
			There are two independent factors that contribute to this: 
			firstly, as the density of a stellar wind is inversely proportional to its terminal speed, the density decreases by having a faster wind.  
			Then, as the cooling is proportional to $\rho^2$, its efficiency is reduced. 
			Secondly, a faster wind leads to a higher temperature reached by the shocked material. 
			For typical temperature values of shocked material in stellar winds (\hbox{${\sim}10^6$--$10^{7}\rm K$}), at higher temperature the cooling is less efficient. 
			Thus, the slab formed in wind collisions cools down more slowly if we consider faster winds. 
			This is exactly what happens in the case of the model B+10, seen by comparing the density at \hbox{$z=0$} of model B+10 in Figure~\ref{fig:B+10:1} to model B10 in Figure~\ref{fig:B10:3}.  
			Here, the slab remains thick, i.e. supported by thermal pressure, for longer compared to the case of model B10. 
			The model B10 at \hbox{$t=3.3\rm\ yr$} (\hbox{$0.83t_{\rm cross}$}) shows that the slab has already cooled down and that it is starting to become unstable. 
			On the contrary, in model B+10 even at \hbox{$t=8.0\rm\ yr$} (\hbox{$3\ t_{\rm cross}$}) the slab is still thick and hot. 
			Although condensation is observed in its inner part probably due to thermal instabilities, overall the slab remains thick, which means that it has not radiated its thermal energy yet (see Figure~\ref{fig:B+10:1}).
			Only after about ten years, the slab ends up collapsing initially at the centre and, at the same time the instability starts to develop. 
			Figure~\ref{fig:B+10:2} presents the state of the model B+10 at \hbox{$t=34.8$} (\hbox{$13t_{\rm cross}$}), which corresponds to the end of the simulation. 
			It is important to remark that in this case the system was not completely radiative or adiabatic, instead it was in the transition of both regimes, i.e. \hbox{$\chi\approx1$} (see Figure~\ref{fig:chi}). 
			This is why the slab does not cool down easily, but with the help of thermal instabilities it manages to radiate its energy to become thin and, therefore subject other instabilities.
			Moreover, notice that in this model once the slab is completely unstable, its structure looks more violent compared to model B10. 
			From a theoretical point of view this is expected because the Mach number is larger in this case (\hbox{$\mathcal{M}\approx75$}) compared to the model B10 (\hbox{$\mathcal{M}\approx50$}). 
			This means that the fluid can become more turbulent and, therefore, the density contrast is larger \citep{P11}. 
			Such behaviour will also be present when studying in detail the structure of the unstable slab.
			
			In models with smaller stellar separations (and also smaller domains), faster winds do not produce differences as significant as in the B models . 
			This is due to the cooling efficiency not changing as dramatically between C10 and C+10 compared to the B10 and B+10 models. 
			The winds of both models C10 and C+10 are in the radiative wind regime (see Figure~\ref{fig:chi}), thus we expect their slabs to cool down relatively fast. 	
			Even though model C+10 considers a faster wind speed, which makes cooling more inefficient, at the same time the winds are denser at the collision given that the stellar separation is shorter. 
			Figure~\ref{fig:CC+} shows a comparison between systems C10 and C+10 at exactly the same simulation time. 
			Here we can observe that the slab of model C10 at \hbox{$t=2.2\rm\ yr$} is already completely unstable (see Figure~\ref{fig:CC+:1}). 
			However, in model C+10 the instability is still growing, even though some parts of the slab have not cooled down yet as they remain thick (see Figure~\ref{fig:CC+:2}).
			If we consider even shorter stellar separations, namely models D10 and D+10, the differences are even smaller (not shown here). 
			Both systems consider winds, which can cool more efficiently compared to C or B models (see Figure~\ref{fig:chi}).  
			This is because the winds are denser at the collision due to the shorter stellar separation, which translates into an increase of the cooling efficiency. 
			Increasing the wind from \hbox{$500\rm\ km\ s^{-1}$} to \hbox{$750\rm\ km\ s^{-1}$} does not seem to be enough for overcoming such effect. 
			Here, in both cases the slabs cool down very rapidly, and they become unstable very easily. 
			Therefore, we do not see significant differences between models D10 and D+10. 
		
		\subsubsection{Asymmetric models}
		
			Figure~\ref{fig:asym} presents density maps at \hbox{$z=0$} showing the evolution of models B10, BA10, and BA+10 along the upper, central, and lower rows, respectively. 
			Each column contains each system at the same simulation time, specifically from left to right \hbox{$t=1.2,2.6,3.3,11.2\rm\ yr$}. 
			As the winds are not identical, the slab formed after the collision is not at the midpoint between the stars. 
			In general, the equilibrium position of the slab is determined by balancing the wind momenta. 
			Based on this, the slab should be at a distance \hbox{$[a\sqrt{\eta}/(1+\sqrt{\eta})]$} from the star with the weaker wind \citep{S92}, which in our case also corresponds to the slower wind located on the left side of the domain . 
			Thus, in models BA10 and BA+10 the slab is centred at \hbox{${\sim}0.41a$} and \hbox{${\sim}0.37a$} from the slow wind star, respectively. 
			Although the difference is small compared to the symmetric cases, this causes the slow wind to be denser at the collision and, therefore, being more radiatively efficient. 
			This picture is even more dramatic when winds collide for the first time in the simulation. 
			Mainly, because the location of such encounter is determined by the speed of the winds which does not necessarily coincide with the equilibrium position of the slab. 
			As in the model BA10 (BA+10) the stronger wind is twice (three times) as fast the initial collision occurs at \hbox{${\sim}0.33a$} (\hbox{$0.25a$}) from the weak wind star. 
			Notice that both separations are shorter compared to the distance to the slab equilibrium position (see first column of Figure~\ref{fig:asym}).  
			Therefore, the weaker wind collides being even denser and the radiative cooling takes place even faster compared to the case when the slab is located on its rest position. 
			On the contrary, the opposite applies for the faster wind, i.e. it is more diluted when it collides for the first time causing the cooling to be less efficient.
			
			The asymmetric models also show that instabilities are triggered at earlier times. 
			In less than a year after the initial wind collision, we can visually recognise patterns consistent with the KHI in the interaction region located away from the apex.  
			This is not surprising as \citet{L11} had already observed such behaviour finding that even a very small speed difference between the winds could excite this instability. 
			Nevertheless, in these cases this instability is not necessarily the only one acting like in the adiabatic models of \citet{L11}.  
			Instead, as our models consider radiative cooling it is possible that the slab loses its thermal support, at least partially, so that the KHI can develop simultaneously along with the thin-shell instabilities, most likely with the Vishniac instability \citep{V83}. 
			This is the reason behind the wind interaction becoming even more complex after a couple of years, especially if we compare its evolution with the symmetric model B10 (see second column of Figure~\ref{fig:asym}). 
			
			There are two more important observations we can infer from the evolutionary sequences of Figure~\ref{fig:asym}. 
			Firstly, notice that with decreasing $\eta$, i.e. larger wind speed difference, the instabilities seem to grow faster. 
			For instance, the slab of model B10 looks completely unstable only in the last panel of the sequence which is at \hbox{$t=11.2\rm\ yr$} (see Figure~\ref{fig:evol:B10-4}). 
			On the other side, the slabs of the asymmetric models are already unstable in the second panel of the sequence, i.e. at \hbox{$t=2.6\rm\ yr$} (see Figure~\ref{fig:evol:BA10-2} and~\ref{fig:evol:BA+10-2}). 
			Also, the instabilities grow faster in BA+10 than in BA10 (see central and lower rows of Figure~\ref{fig:asym}). 
			At \hbox{$t=3.3\rm\ yr$} there are larger amplitude modes excited in BA+10 (see Figure~\ref{fig:evol:BA+10-3}) than in BA10 (see Figure~\ref{fig:evol:BA10-3}). 
			Even at earlier times longer wavelength modes seem to be excited in BA+10 (see Figure~\ref{fig:evol:BA+10-2}) but not yet triggered in BA10 (see Figure~\ref{fig:evol:BA10-2}).
			The KHI is very likely responsible for this as its growth timescale in the linear regime is given by \hbox{$t_{\rm KHI}=\lambda/(2\pi\Delta v)$}, where $\lambda$ is the wavelength and $\Delta v$ is the velocity difference between the fluid layers. 
			Thus, a given mode $\lambda$ grows faster the larger the velocity difference is. 
			In model BA+10, $\Delta v$ is twice that of model BA10, so an arbitrary mode should grow twice as fast. 
			Secondly, it is important to remark that once all of the systems reached their stationary state, the slab of model BA+10 shows the largest density contrast compared to the rest. 
			Additionally, the structure of the unstable shell looks more clumpy and less filamentary than BA10 and B10 (see last column of Figure~\ref{fig:asym}). 
			The explanation of such features could be the degree of the supersonic nature of winds. 
			In B10, BA10, and BA+10, the faster winds have a Mach number of \hbox{$\mathcal{M}\approx50,100,150$}, respectively. 
			As this quantity reflects the compression of the material of the slab, it is natural to expect higher compression, and therefore a denser slab, with a higher Mach number. 
			Also, let us remember that at lower $\eta$ the collision takes place closer to one of the stars, which also translates into a higher density in the shell. 
			Therefore, these could explain the larger density contrast observed in the density maps, especially in BA+10. 
			
			In order to describe the thermodynamic state of the wind interaction in the asymmetric models, Figures~\ref{fig:BA10-xz} and~\ref{fig:BA+10-xz} present density and temperature maps of models BA10 and BA+10, respectively.  
			The upper row of each figure contains density maps while the lower row presents temperature maps. 
			The left and right columns show maps at the \hbox{$z=0$} and \hbox{$x=x_{\rm slab}$} planes, respectively, where $x_{\rm slab}$ corresponds to the slab rest position of a given model. 
			Analysing the temperature maps, it is possible to observe that most of the domain is kept at low temperatures (\hbox{$T\lesssim10^5\rm K$}). 
			Bear in mind that mostly these regions correspond to the free-wind regions, which are initially blown at ${\sim}10^4\rm\ K$. 
			However, there are some regions in the slab that are also at this temperature. 
			Observing carefully it is possible to see that each of these regions corresponds to the densest parts of the slab. 
			Furthermore, notice that they all are surrounded by hotter material due to the presence of the adiabatic shock of the faster winds. 
			As this material cannot cool down efficiently, the condensed regions in the slab must originate solely from the slowest wind. 
			This description resembles very well the description of the Vishniac instability. 
			Here, the dense slab is confined on one side directly by ram pressure and on the other by the thermal pressure of the adiabatic shock. 
			This could be the reason why the unstable slab is very different in the asymmetric cases compared to the symmetric models. 
			Let us remember that in our symmetric models both winds were radiatively efficient, by construction, therefore the slab ended up being confined by ram pressure on both sides. 
			In that case the unstable shell is better described by the NTSI \citep{V94}. 
			
			Finally, let us analyse differences between the density and temperature maps of models BA10 and BA+10 (see Figures~\ref{fig:BA10-xz} and~\ref{fig:BA+10-xz}). 
			At $x=x_{\rm slab}$ the structures present are denser in model BA+10 (Figure~\ref{fig:BA+10-xz:2}) compared to BA10 (Figure~\ref{fig:BA10-xz:2}). 
			This fact is simply explained by the stronger pressure confinement of the winds. 
			On one side the stronger wind is faster, so the ram pressure is larger. 
			On the other side, the slab is being pushed closer to the weaker wind star, so this wind is denser at the collision, which also enhances the ram-pressure strength.
			
			Another important observation is related to the temperature reached in certain regions of the slab. 
			As expected, model BA+10 reaches higher temperatures, in general, than BA10 since the fast wind of the former is 50 per cent larger.
			More important it is the temperature differences along each of the slabs (see Figures~\ref{fig:BA10-xz:3} and~\ref{fig:BA+10-xz:3}). 
			Higher temperatures in the slab are found closer to the apex. 
			This is due to the fact that the shocks are closer to be normal to the slab in this region. 
			Away from the apex, ram pressure decreases with density and the velocity of the winds is not entirely compressing material; instead it helps to advect the slab away from the domain. 		
			On top of this, once the slabs become unstable the shocks do not hit, in general, perpendicular to the slab, so the shocks unlikely generate the maximum compression expected in a plane-parallel setup.  
			
	\subsection{Structure search and characterisation}
	\label{sec:clumps}
	
		Having described the general evolution of the systems, we proceed to study the properties of the overdensities formed in the slab. 
		Firstly, we describe our structure finder algorithm. 
		Then, we study the physical properties and dynamics of the clumps, and how the wind speed and stellar separation determine such properties.
	
		\subsubsection{Identification criteria}
	
			In each simulation run we searched for overdense regions (hereafter clumps).
			To do so, we made use of a clump finder algorithm that was applied to every single snapshot of the simulations.
			Typically, these type of algorithms receive (at least) two input parameters: a density threshold $\rho_{\rm tr}$ and a minimum number of cells for defining a clump $N_{\rm cell}$. 
			Here, we used a density threshold of \hbox{$\rho_{\rm tr}=\bar{\rho}+5\sigma_{\rho}$}, where $\bar{\rho}$ and $\sigma_{\rho}$ correspond to the mean density and density dispersion, respectively.
			The value of $\rho_{\rm tr}$ is about \hbox{${\sim}10^{-17}\rm\ g\ cm^{-3}$} in models with a large domain like B10 (see Figure~\ref{fig:B10} as reference). 
			We tested several threshold values, and noticed this value gave the most reasonable result maximising the number of selected cells in the slab while minimising the cells of the free wind region. 
			The minimum number of cells for detecting a clump was set to \hbox{$N_{\rm cell}=10$} (for the standard resolution runs). 
			Making use of these parameters the algorithm executes the following tasks. 
			Firstly, it applies a density cut on the cells of the snapshot, i.e. it ignores every cell whose density is lower than the threshold. 
			Then, it searches for physically connected structures within the remaining cells. 
			It continues analysing the substructure of each previously identified structure. 
			By doing so, it defines a clump per each density local maximum found. 
			The algorithm iterates until every cell was assigned to a (sub)structure. 
			The output is a list of structures with their associated substructures. 
			For each substructure we extracted its physical properties, such as total mass $m$, centre-of-mass position $\mathbf{R}_{\rm cm}$, and velocity $\mathbf{v}_{\rm cm}$. 
			For a more detailed description of the algorithm we refer the reader to the Appendix~\ref{sec:app1}. 
			It is important to remark that these clumps are not gravitationally bound.
			As we expect them to have, at most, a mass of the order of the Earth, self-gravity is negligible compared to the wind ram-pressure confinement. 
			Therefore, the clumps correspond to pressure confined overdensities.
			
\begin{figure*}
				\centering
				\includegraphics[width=0.9\textwidth]{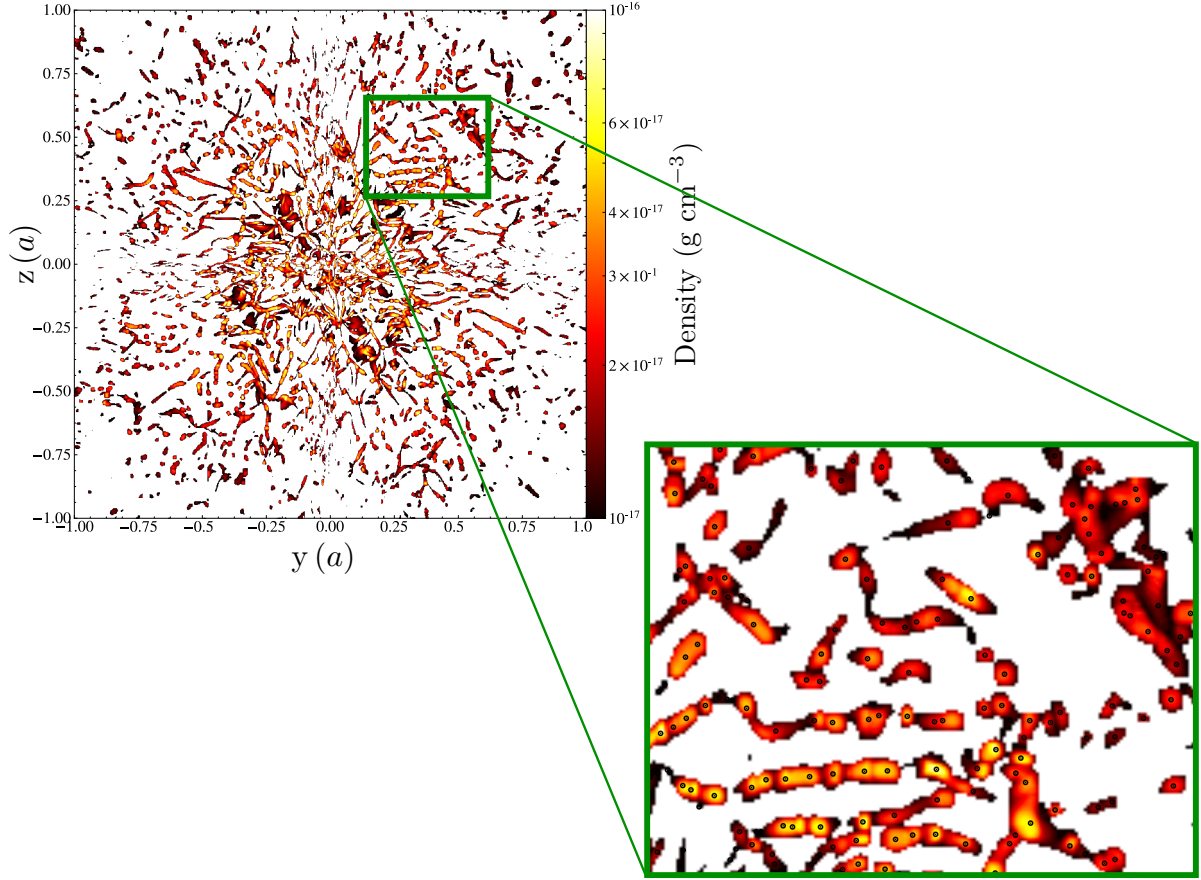}
				\caption{Density projection along the $x$-axis of the model B10 at \hbox{$t=11.2\rm\ yr$} after performing a density cut at \hbox{$\rho_{\rm tr}=\bar{\rho}+5\sigma_{\rho}$}, i.e. not considering cells whose \hbox{$\rho\leq\rho_{\rm tr}$}. 
				The zoomed region shows an example of the overdensities our clumpfinder algorithm identifies as clumps. 
				Black circles highlight the centre-of-mass of the identified overdensities. 
				This snapshot analysis corresponds to the same model at the same simulation time as the one shown in Figure~\ref{fig:B10_3d}.
				}
				\label{fig:clumps_example}
			\end{figure*}

			An example of the analysis performed by the algorithm is shown in Figure~\ref{fig:clumps_example}. 
			It corresponds to the model B10 at time \hbox{$t=11.2\rm\ yr$}.
			It presents a density projection along the $x$-axis (weighted by density) but considering only cells above the density threshold. 
			The zoomed-in region marks the centre-of-mass of clumps identified by the algorithm. 
			
			\begin{figure}
	 			\centering
				\includegraphics[width=0.45\textwidth]{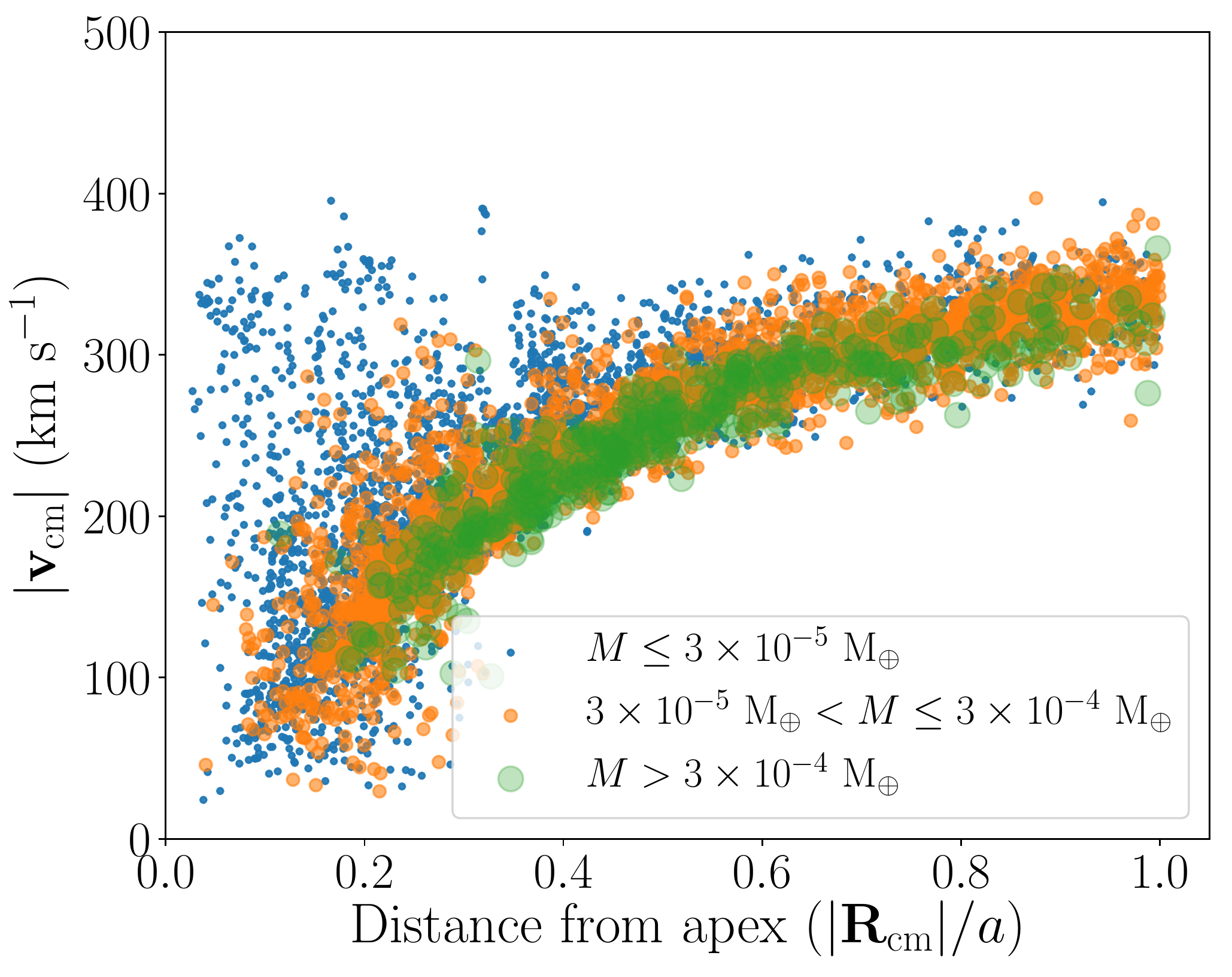}
				\caption{Magnitude of the clump centre-of-mass velocity \hbox{$|\mathbf{v}_{\rm cm}|$} as a function of their $3D$ distance from the apex (which in symmetric models coincides with the centre of the domain) \hbox{$|\mathbf{R}_{\rm cm}|$} at \hbox{$t=11.2\rm\ yr$} in model B10. 
				Clumps located further than \hbox{$|\mathbf{R}_{\rm cm}|=a$} are not shown. 
				Each point represents a single clump. 
				The size and colour encode their mass. 
				The most massive clumps are shown as big green dots. 
				The lightest clumps appear as small blue dots. 
				Notice that clumps tend to follow a clear trend, specially as they move away from the centre.}
				\label{fig:b10_rv}
			\end{figure}
			
			\begin{figure*}
	 			\centering
				\includegraphics[width=0.7\textwidth]{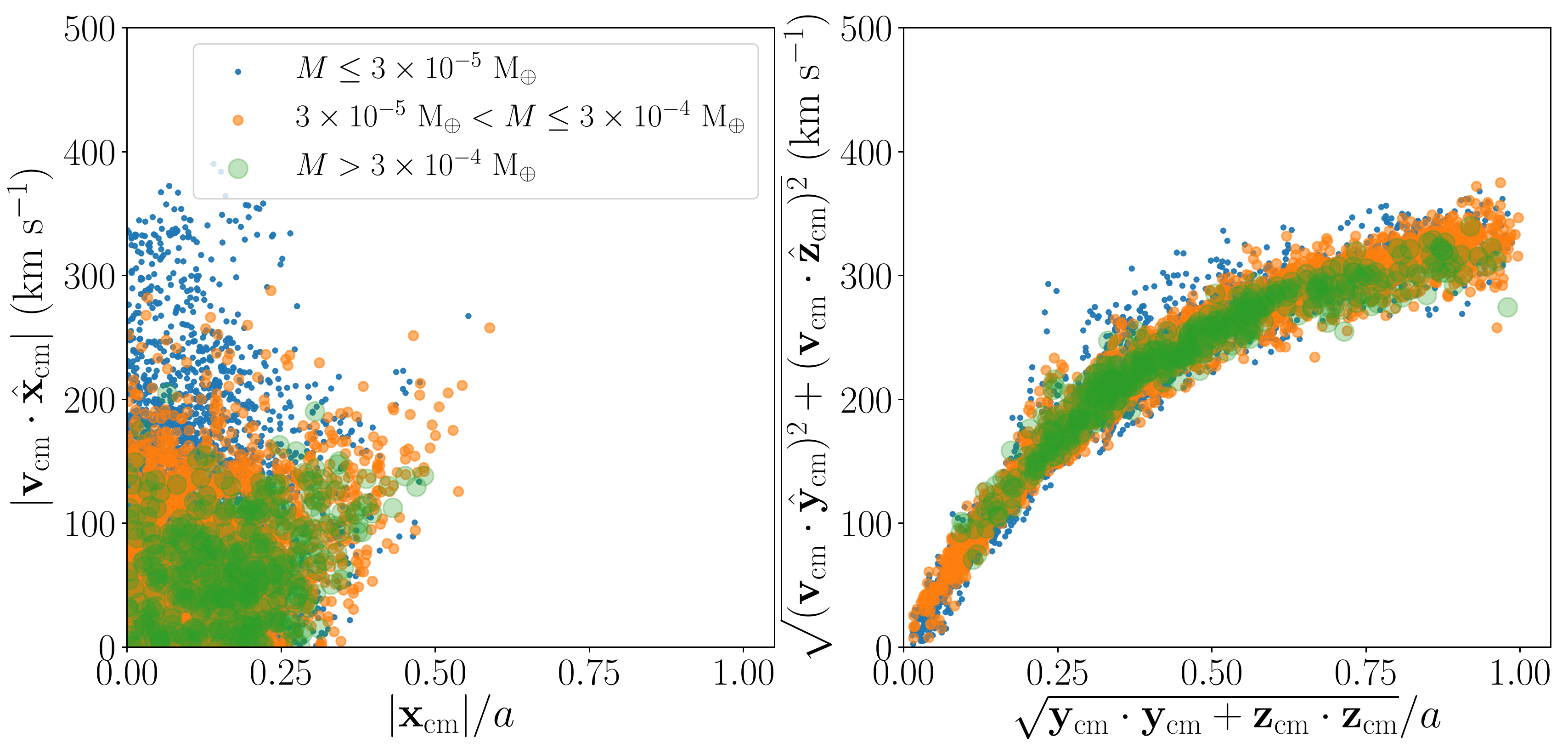}
				\caption{Figure analogous to Figure~\ref{fig:b10_rv}, however, here each panel shows the components of velocity and position.
				Left panel shows the velocity and distance component along the $x$-axis, while the right panel contains the velocity and distance projections onto the $yz$-plane.
				Bear in mind that the line connecting both stars is parallel to the $x$-axis, and the slab is initially located at \hbox{$x=0$}.}
				\label{fig:b10_rvxyz}
			\end{figure*}
			
		\subsubsection{Clump masses and motion}
		
			Now, we proceed to analyse the simulations in order to search and characterise clumps. 
			As we described in Section~\ref{sec:hydro}, in each model there is a point at which the system reaches an approximate stationary state. 
			Under this regime the slab shape is completely determined by the instabilities.
			In general, we noticed that this state starts, approximately, after two wind crossing timescales (\hbox{$t>2t_{\rm cross}$}). 
			Based on this, we decided to analyse the simulation only during this self-regulated state, as we aim to characterise the long-term behaviour of the system. 
			Firstly, we describe in detail the analysis of model B10 so that we present a general view of the clump properties and dynamics. 
			Then, we analyse how these results change if we modify the properties of the stellar winds. 
			
			Figure~\ref{fig:b10_rv} presents the velocity of the clumps as a function of their distance from the apex, which coincides with the centre of the domain in symmetric models.
			On top of this, clumps are shown as dots (each dot corresponds to a single clump). 
			Their mass are colour- and size-coded. 
			Larger green dots represent more massive clumps while smaller blue dots stand for lighter clumps. 
			This analysis corresponds to the model B10 at time \hbox{$t=11.2\rm\ yr$}.  
			Overall, notice that the parameter space where most clumps are located is well-defined. 
			The shape of this diagram does not change significantly with time in the stationary regime. 
			This fact points to a sequence that clumps seem to follow since they are formed at small \hbox{$|\mathbf{R}_{\rm cm}|$} until they leave the domain \hbox{$|\mathbf{R}_{\rm cm}|/a\to1$}.
			Notice that at short distances the scatter on their speed is large. 
			Here, clumps can have almost null speed up to \hbox{$400\rm\ km\ s^{-1}$}, which is about $80$~per~cent of the wind speed. 
			Nevertheless this description does not apply to the most massive clumps given that only few of them are present at \hbox{$|\mathbf{R}_{\rm cm}|/a<0.25$}. 
			Once the distance goes beyond this value, we can observe that the most massive clumps start to populate the diagram. 
			At longer separations the scatter of the clump speed distribution tends to decrease. 
			Furthermore, as clumps are getting closer to the boundaries of the domain (\hbox{$|\mathbf{R}_{\rm cm}|/a\to1$}) their velocity seems to be converging. 
			
			In order to analyse this in more detail we divided the velocity and position vectors into components. 
			Such a description is presented in Figure~\ref{fig:b10_rvxyz}, where each panel shows the clump velocity as a function of distance along the $x$-axis (left panel), and within the $yz$-plane (right panel), bearing in mind that the $x$-axis is parallel to the line connecting the stars. 
			Here we can clearly observe that the dispersion seen at small $\mathbf{R}_{\rm cm}$ in Figure~\ref{fig:b10_rv} is shown solely in the $x$-component (left panel of Figure~\ref{fig:b10_rvxyz}), and not along the other components. 
			This dispersion seems to be caused by the instability which produces significant displacements of the slab towards one of the stars before the structure has time to be advected away from the domain. 
			Thus, initially overdensities are pushed to either of the stars, and only when they reach certain distance from the centre they start to be accelerated steadily along the $x$-axis. 
			However, the other components do not show the same behaviour. 
			In the right panel of Figure~\ref{fig:b10_rvxyz}, we can observe that the clump speed projected onto the $yz$-plane increases steadily with distance. 
			This means these components are responsible for driving the clumps away from the system. 
			Furthermore, it is possible to observe that the acceleration decreases with distance, which also implies that the velocity is converging. 
			 This is due to the change of the stellar wind ram-pressure strength as a function of distance, \hbox{$P_{\rm w}\propto r^{-2}$}. 
			Unfortunately, our domain size does not allow us to observe what occurs further away. 
			In general, this will depend largely on the environment this system is immersed in. 
			This is why we preferred not to go beyond this range with our models. 
			By doing so, these high-resolution simulations provide a detailed view of the initial properties and behaviour of clumps immediately after being formed. 
			
\begin{figure*}
	 			\centering
				\subfloat[][Model B10: clump mass histogram]{\includegraphics[width=0.45\textwidth]{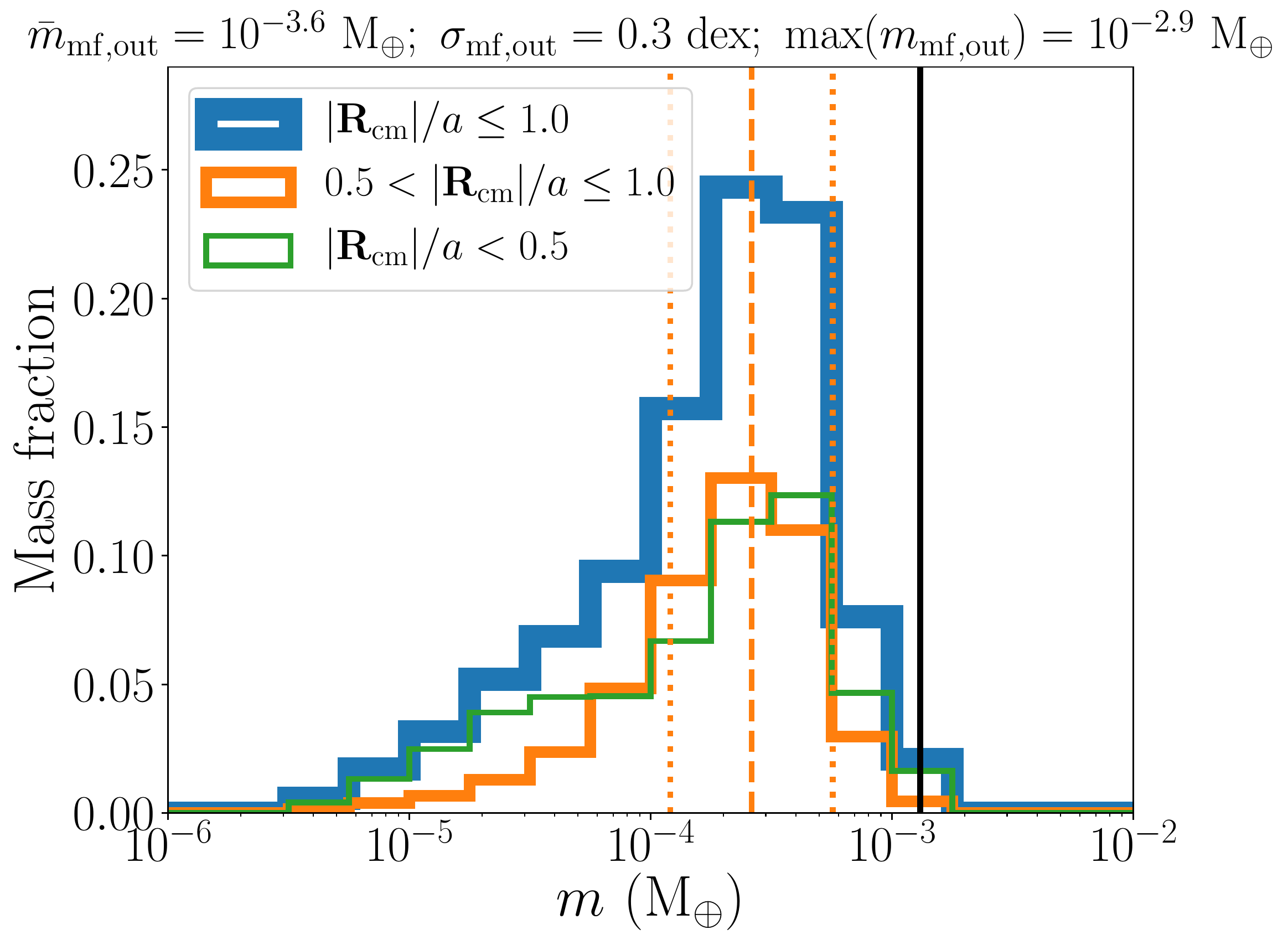}\label{fig:out:1}}
				\hspace{0.25cm}
				\subfloat[][Model B10: clump velocity histogram]{\includegraphics[width=0.45\textwidth]{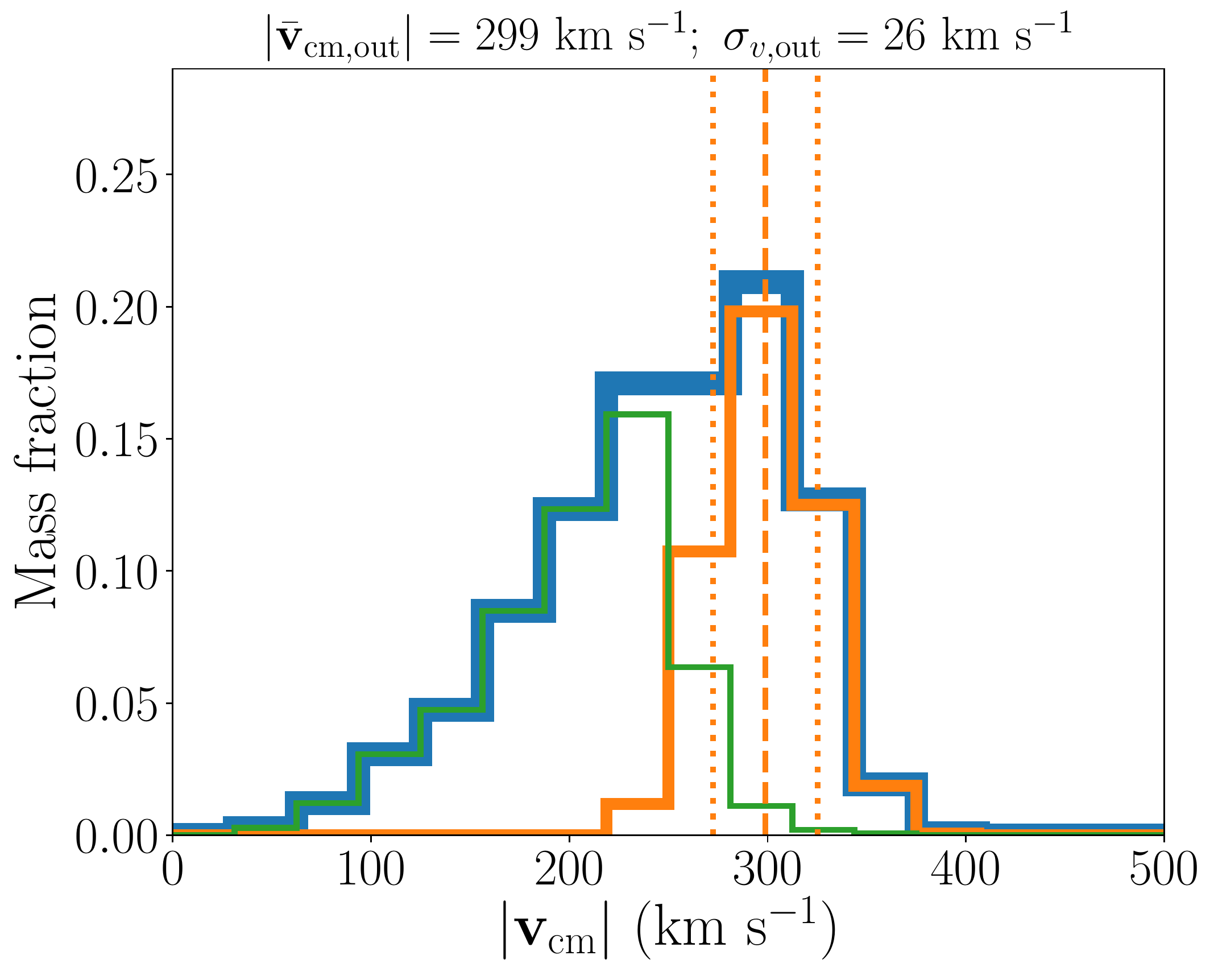}\label{fig:out:2}}
				\caption{Clump mass fraction in bins of mass (left panel) and velocity (right panel) in different regions at \hbox{$t=11.2\rm\ yr$} in model B10. 
				In both panels, the thin green and thicker orange lines show clumps located inside a sphere of radius $0.5a$ (inner region), and a spherical shell delimited by radii $0.5a$ and $a$ (outer region), respectively. 
				The thickest blue lines represent the sum of the green and orange line histograms, i.e. clumps enclosed within a sphere of radius $a$. 
				Notice that the $x$-axis scale is logarithmic on the left panel but linear on the right panel. 
				Both panels also show the mean and standard deviation of the distribution for clumps in the outer region (orange lines) as dashed and dotted orange vertical lines, respectively. 
				The vertical solid black line in the left panel corresponds to the mass of the most massive clump.
				The numerical values plotted with vertical lines are shown on top of each panel. 
				}
				\label{fig:out}
			\end{figure*}

			To conduct a quantitative description of the clump mass and velocity distribution, we divided the clumps located in two groups according to their spatial location: an inner and an outer region. 
			The former is defined as a sphere of radius $0.5a$ centred in the apex of the wind interaction.
			The latter was defined as a concentric spherical shell whose inner and outer radii are $0.5a$ and $a$, respectively. 
			Figure~\ref{fig:out} presents clump mass fraction distribution in bins of mass (left panel) and velocity (right panel) of model B10 at $t=11.2\rm\ yr$. 
			The mass fraction is calculated as the mass contribution of each bin relative to the total mass in clumps. 
			The line colour highlights the region where clumps are located. 
			The green histogram represents clumps enclosed in the sphere of radius $0.5a$ (inner region). 
			The orange histogram shows objects inside the spherical shell (outer region). 
			The sum of the two is shown as a blue line. 
			Vertical dashed lines stand for the mean value for clumps of the outer region. 
			Meanwhile, vertical dotted lines show the standard deviation of the distributions. 
			Finally, the most massive clump is shown with a vertical solid black line.  
			Although some clumps are of fairly low mass \hbox{$m\approx10^{-5}\rm\ M_{\oplus}$}, the most significant mass contribution is in clumps of \hbox{${\sim}3\times10^{-4}\rm\ M_{\oplus}$}. 
			Even there are clumps that reach masses of \hbox{$m\approx10^{-3}\rm\ M_{\oplus}$}. 
			Here it is important to consider that the lower mass end is a direct consequence of the parameters of our clump finding algorithm: minimum cell size and density threshold. 
			Therefore, we should not interpret it as the physical lower limit of the distribution. 
			On the contrary, the upper mass limit is set purely by the hydrodynamics and radiative properties of the system.  
					
			Figure~\ref{fig:out:1} also shows that in the inner region there is more mass in very light clumps (\hbox{$m<5\times10^{-5}\rm\ M_{\oplus}$}) compared to the outer region. 
			In stationary state this histogram does not change significantly. 
			A possible explanation for this behaviour is that small clumps are formed in the inner region, some are destroyed while others merge into larger ones. 
			Thus, only a fraction of them manage to reach the outer region, which now contains less smaller clumps than the inner region. 
			In the case of more massive clumps, they lose mass as their ram-pressure confinement decreases while escaping from the system. 
			However, as the mass in massive clumps is the same between the inner and the outer region, more massive clumps should be constantly forming. 
			Therefore, on one side clumps are losing mass, and moving to smaller mass bins, while at a similar rate lighter clumps are either growing or merging to form more massive ones. 
			Section~\ref{sec:life} presents a detailed study tracking the properties of a single clump. 
			
			Figure~\ref{fig:out:2} shows that clumps in the inner and outer regions have very different speed distributions. 
			In the inner region clumps have lower speeds and also span a wider range of values. 
			On the contrary, the outer region clumps tend to have larger speeds and smaller dispersion. 
			This properties can be easily explained by observing the evolution of the system. 
			Clumps are formed at short distances from the domain center where the wind collision cancels most of their linear momentum. 
			This causes overdensities not to carry much momentum, at least initially. 
			Then the instability growth displaces overdensities towards one of the stars increasing their speed, specifically, along the $x$-axis (see left panel of Figure~\ref{fig:b10_rvxyz}). 			
			Simultaneously, clumps start being advected away from the apex steadily increasing their velocity (see right panel of Figure~\ref{fig:b10_rvxyz}). 
			At this point, clumps stop being pushed toward any of the stars, and instead they are only gaining momentum for escaping from the system. 
			After this transition, most clumps already have a well-defined velocity as they have spent roughly the same amount of time being accelerated by the winds. 
			The latter regime corresponds to objects which are already in the outer region where their speed is about ${\sim}60$~per~cent of the wind speed and its dispersion is only ${\sim}10$~per~cent.  	 
	
			Recapping, model B10 shows that clumps in the inner region are still being formed, because they are part of the ram-pressure confined slab. 
			Here, there is a larger mass fraction of lighter clumps, they are being accelerated, and, in general their speeds are small but show a large dispersion. 
			On the contrary, in the outer region there are not as many light clumps due their destruction and/or merging events. 
			Their acceleration is decreasing given that ram pressure loses strength with distance. 
			Therefore, they seem to have reached their terminal speed that is about ${\sim}3/5$ of the wind speed. 
			Notice that this description only applies to clumps located not further than the length of the stellar separation \hbox{$|\mathbf{R}_{\rm cm}|=a$}. 
			Clump properties can be affected by the medium into which they are ejected. 
			
			\subsubsection{The life of a clump} 
			\label{sec:life}
				
\begin{figure*}
				\centering
				\includegraphics[width = 0.45\textwidth]{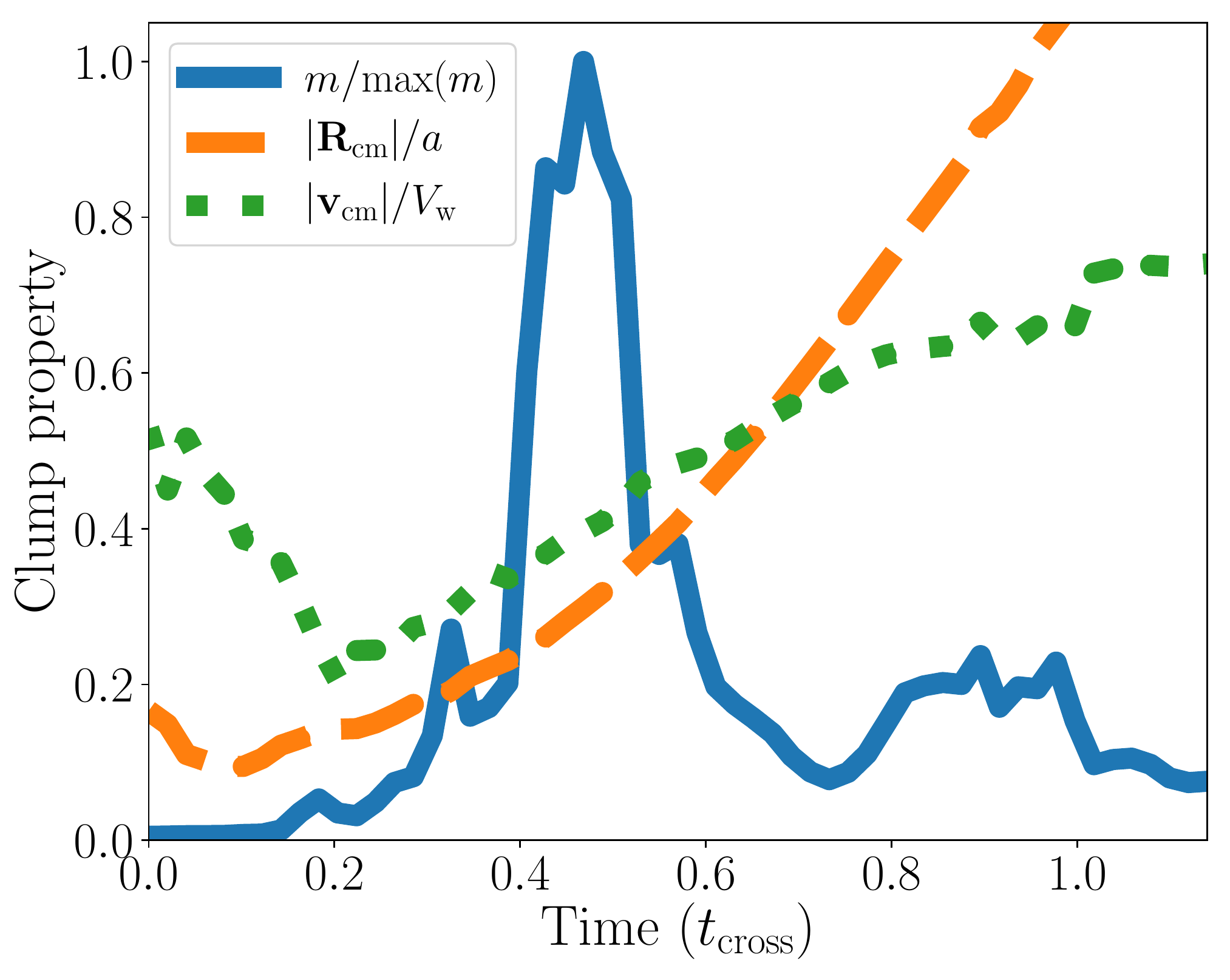}
				\hspace{0.25cm}
				\includegraphics[width = 0.45\textwidth]{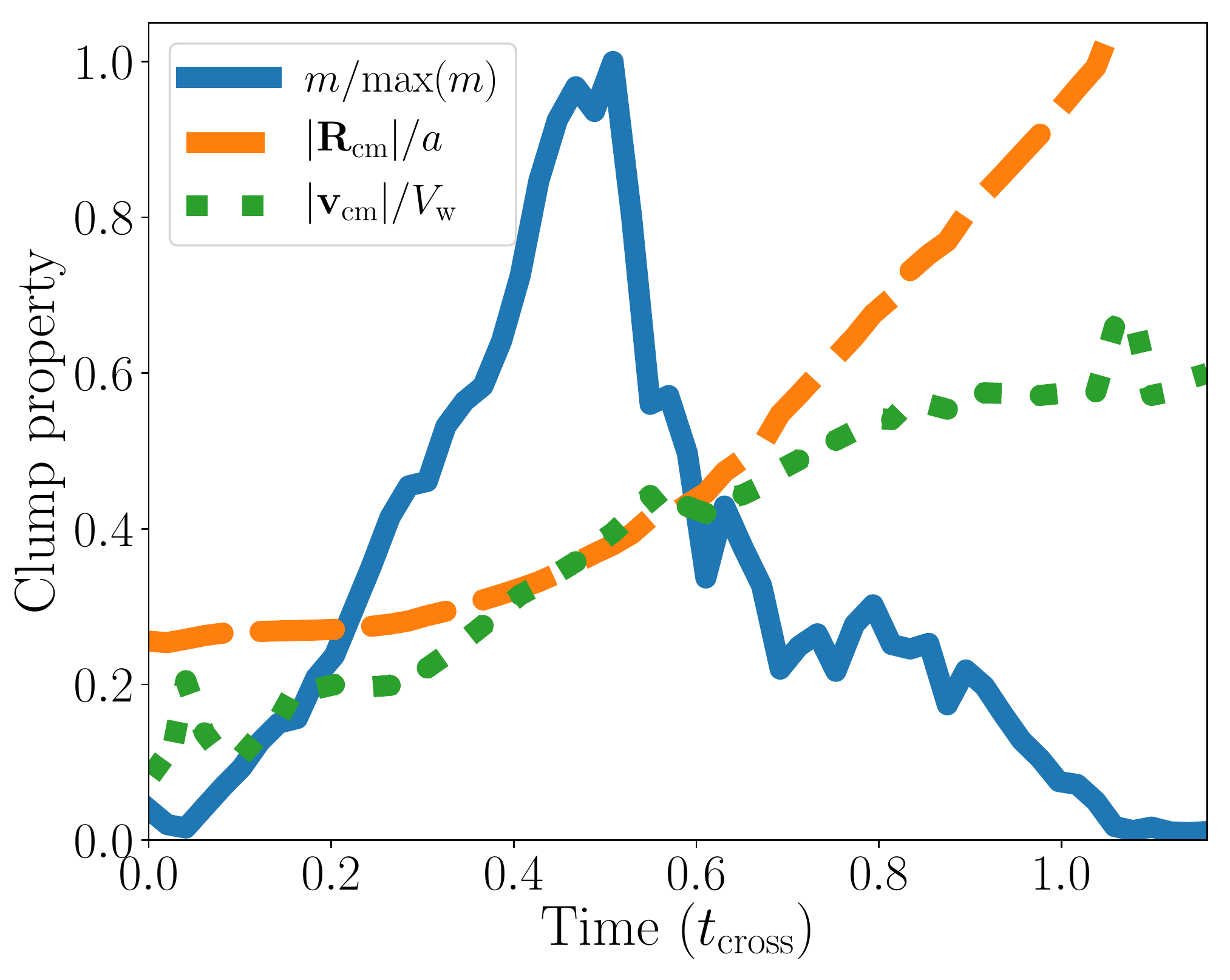}
				\caption{Clump evolution as a function of time. 
				Each panel represents a different clump tracked through the simulation. 
				Solid blue, dashed orange, and dotted green lines stand for clump mass, distance from the apex, and velocity magnitude, respectively. 
				Notice these clumps tend to reach their maximum mass at \hbox{$|\mathbf{R}_{\rm cm}|/a\approx0.3$}, being in both cases about $10^{-3}\rm\ M_{\oplus}$. 
				The process of formation, growth and advection take about one wind crossing timescale.}
				\label{fig:tracking}
			\end{figure*}

\begin{figure}
				\centering
				\includegraphics[width = 0.4\textwidth]{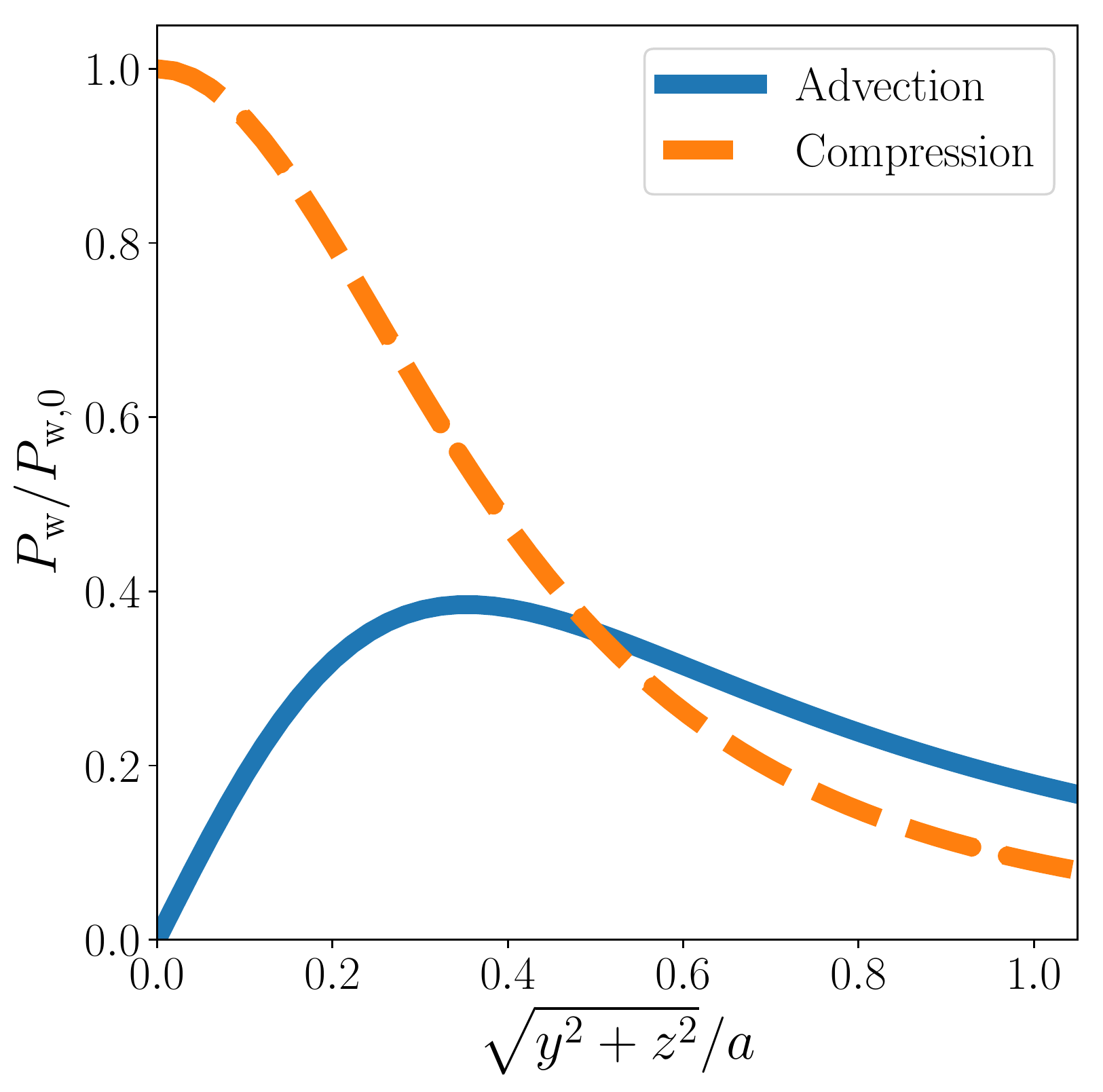}
				\caption{Ram pressure exerted by a single wind in the $yz$-plane as a function of the projected distance along the plane. 
				In our symmetric models this plane coincides with the rest position of the slab. 
				The pressure was scaled by the ram pressure at the apex and the distance by the stellar separation. 
				Also, it is divided in two components: the advective which is projected into the $yz$-plane (solid blue line) and the compressive that is parallel to the $x$-axis (dashed orange line). 
				Notice that there is a regime transition at \hbox{$\sqrt{y^2+z^2}/a=0.5$}.}
				\label{fig:ram}
			\end{figure}

				In some models it was possible to follow the evolution of clumps in detail.
				However, this analysis was performed only to the most massive clumps as the problem of tracking overdensities is not straightforward in Eulerian hydrodynamics. 
				The criterion for tracking clumps were based on extrapolating the position of a clump at a given snapshot into the next one, and then performing a search in it using the predicted position, as well as taking into account the mass of the clump. 
				In order to test that the algorithm was able to follow the same clump, we inspected visually the density maps of its vicinity to observe a coherent evolution of the overdensity. 
				Figure~\ref{fig:tracking} presents two examples of this analysis performed on model B10. 
				Specifically, it shows the time evolution of the clump physical properties, i.e. mass (solid blue line), distance from the apex (dashed orange line) and velocity magnitude (dotted green line). 
				Notice that all values were scaled to representative values in order to make a fair comparison between different objects. 
				The left panel shows that mass increases relatively fast, which seems to be related with the merging of similarly massive clumps. 
				On the other hand, the right panel shows how mass accumulates in a given object at a lower rate. 
				Both cases show that the maximum mass is reached at a distance of \hbox{$|\mathbf{R}_{\rm cm}|/a\approx0.3$} from the apex.
				After this point, clumps seem be to losing mass constantly and relatively fast. 
				This occurs as the result of the decrease of the ram-pressure confinement of the clumps as they are advected. 
				In particular, there is a transition of the pressure confinement regime around the point where the mass of clumps is maximal. 
				Figure~\ref{fig:ram} shows the ram pressure of the winds as a function of the distance from the apex, although projected onto the $yz$-plane. 
				Here we divided the pressure into two components: compressive and advective. 
				We defined the compression as the component parallel to the $x$-axis, i.e. perpendicular to the slab at the apex, which causes the slab confinement.
				Meanwhile, the pressure responsible for the advection of the clumps away from the domain is the component on the $yz$-plane. 
				Both values were scaled by the value of the wind ram pressure at the apex, which is given by \hbox{$P_{\rm w,0}=\dot{M}_{\rm w}V_{\rm w}/(\pi a^2)$}. 
				This analysis shows how the compression of the slab decreases with distance and, more specifically, that it starts being smaller than the advection component at about \hbox{$\sqrt{y^2+z^2}/a\approx0.5$}. 
				After this point the slab, and clumps, lose a significant fraction of their confinement, which results into their mass loss and, in some cases, in their eventual destruction. 
				Furthermore, the advection strength, although slower, decreases constantly, which also can explain the fact that most of the clump momenta is gained at \hbox{$|\mathbf{R}_{\rm cm}|/a\lesssim0.5$}. 
				Finally, clumps leave the domain (\hbox{$|\mathbf{R}_{\rm cm}|/a\approx1$}) at about \hbox{$t\approx t_{\rm cross}$}, which means they travel in average at about half of the wind speed. 
				
			\begin{table}
			\begin{adjustbox}{width=\columnwidth,center}
			\begin{threeparttable}
            			\caption{Ejected clump properties}
            			\begin{tabular}{|l|c|c|c|c|c|}
				\hline
				Name	&	$\bar{m}_{\rm out}$	& $\sigma_{m,\rm out}$	&	${\rm max}(m_{\rm out})$	&	$\bar{v}_{\rm out}$	&	$\sigma_{v,\rm out}$\\
						&	$\rm M_{\oplus}$						&	($\rm dex$)			&	$\rm M_{\oplus}$	&	($\rm km\ s^{-1}$)	& ($\rm km\ s^{-1}$)\\
				\hline	
				B10		&	$10^{-3.6}$	&	$0.3$	&	$10^{-2.8}$	&	$299$	&	$26$\\
				B+10	&	$10^{-3.5}$	&	$0.5$	&	$10^{-2.7}$	&	$483$	&	$53$\\
				C10		&	$10^{-4.0}$	&	$0.3$	&	$10^{-3.5}$	&	$279$	&	$25$\\
				C+10	&	$10^{-4.1}$	&	$0.5$	&	$10^{-3.3}$	&	$438$	&	$45$\\
				D10		&	$10^{-4.6}$	&	$0.4$	&	$10^{-3.9}$	&	$303$	&	$25$\\
				D+10	&	$10^{-4.7}$	&	$0.4$	&	$10^{-4.0}$	&	$403$	&	$36$\\
				\hline
				B9		&	$10^{-3.0}$	&	$0.5$	&	$10^{-2.2}$	&	$293$	&	$24$\\
				B11		&	$10^{-3.9}$	&	$0.4$	&	$10^{-3.0}$	&	$296$	&	$24$\\
				\hline
				BA10	&	$10^{-3.4}$	&	$0.5$	&	$10^{-2.7}$	&	$409$	&	$84$\\
				BA+10	&	$10^{-3.2}$	&	$0.5$	&	$10^{-2.3}$	&	$388$	&	$79$\\
				\hline
		        		\end{tabular}
				\label{tab:results}
				\begin{tablenotes}
					\item \textit{Notes.} 
					Summarised results: clump statistical physical properties per model. 
					Column~1: name of simulation run.
					Column~2: mean mass fraction in clumps at \hbox{$0.5<\mathbf{R}/a\leq 1$} in Earth masses. 
					Column~3: standard deviation of the clump mass distribution in $\rm dex$. 
					Column~4: maximum clump mass in Earth masses. 
					Column~5: mean clump speed at \hbox{$0.5<\mathbf{R}/a\leq 1$}. 
					Column~6: standard deviation of the clump speed at \hbox{$0.5<\mathbf{R}/a\leq 1$}.
				\end{tablenotes}
			\end{threeparttable}
			\end{adjustbox}
        			\end{table}
			
		\subsubsection{The effects of wind speed and stellar separation}
		
			Although so far we have focused on describing the model B10, the qualitative behaviour of symmetric models is very similar.  
			Hence its description also applies for the rest of those models. 
			Now, we will present the results on how the clump properties are affected by changing the wind speed and/or the stellar separation. 
			Summarised results are shown shown in Table~\ref{tab:results}. 
			Here we included the mean mass fraction in clumps and mean velocity, their dispersion and the maximum clump mass of each model studied. 
			In model B+10 we increased the velocity of each wind by $50$~per~cent compared to B10, by doing so we found that both the clump mass and velocity distributions have a slightly larger dispersion. 
			Notice that this is consistent with the fact that the density maps of this model also showed starker contrasts compared to B10. 
			Nevertheless, the mean values do not seem to change significantly, which might be the net effect of a higher compression and, at the same time, a more diluted wind caused by the presence of faster winds. 
			The clump velocity ejection is also observed to be of ${\sim}3/5$ of the wind speed.
			In the case of models with a smaller stellar separation, the mean mass fraction in clumps decreases by \hbox{${\sim}0.5\rm\ dex$} in model C10, and becomes slightly smaller than this in C+10 (see Table~\ref{tab:results}). 
			However, the outer clump velocities follow the same previous relations where the mean speed is about 60~per~cent of the wind speed and its dispersion is $10$~per~cent of the mean velocity. 
			If we consider the even smaller stellar separation in models D10 and D+10, we observe the same evolution overall, although clumps are now ${\sim}1\rm\ dex$ lighter than B10 and B+10. 
			Furthermore, clumps seem to be accelerated and ejected following the same proportion observed in the other models. 
			In summary, higher speeds cause clump mass and velocity distributions having larger dispersions, while clump masses seem to be correlated with the stellar separation. 
		
\begin{figure}
				\centering
				\includegraphics[width = 0.45\textwidth]{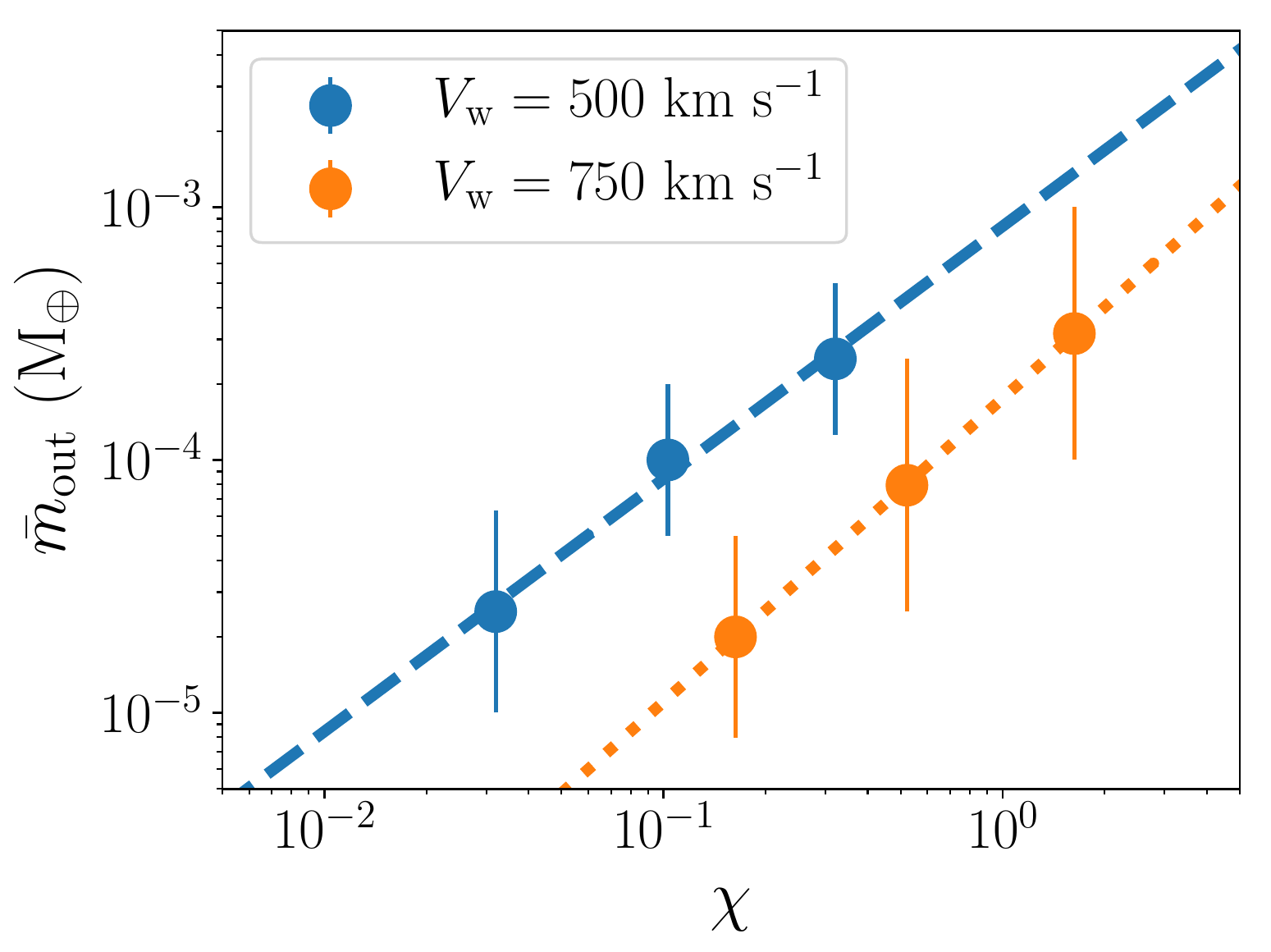}
				\caption{Mean mass fraction in ejected clumps $\bar{m}_{\rm out}$ as a function of the cooling parameter $\chi$. 
				Each point represents each symmetric model simulated. 
				Error bars show the standard deviation of the distribution which is about \hbox{$0.6\rm\ dex$}. 
				Blue and orange points stand for models whose wind speed is $500$ and \hbox{$750\rm\ km\ s^{-1}$}, respectively. 
				The dashed and dotted lines correspond to the best linear fit to each set of models.}
				\label{fig:results}
			\end{figure}

			Now we proceed to analyse the systematic differences observed in clump masses with stellar separation.  
			Figure~\ref{fig:results} shows the mean mass fraction of the outer clumps (with their respective dispersion) as a function of the cooling parameter of the winds estimated at the apex of each system. 
			Notice that this analysis only considers the symmetric models. 
			Here we can separate them into two families according to their wind speed. 
			The blue and orange points show models with $500$ and \hbox{$750\rm\ km\ s^{-1}$}, respectively. 
			We fit the best linear function (in logarithmic scale) to each family of points:
			
			\begin{eqnarray}
				\log\left(\frac{\bar{m}_{\rm out}}{\rm M_{\oplus}}\right) & = & 1.0 \log\chi - 3.1;\ V_{\rm w} = 500{\rm\ km\ s^{-1}}, \\
				\log\left(\frac{\bar{m}_{\rm out}}{\rm M_{\oplus}}\right) & = & 1.2 \log\chi - 3.8;\ V_{\rm w} = 750{\rm\ km\ s^{-1}}. 
			\end{eqnarray}
			
			\noindent Such relations confirm the fact that the slower the slab cools down the more massive the clumps can be.
			
			Recalling the definition of the cooling parameter, we can recover the dependence of the mean mass fraction in clumps as a function of the stellar separation. 
			This results in \hbox{$\bar{m}_{\rm out}\propto a$} for symmetric systems. 
			Thus, as long as the winds are radiative, the mass of the clumps, in general, increases with the stellar separation.
			
			\begin{figure*}
				\centering
				\subfloat[][Model BA+10: clump mass histogram]{\includegraphics[width=0.485\textwidth]{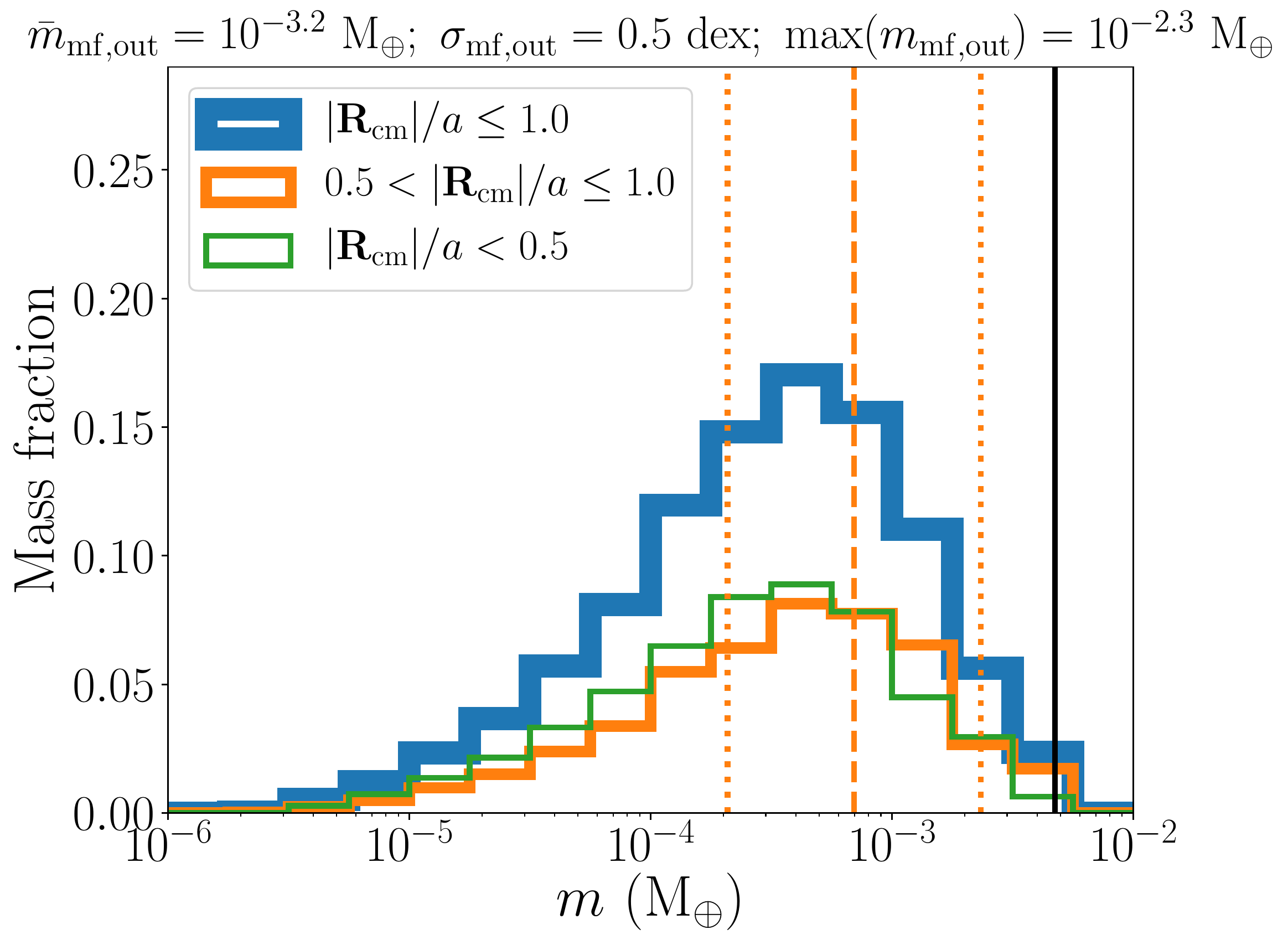}\label{fig:out_asym:1}}
				\hspace{0.25cm}
				\subfloat[][Model BA+10: clump velocity histogram]{\includegraphics[width=0.45\textwidth]{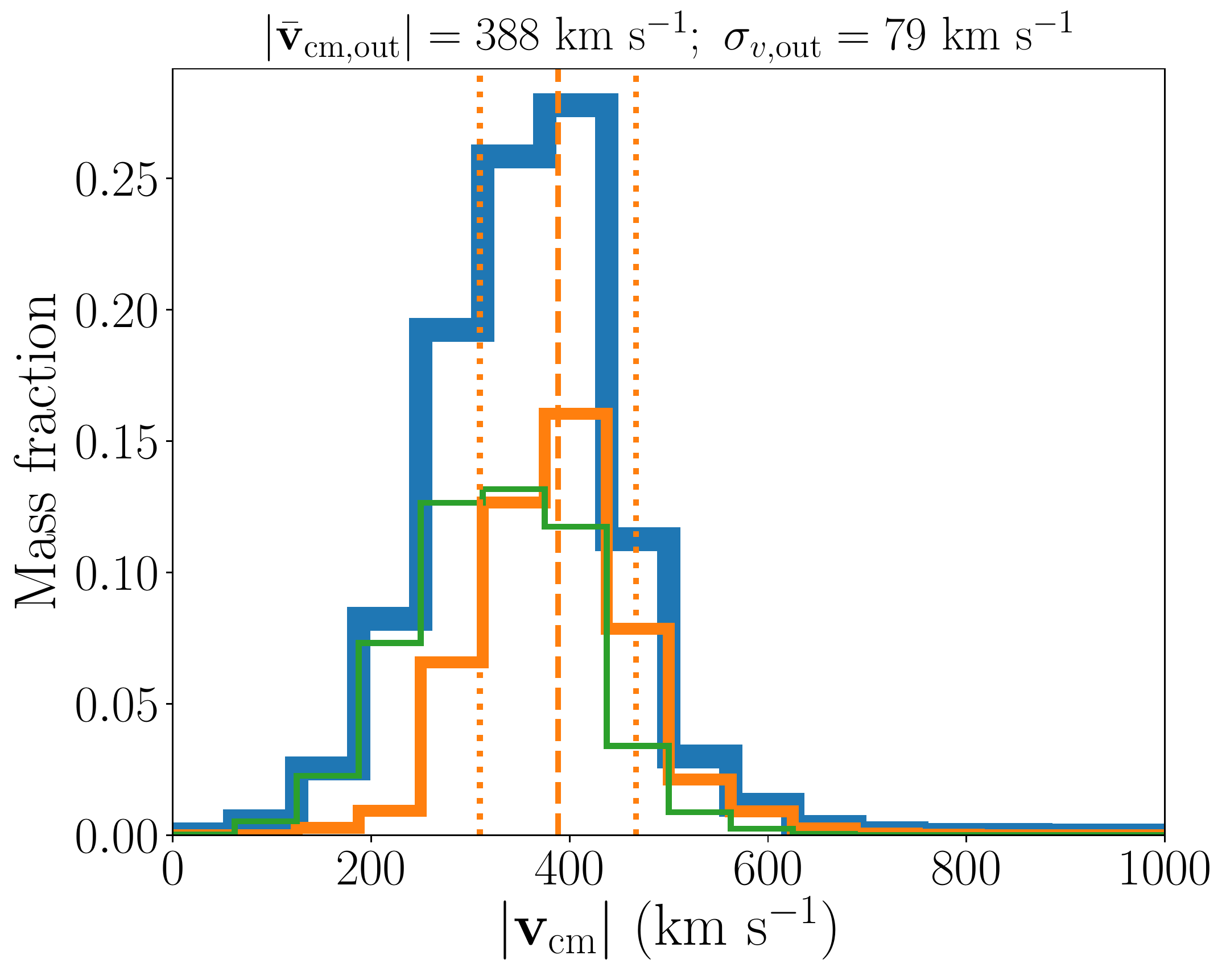}\label{fig:out_asym:2}}
				\caption{Analogous to Figure~\ref{fig:out} but for model BA+10. 
				Notice that the dispersion of each distribution is larger in this case compared to B10.}
				\label{fig:out_asym}
			\end{figure*}	
			
		\subsubsection{Clumps in asymmetric systems}
			
			Finally, we present the properties of clumps in the asymmetric models.
			Figure~\ref{fig:out_asym} shows clump mass fraction in bins of mass and speed of model BA+10. 
			Although the mean mass fraction value is roughly the same, the mass distribution spans a larger mass range compared to model B10  (see Figure~\ref{fig:out_asym:1}). 
			For instance, the most massive clump formed reaches a significantly higher mass than model B10 of about 1~$\rm dex$. 
			This feature can be attributed to the higher Mach number of one of the winds, which compresses the material stronger and, at the same time, generates a more turbulent state in the unstable slab. 
			It is also important to notice the differences in the distribution of clump masses.  
			In model BA+10 the mass of clumps in the inner and outer regions are extremely similar, both having peaks at low mass. 
			However, in model B10 the inner and outer clump masses are much more different. 
			Inner clumps are mostly very light, while outer clumps in general are more massive (see Figure~\ref{fig:out:1}). 
			The presence of different instability mechanisms could be the explanation of these differences.
			Let us remember that in asymmetric models the NTSI is acting along with the KHI. 
			Thus, it is possible that the growth of clumps created by the former process is being limited by the latter.
			
			In the case of the velocity distribution, the mean ejection speed of outer clumps is larger than in symmetric models. 
			For instance, clumps in model BA+10 reach about \hbox{${\sim}430\rm\ km\ s^{-1}$}, which is about ${\sim}90$~per~cent of the weaker wind speed. 
			We also notice that the dispersion is larger than in symmetric cases (see Figure~\ref{fig:out_asym:2}) being roughly ${\sim}20$~per~cent of the mean speed. 
			Therefore, the stronger (faster) wind contributes to accelerate the material to higher speeds, but the fact that the slab is subject to the KHI also causes larger fluctuations in the velocity field. 
			Although not shown here, the properties of clumps of model BA10 also show differences compared to model B10.
			The clumps of model BA10 show a larger dispersion in the clump mass and velocity distribution compared to model B10, though not as much as BA+10 (see Table~\ref{tab:results}). 
			This points to the fact that the degree of asymmetry and, more importantly, the wind velocity difference, is the cause behind such differences. 
		
\section{Discussion}
\label{sec:discussion}

	In this section we proceed to compare the results of this work with previous analytical estimates mainly from \citet{C16}. 
	Furthermore, we discuss the choice of the resolution employed in our simulations, and study the convergence of the results. 
	Finally, we discuss the implications of the results on the Galactic Centre hydro- and thermodynamics.

	\subsection{Comparison with analytical estimates}
	\label{sec:comparison}
	
		\citet{C16} estimated the range of clump masses expected in symmetric wind collisions. 
		For systems like model B10, the analytic estimates show that clumps could have masses up to the order of ${\sim}{\rm M}_{\oplus}$. 
		The results of the model B10 hydrodynamic simulations do adhere to this analytical upper limit since clumps are formed with \hbox{${\sim}10^{-3}\rm\ M_{\oplus}$} at most (see Table~\ref{tab:results}). 
		However, there is a difference of three order of magnitudes between the theoretical and the numerical clump mass upper limit, and this difference does not change if we analyse other models either. 
		To explain this difference we have to bear in mind that there are differences in the geometry between the analytical and numerical models: $i)$ the planar and spherical winds, and $ii)$ the ``0D" and 3D approaches. 
		Stellar winds are naturally closer to being spherical rather than planar. 
		Thus, in a collision of spherical stellar winds the maximum compression of the slab occurs solely at the apex, and not through the entire slab, which is one of the implications of the planar wind assumption. 
		Furthermore, the density of a spherical wind changes with the distance from the star while in the case of a planar wind it is assumed to be constant. 
		Therefore, if the amplitude of the NTSI increases, it suffers from a damping effect as the ram pressure is stronger approaching to any of the stars, which does not act in the case of colliding planar flows. 
		Secondly, modelling the complete system with the appropriate geometry in 3D can differ significantly from the simple ``0D" approach of considering an infinite slab. 
		The fact that the gas in the slab can be advected away from the apex can affect how the gas can concentrate on the knots of the perturbed shell. 
		As discussed in previous sections, the material is advected quicker as it moves away from the apex due to the acceleration caused by the wind ram pressure. 
		Thus, it is possible that the idealised geometry of the analytical model could account, at least partially, for the difference in the clump mass upper limits. 
		
		\begin{figure*}
			\centering
			\subfloat[][Model B9: density map at $z=0$]{\includegraphics[width=0.33\textwidth]{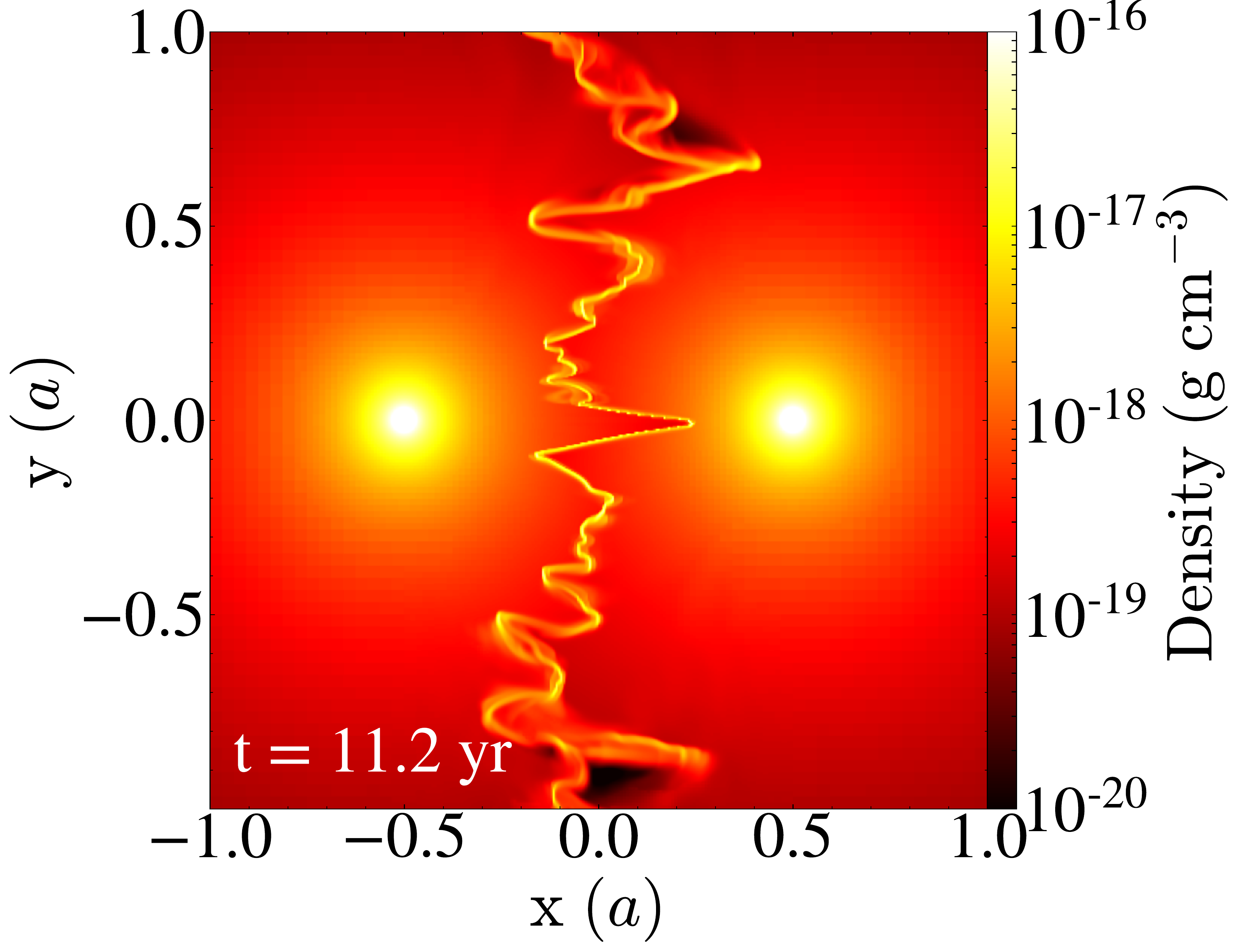}\label{fig:res:B9_z}}
			\subfloat[][Model B10: density map at $z=0$]{\includegraphics[width=0.33\textwidth]{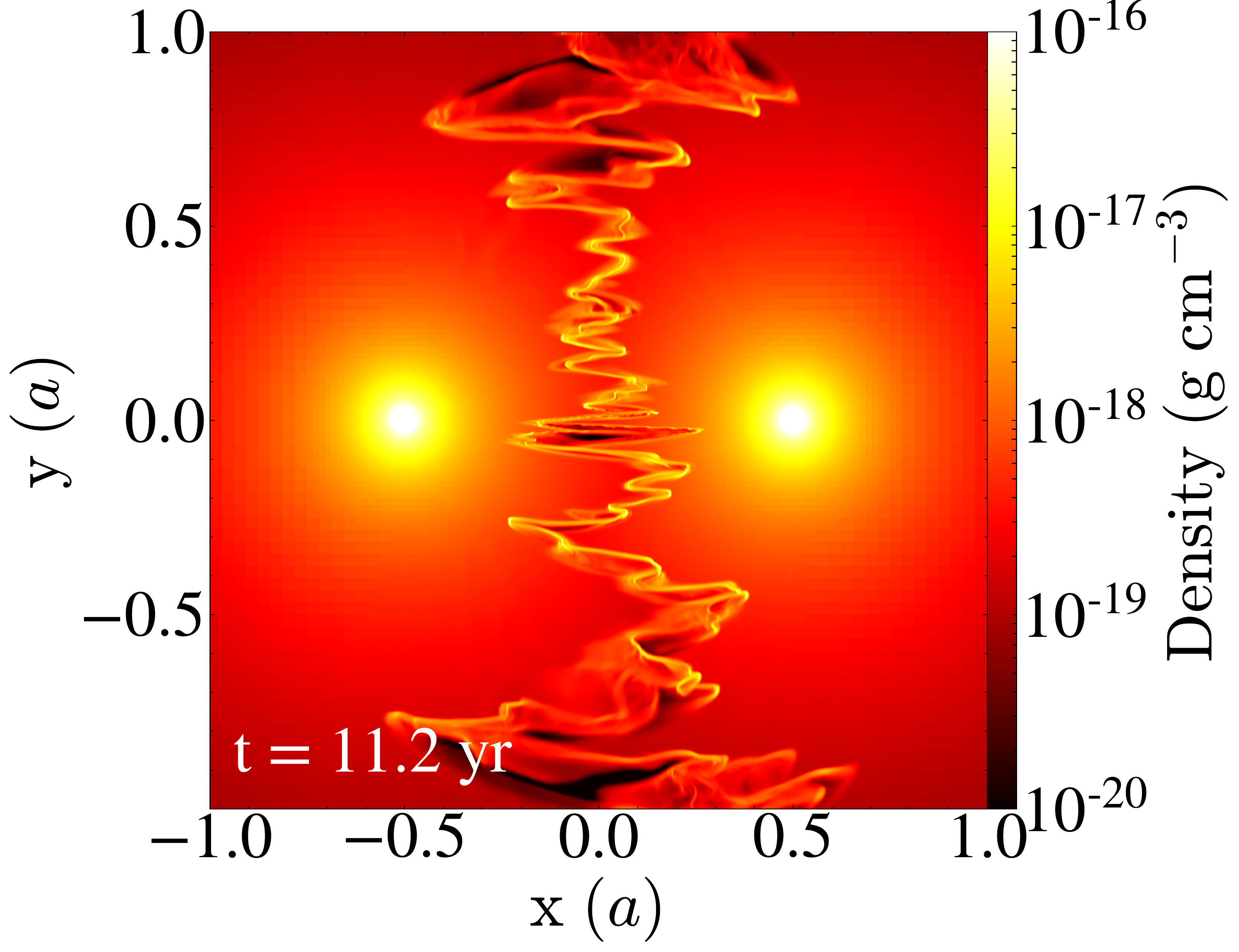}\label{fig:res:B10_z}}
			\subfloat[][Model B11: density map  at $z=0$]{\includegraphics[width=0.33\textwidth]{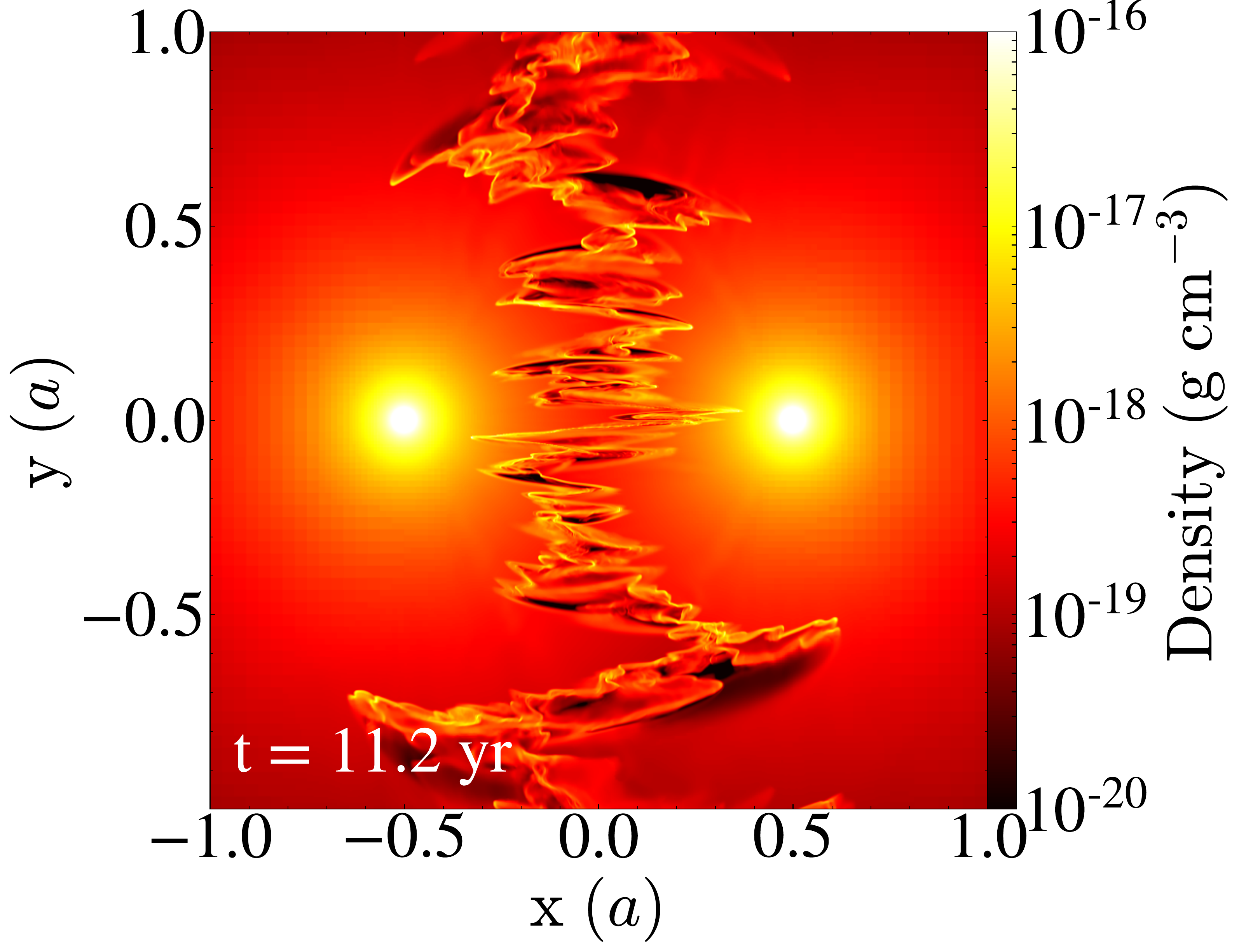}\label{fig:res:B11_z}}
			\newline
			\subfloat[][Model B9: density map at $x=0$]{\includegraphics[width=0.33\textwidth]{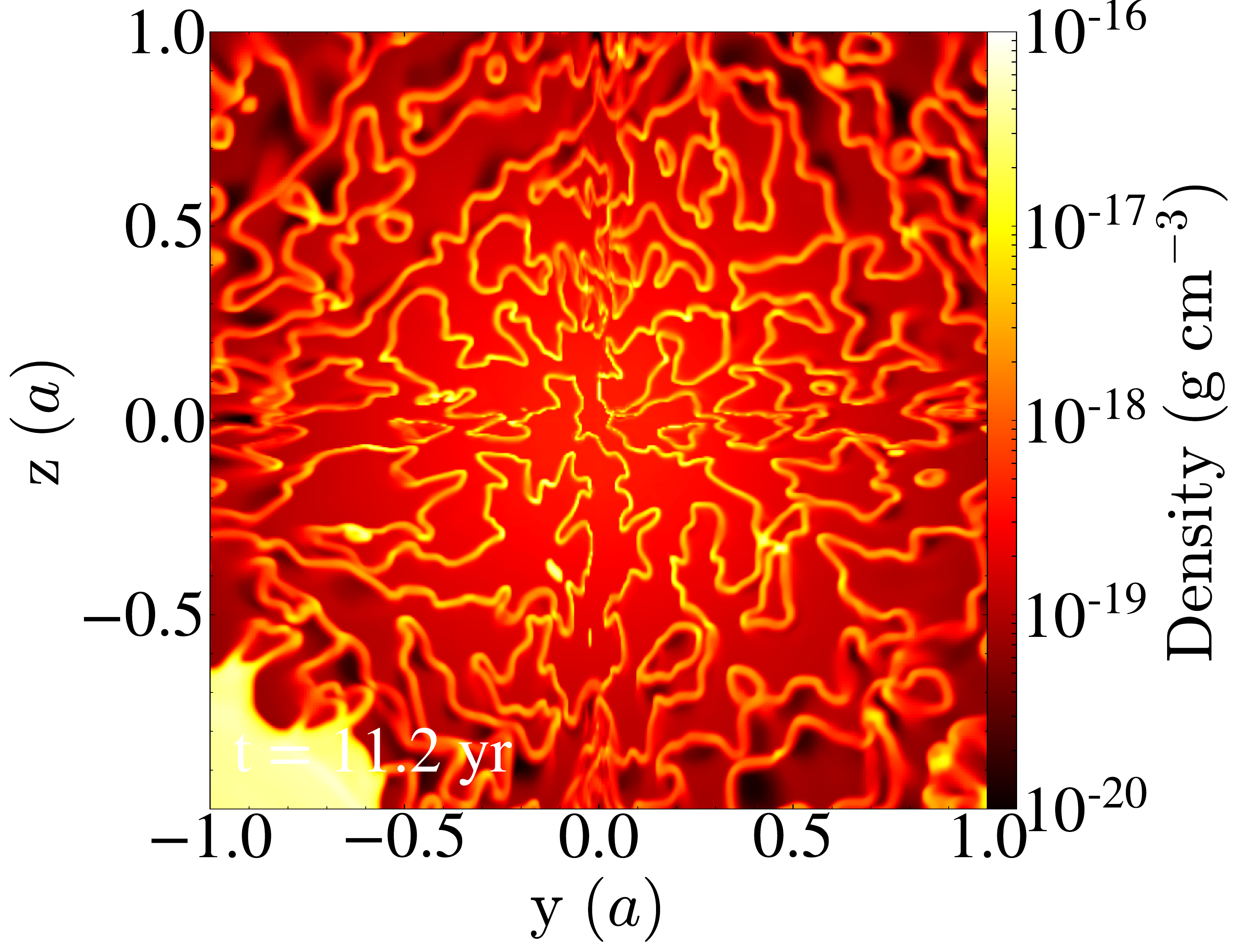}\label{fig:res:B9_x}}
			\subfloat[][Model B10: density map at $x=0$]{\includegraphics[width=0.33\textwidth]{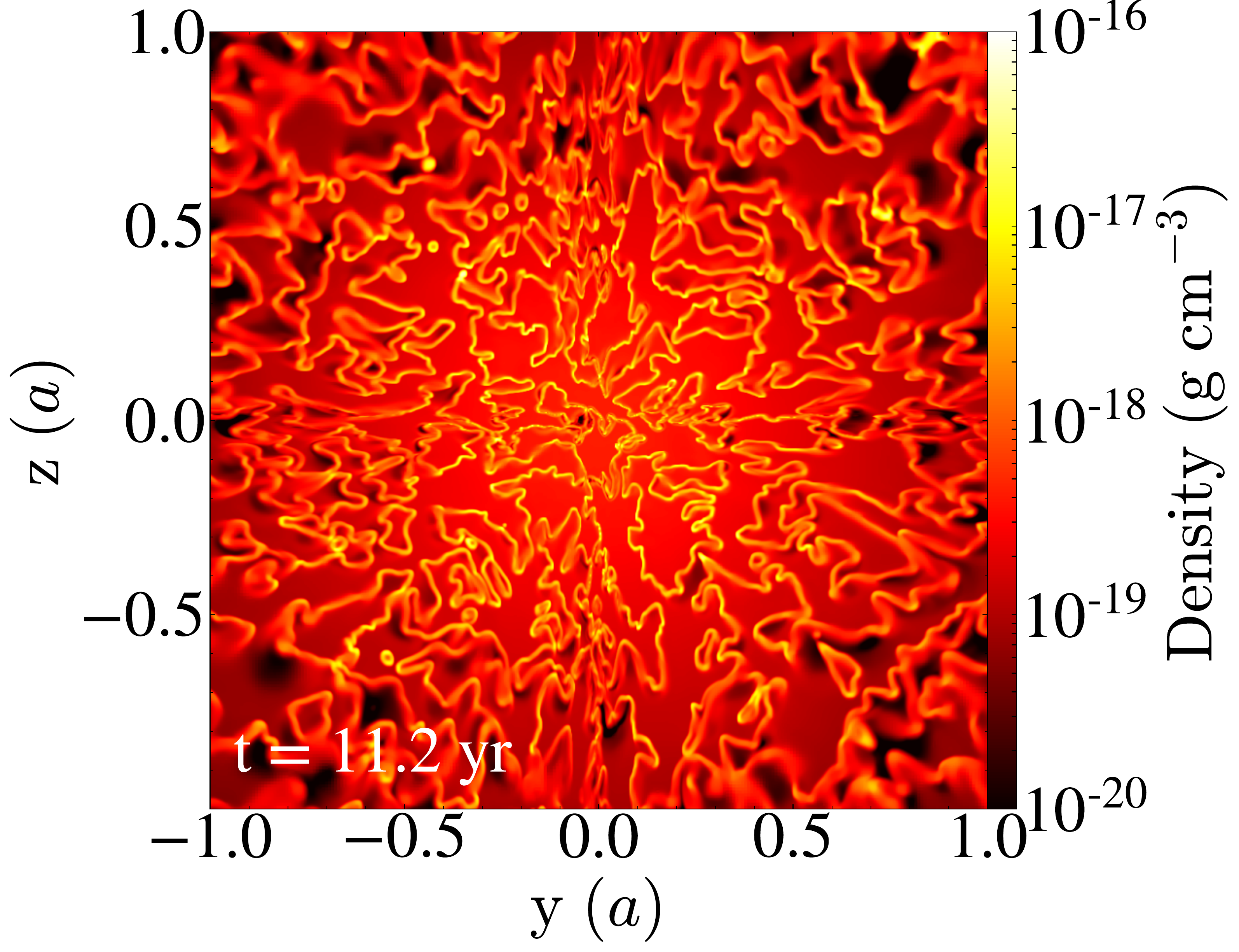}\label{fig:res:B10_x}}
			\subfloat[][Model B11: density map at $x=0$]{\includegraphics[width=0.33\textwidth]{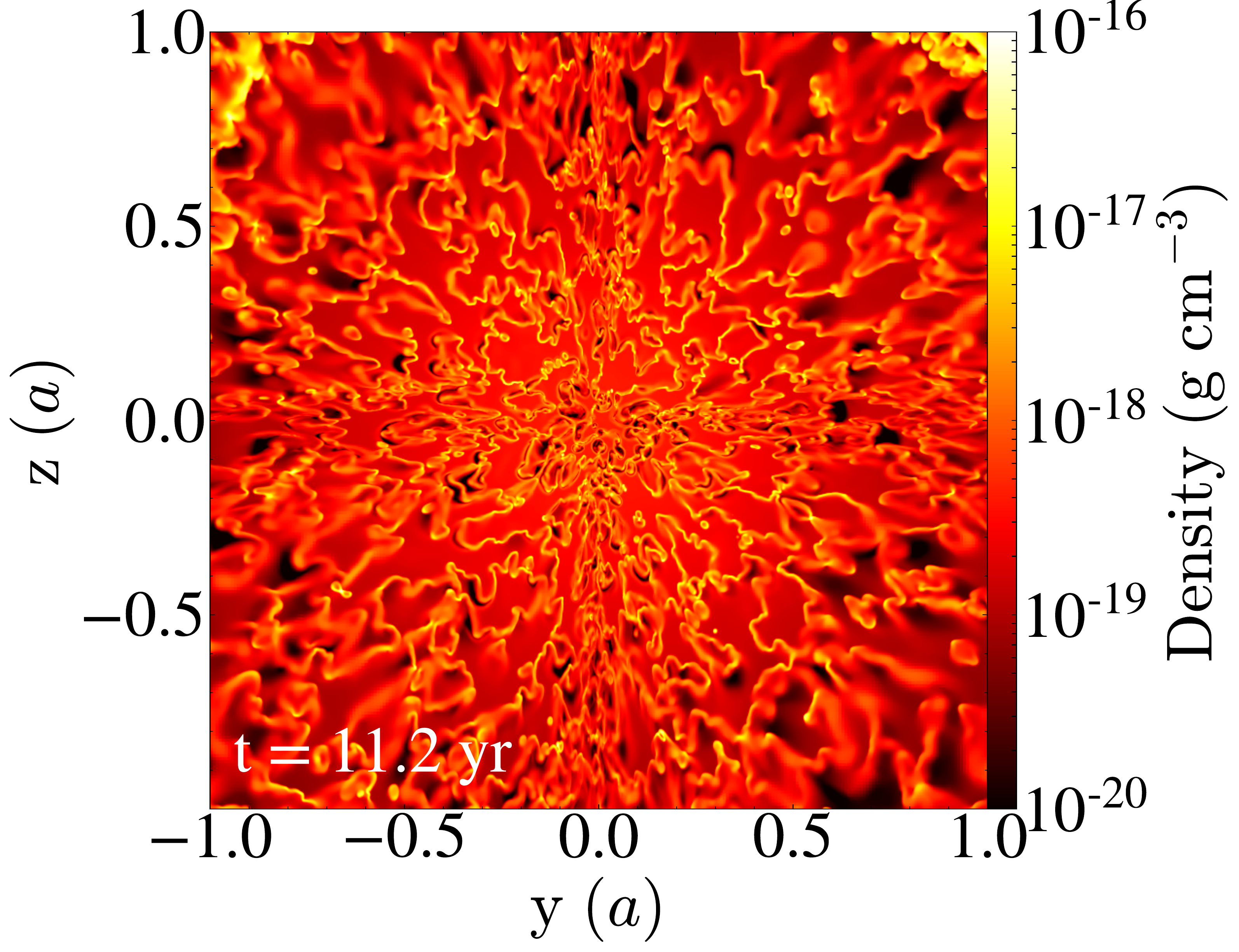}\label{fig:res:B11_x}}
			\newline
			\caption{
			Density maps of models B9, B10 and B11 along the left, central and right columns, respectively.   
			The upper row shows maps at $z=0$ while the lower row presents maps at $x=0$. 
			Every map corresponds to exactly the same simulation time \hbox{$t=11.2\rm\ yr$}.
			Notice that at large scale models look very similar. 
			However, it is possible to recognise differences at small scales as the higher the resolution the finer structure can be seen due to shorter unstable modes excited. 
			The dense feature observed in left lower corner of Figure~\ref{fig:res:B9_x} is related to the initial wiggles observed away from the apex, which have not been completely advected away from the domain yet.}
			\label{fig:res}
		\end{figure*}
		
		\begin{figure*}
			\centering
			\subfloat[][Model B9: clump mass histogram]{\includegraphics[width=0.33\textwidth]{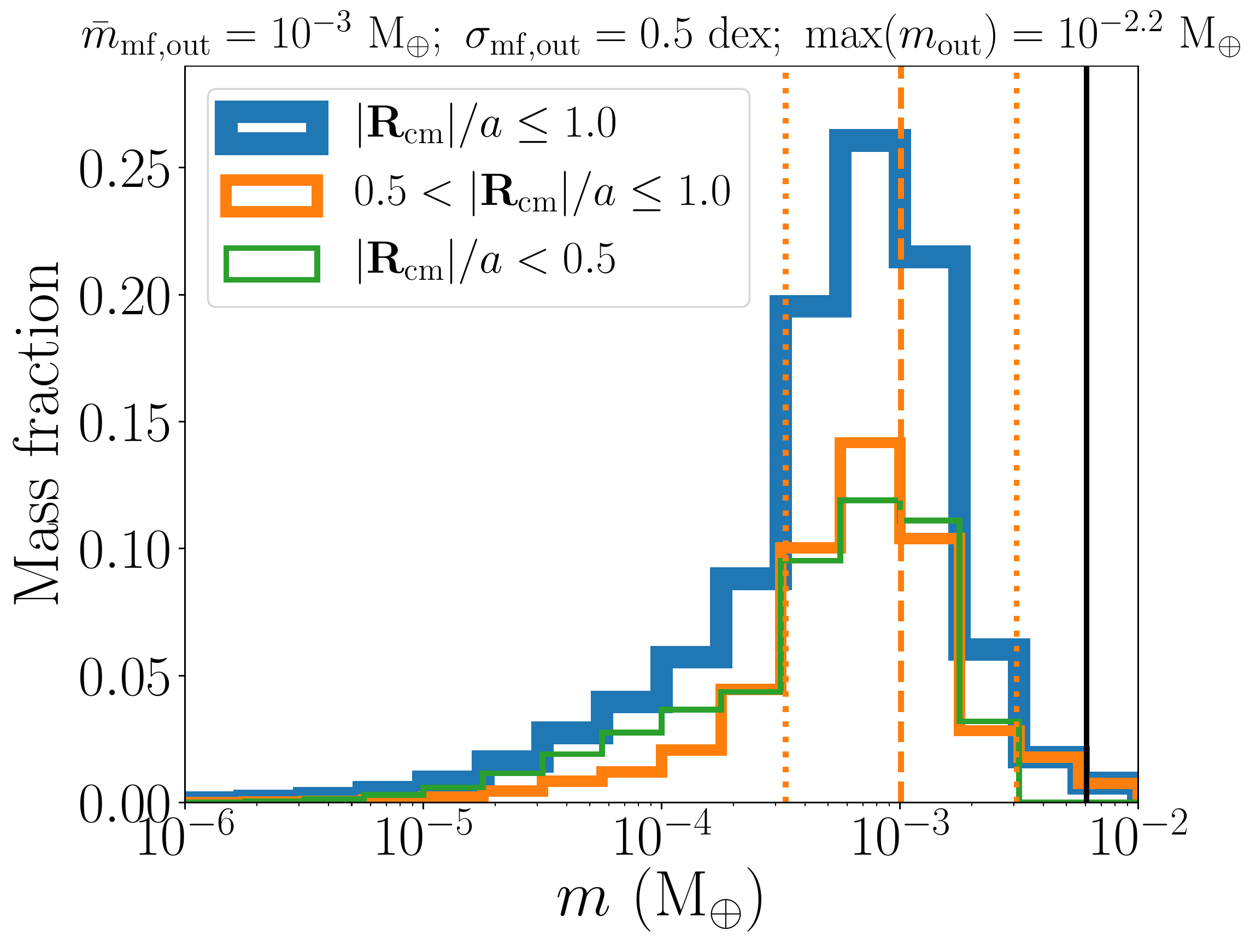}\label{fig:res_hMs:B9}}
			\subfloat[][Model B10: clump mass histogram]{\includegraphics[width=0.33275\textwidth]{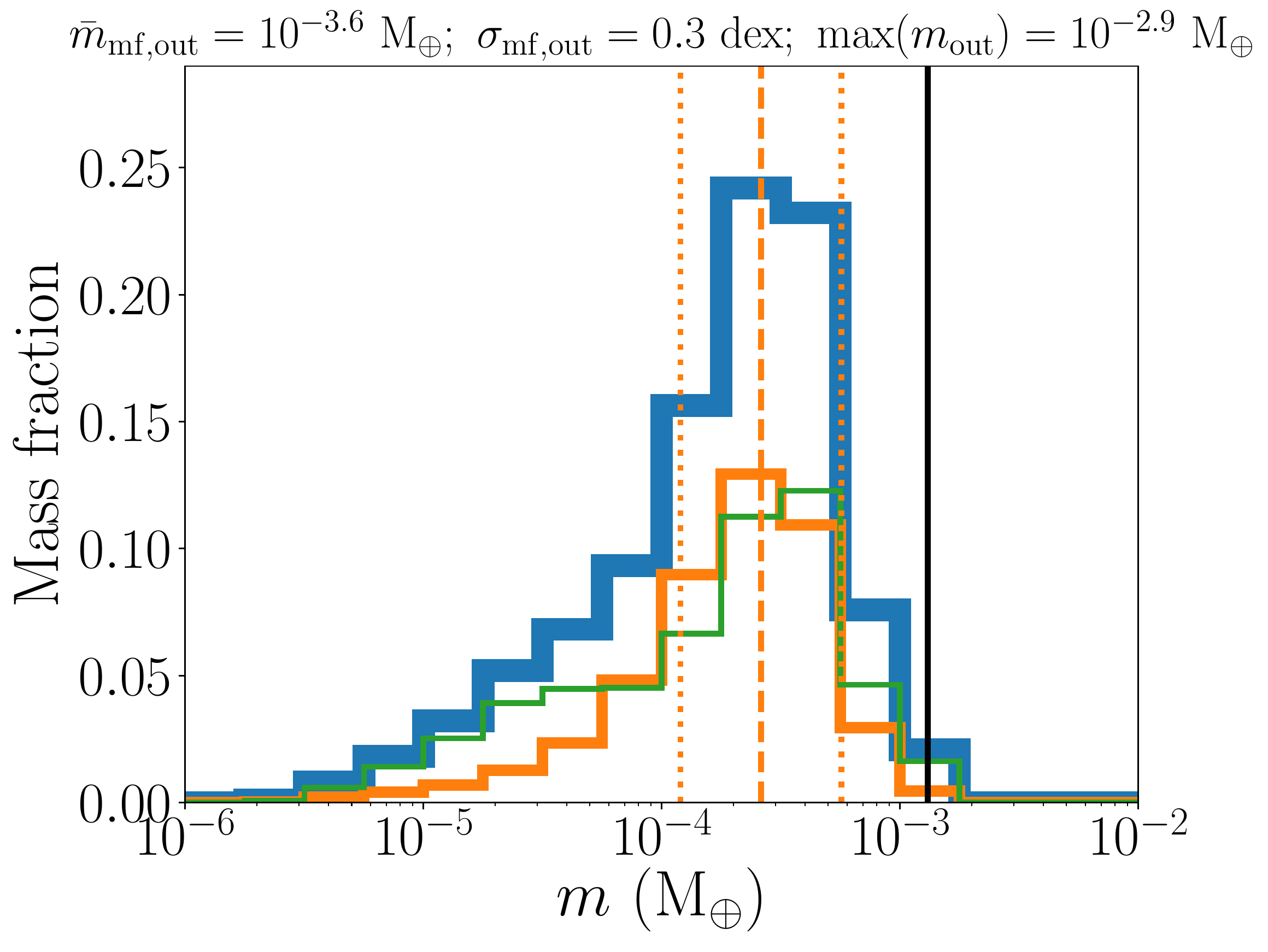}\label{fig:res_hMs:B10}}
			\subfloat[][Model B11: clump mass histogram]{\includegraphics[width=0.33\textwidth]{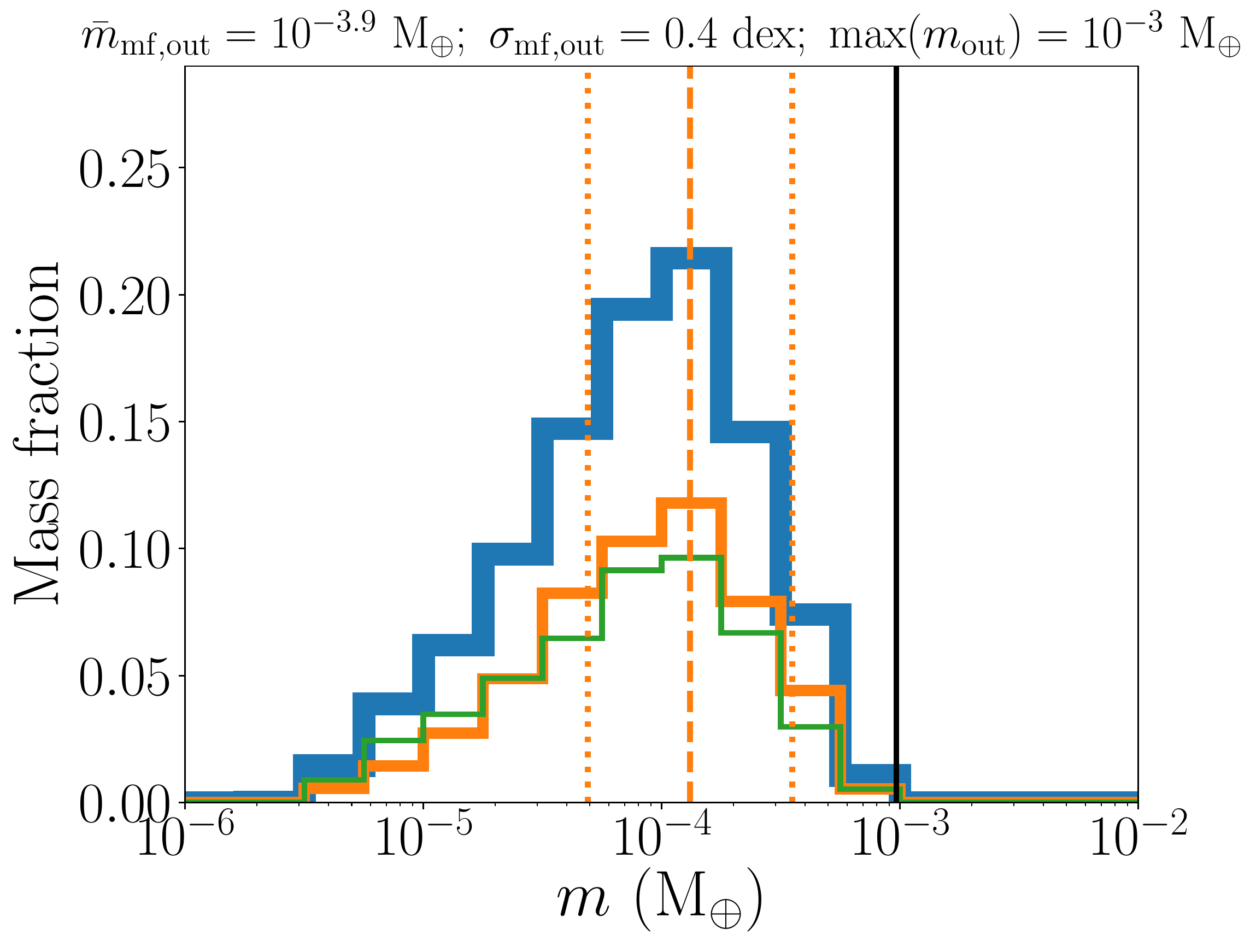}\label{fig:res_hMs:B11}}
			\newline
			\subfloat[][Model B9: clump velocity histogram]{\includegraphics[width=0.33\textwidth]{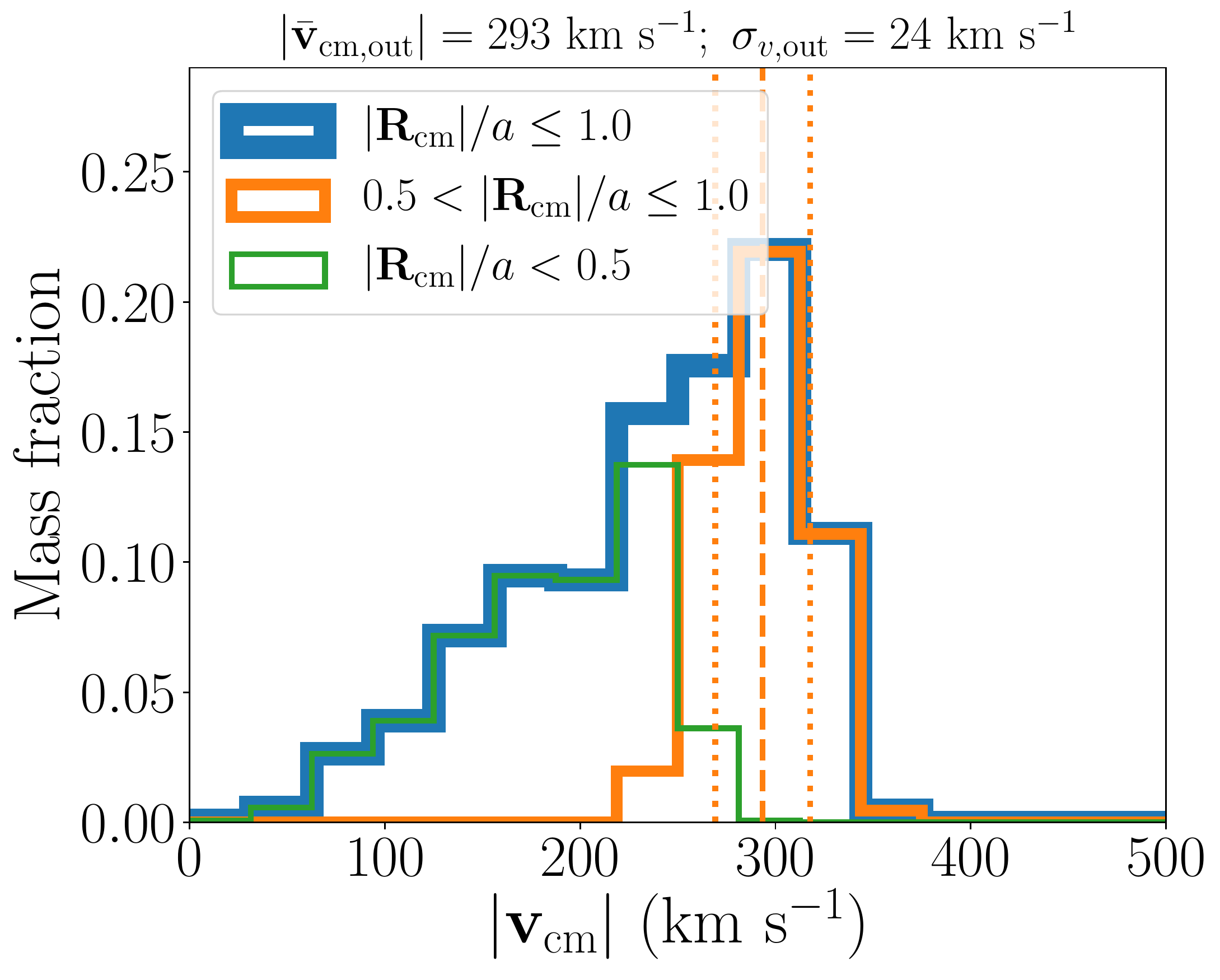}\label{fig:res_hV:B9}}
			\subfloat[][Model B10: clump velocity histogram]{\includegraphics[width=0.33\textwidth]{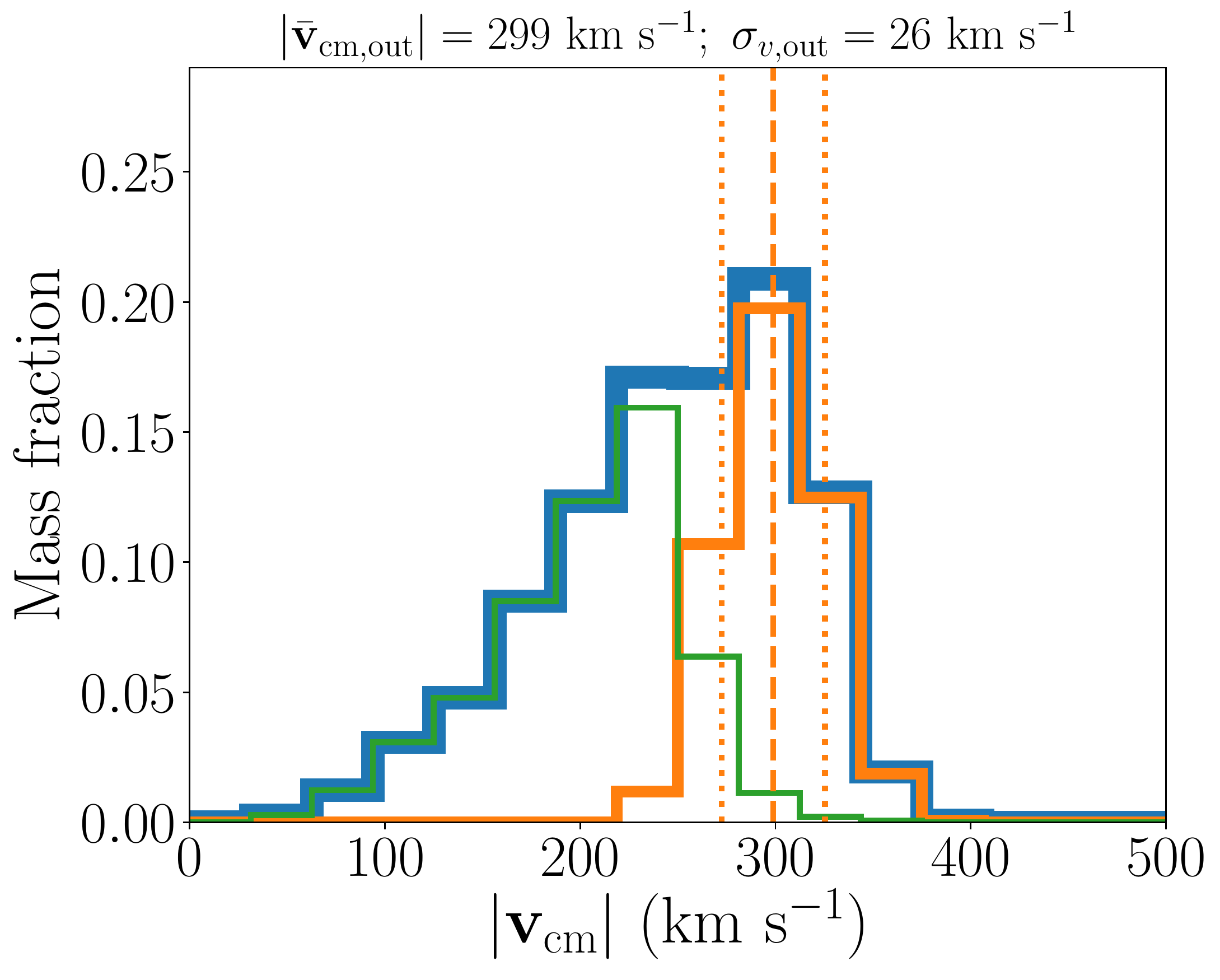}\label{fig:res_hV:B10}}
			\subfloat[][Model B11: clump velocity histogram]{\includegraphics[width=0.33\textwidth]{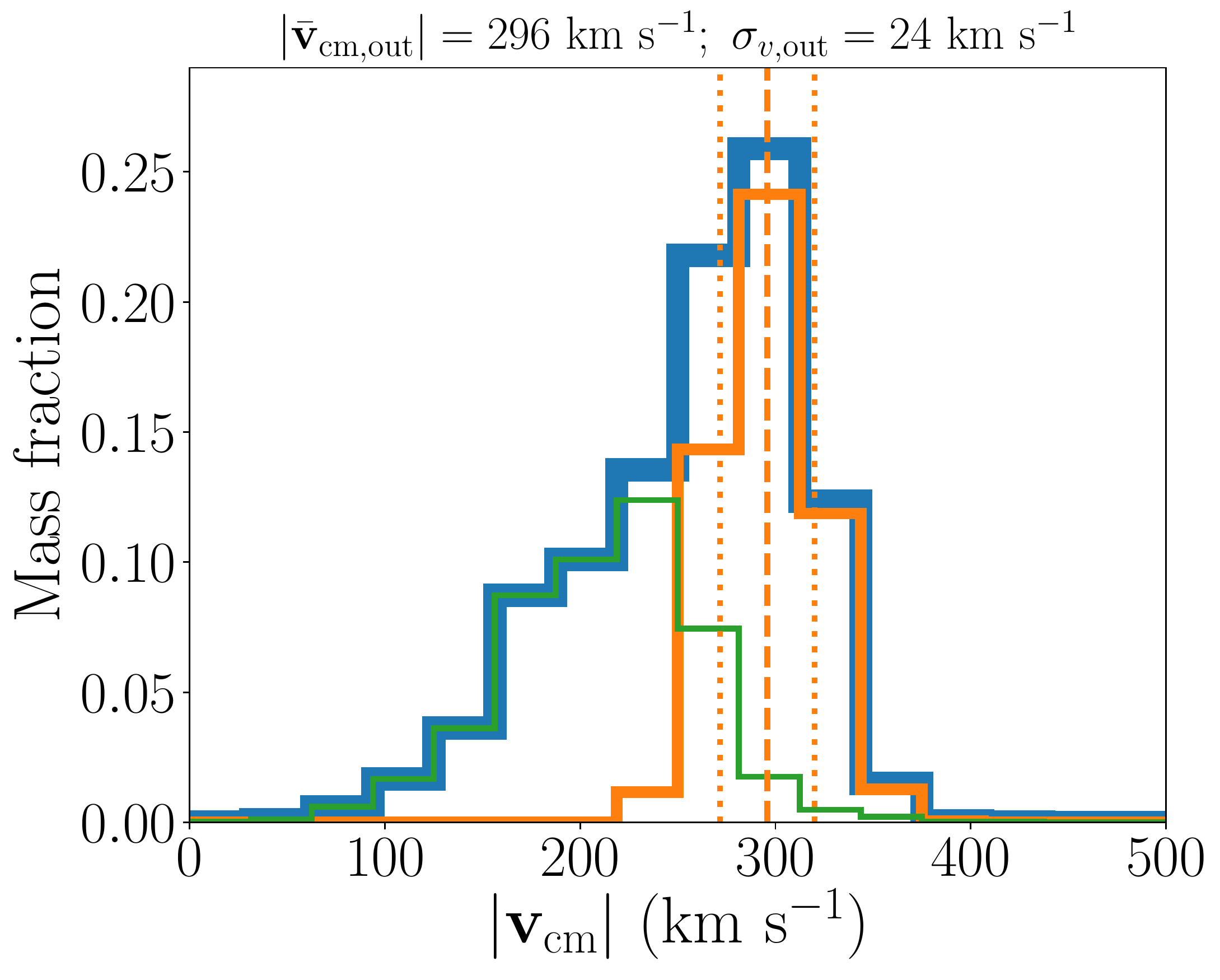}\label{fig:res_hV:B11}}	
			\caption{Clump mass fraction in bins of mass and velocity of ejected clumps of models B9, B10 and B11. 
			The upper and lower rows show the clump mass fraction in bins of mass and velocity, respectively. 
			From left to right, each column contains the analysis of models B9, B10 and B11. 
			In order to make a fair comparison between different resolution models we selected clumps only above a certain physical size.  
			}
			\label{fig:res_h}
		\end{figure*}

	\subsection{Impact of spatial resolution}
	\label{sec:converge}
		
		We have studied the properties of clumps formed in stellar winds collisions namely their initial mass and velocity. 
		In order to follow the clump formation process in detail, ideally we would need to resolve sizes comparable with the unstable wavelengths excited in the slabs. 
		As the shortest wavelength of the NTSI is set by the width of the slab, the challenge is to be able to resolve such lengths as well as possible. 
		In a radiative wind collision once the slab cools down it can become extremely thin, as seen in model B10 (see Figure~\ref{fig:B10:2}). 
		However, this situation changes rapidly once instabilities are excited in the slab as it becomes slightly wider as compression diminishes because shocks are not strictly perpendicular to the slab anymore.  
		Previously, \citet{L11} warned about the computational difficulty of resolving wind confined thin shells. 
		In their study, they ensured to have at least eight cells for resolving the slab in order to capture the development of the instabilities, this was possible given that such models considered an isothermal equation of state in 2D, which significantly decreases the computational cost. 
		Unfortunately, in 3D even with the aids of the AMR technique it is difficult to afford such resolution. 
		The main problem is related to the size of the slab which is usually refined up to the maximum level increasing significantly the computational cost of the simulations. 
		Therefore, it is not possible to enhance the resolution much more from our standard setup. 
		Nevertheless, we ran a couple of tests at lower and higher resolution to illustrate how the results could be affected by the resolution used in this study. 
		Specifically, we investigated models B9 and B11, whose effective maximum resolution are $512^3$ and $2048^3$ cells, respectively. 
		
		In Figure~\ref{fig:res} we present a comparison of models B9 (left column), B10 (central column) and B11 (right column) showing density maps at the \hbox{$z=0$} (upper row) and \hbox{$x=0$} (lower row) planes. 
		Firstly, focusing on the shape of the slabs at \hbox{$z=0$} (see Figures~\ref{fig:res:B9_z}, \ref{fig:res:B10_z} and~\ref{fig:res:B11_z}), notice that in general the unstable slab seems very similar, though there are differences especially at small scales. 
		More complex substructure appears in the slab as the resolution increases.
		This is even clearer in the maps at \hbox{$x=0$} (see Figures~\ref{fig:res:B9_x}, \ref{fig:res:B10_x} and~\ref{fig:res:B11_x}). 
		The finer structure is very likely due to the excitation of shorter unstable modes, which can be resolved only with high enough resolution. 
		Evidence supporting this idea is related to the time the slab takes to become unstable. 
		It is known that the shorter scale modes of the NTSI grow faster than longer ones \citep{V94}.
		Therefore, this explains why the instability starts growing at earlier times as the resolution increases, i.e. the slab in B11 becomes unstable faster than B10, and this also applies comparing B10 and B9.  
		 
		Now, let us analyse the statistical properties of clumps at different resolutions. 
		In order to make a fair comparison between these simulations, we decided to study only clumps above a fixed physical size. 
		To do so, we set the minimum number of cells to define a clump as \hbox{$N_{\rm cell}=1,8,64$} in runs B9, B10, and B11, respectively. 
		Figure~\ref{fig:res_h} shows clump mass fraction in bins of mass and velocity along the upper and lower rows, respectively. 
		Each column represents different resolution models increasing from left to right. 
		The different colour solid lines represent the inner (green) and outer (orange) clumps, and the sum of the two (blue). 
		Firstly, notice that the clump mass histograms change appreciably when changing the resolution. 
		Notice that increasing the numerical resolution the mean mass fraction shifts towards lower values with a difference of \hbox{$0.6\rm\ dex$} between B9 and B10, and  \hbox{$0.3\rm\ dex$} comparing B10 and B11. 
		Furthermore, the clump maximum mass is about one order of magnitude different between B9 and B10 but we observe almost no difference comparing B10 and B11. 

		The clump velocity histograms (see lower panels of Figure~\ref{fig:res_h}) look relatively similar regardless of resolution.
		Even the mean clump speed and the dispersion do not change with resolution. 
		Probably, the most important difference is the fast speed tail of the inner clump distribution (green line). 
		Observe that it does not span to high speed in the model B9, yet it does in B10 and B11, meaning that the velocity dispersion of the inner clumps is smaller. 
		This could be attributed to the fact that inner clumps, and also clumps in general, are more massive in the low resolution run, so it is more difficult for them to reach higher speeds. 
		On the other hand, at higher resolution the inner clumps that have high speed are typically the least massive ones (see Figure~\ref{fig:b10_rv}). 
		From this analysis we can conclude that the ranges in mass and velocity covered by the clumps seems to be converging for our standard resolution but the detailed shape of distributions has not. 
		
	\subsection{Implications for the Galactic Centre hydro- and thermodynamic state}
	\label{sec:gc}
		
		As we shown in \citet{C16} close encounters (\hbox{$<2000\rm\ au$}) between the WR stars in the Galactic Centre, although not very frequent (\hbox{${\sim}10^{-3}\rm\ yr^{-1}$}), can take place. 
		Such encounters, in general, correspond to fairly asymmetric wind collisions. 
		In some cases they have very low $\eta$ so that the cooling parameter of the weaker wind can become of the order of unity. 
		According to our results, clump formation can take place under these conditions (e.g. model B+10). 
		However, given their low masses (\hbox{$\lesssim0.01\rm\ M_{\oplus}$}) they would be destroyed very rapidly. 
		The fact that they are ejected into a hot, diffuse and dynamic medium makes them susceptible to ablation and, more importantly, to thermal conduction \citep{B12}. 
		Specifically, for such light clumps they would evaporate in less than 10 years \citep{C18}. 
		Comparing this timescale with the free-fall timescale of the region, such clumps could be captured by Sgr~A* if they were ejected at a distance of \hbox{$\lesssim0.004\rm\ pc$} (\hbox{$0.1\rm\ arcsec$}); however, the WR stars orbits are located at least one order of magnitude further \citep{P06}. 
		This makes it very unlikely that clumps formed in stellar wind collisions and even in colliding wind binaries, e.g. IRS 16NE, IRS 16SW, E60, have chances of being accreted by the super-massive black hole. 
		Furthermore, these results are an extra piece of evidence against the hypothesis of the dusty G2-like objects \citep{G12} being born in wind collisions \citep{C18}. 
	
\section{Conclusions}
\label{sec:conclusions}

	We present a set of idealised 3D hydrodynamical simulations of stellar wind collisions aiming to characterise the clumps that form as a result of such an interaction. 
	Motivated by the WR stars in the Galactic Centre, we conduct a parameter study of systems of two stars with powerful outflows \hbox{$\dot{M}=10^{-5}\rm\ M_{\odot}\ yr^{-1}$}, wind terminal speeds of \hbox{${\sim}500$--$1500\rm\ km\ s^{-1}$}, and stellar separations in the range \hbox{${\sim}20$--$200\rm\ au$}. 
	We explore models with two identical stellar winds as well as systems with different stellar wind properties. 
	The 3D hydrodynamic evolution of radiative wind collisions confirm the 2D description studied previously \citep[e.g.][]{L11}. 
	Systems with identical radiative winds create hot slabs of material that cools down very rapidly becoming thinner and denser. 
	The resulting slab is susceptible to the NTSI that quickly manages to shape the whole slab, reaching an approximate stationary state at \hbox{$t\gtrsim2t_{\rm cross}$}. 
	As expected, increasing the radiative efficiency of the stellar winds causes slabs to become unstable quicker. 
	Interestingly, systems whose winds are within the transition between the radiative and adiabatic regimes can also generate unstable slabs. 
	Although initially those systems generate a hot, thick slab, thermal instabilities seem to be excited in the innermost part of the slab, which help to destabilise it allowing thin shell instabilities to grow.
	However, it can take a significantly longer time for them to reach a stationary state (\hbox{$t\gtrsim10t_{\rm cross}$}). 
	Symmetric models also display an increase in the density contrast with larger Mach numbers of the stellar winds, i.e. with faster stellar winds. 
	This behaviour is caused by the fact that the flow becomes more turbulent. 
	 
	Asymmetric models show a different behaviour as the wind collisions becomes unstable faster due to the presence of the KHI. 
	The more asymmetric the wind interaction, the faster instabilities are excited and grow. 
	This feature is observed even if only one of the winds is of a radiatively efficient nature. 
	In this case, the cool dense shell and clumps are formed solely of material from the weaker wind while the hot shocked material of the other wind compresses it from the other side of the interaction. 
	Having adiabatic winds with even larger Mach numbers enhances the observed density contrast even more than in the symmetric models studied. 
	
	Overall, the clumps formed in wind interactions through the NTSI have very small masses (\hbox{$m\lesssim10^{-2}\rm\ M_{\oplus}$}). 
	In symmetric models, they are born close to the apex being even lighter (\hbox{${\sim}10^{-6}$--$10^{-5}\rm\ M_{\oplus}$}). 
	At this point most of their momentum is parallel to the line connecting both stars, and spans a wide range. 
	As they are moving outwards (away from the apex), they gain mass and the stellar wind ram pressure accelerates them at the same time. 
	They reach their maximum mass  (\hbox{${\sim}10^{-4}$--$10^{-3}\rm\ M_{\oplus}$}) while escaping from the system at the point where the advective component of the ram pressure starts to dominate over the compressive component. 
	At a distance equal to the stellar separation from the apex, the velocity of clumps is about ${\sim}60$~per~cent of the stellar wind speed with a small dispersion. 
	On average, clumps take about \hbox{${\sim}2t_{\rm cross}$} to be ejected once they are formed. 
	
	The analysis of clumps in symmetric models confirm the fact that their masses are correlated with the cooling parameter, i.e. the less efficient the cooling in the post shocked material the more massive clumps can be. 
	In asymmetric models, although the properties are similar when scaled by the wind properties, the range they span is larger. 
	Furthermore, we found that the clump mass distribution close to the apex is approximately the same shape as the one for clumps further away. 
	This might hint to the KHI limiting the growth of the clumps. 
	
	Although in agreement with previous analytical estimates, clump masses are found to be significantly smaller (a factor ${\sim}1000$) than the theoretical upper limit. 
	Having such small masses means it is very unlikely that clumps formed in stellar wind collisions can be accreted by Sgr~A*, or have an impact onto the Galactic Centre thermodynamics state, especially considering the hostile environment to which they would be subject to. 
	Yet multiple dusty blobs, likely clumps, are observed to be present close to the powerful WR stars of the IRS~13E cluster \citep{F10}. 
	If stellar wind collisions cannot generate such massive clouds, is there another mechanism capable of condensing material from the stars?
	The interaction of multiple stellar winds, or between denser, slower outflows and the ambient medium \citep[e.g. IRS~33E;][]{C19}, are more complex phenomena that are constantly taking place in the region. 
	These scenarios remain as potential explanations worthy of further investigation, although they require much more computational resources and more physical aspects to incorporate into the models.  
		
\section*{Acknowledgments}

	We would like to thank Dr. W.~J.~Henney for reviewing this article as his comments and suggestions helped to improve its quality.
	DC and JC acknowledge the kind hospitality of the Max Planck Institute for Extraterrestrial Physics as well as funding from the Max Planck Society through a ``Partner Group'' grant.
	The authors acknowledge support from CONICYT project Basal \hbox{AFB--170002}. 
	This research was supported by the Excellence Cluster ORIGINS which is funded by the Deutsche Forschungsgemeinschaft (DFG, German Research Foundation) under Germany's Excellence Strategy -- EXC-2094 -- 390783311.
	DC is supported by CONICYT-PCHA/Doctorado Nacional (\hbox{$2015$--$21151574$}). 
	CMPR is supported by FONDECYT grant 3170870. 
	Numerical simulations were run on the HPC systems \textsc{hydra} and \textsc{cobra} of the Max Planck Computing and Data Facility. 
	Data analysis was carried out making use of the \textsc{python} package \textsc{yt} \citep{T11}.

\appendix
\section{Optically-thin radiative cooling}
\label{app:cool}

	The hydrodynamical equations solved using \textsc{Ramses}, specifically the energy equation (see Equation~\ref{eq:energy}) contain the term $\Lambda(T)$ that represents the energy losses of the fluid due to radiative cooling. 
	This function is calculated within the code according to the specified chemical abundances and metallicity. 
	At the beginning of each run the code computes the non-analytic function $\Lambda(T)$ as a function of temperature (and density). 
	This term includes the contribution of the main radiative processes assuming an optically-thin fluid. 
	Given the abundances of H and He, the codes estimates self-consistently their ionisation states in order to compute the density of both H and He nuclei, together with the electron density. 
	Having those, it computes the energy radiated through bremsstrahlung, (inverse-)Compton, as well as recombination lines\footnote{It also allows the possibility of including synchrotron radiation if magnetic fields are considered.} 
	On top of these calculations, it adds the contribution of the metals to the total radiative energy losses. 
	These are taken directly from radiative plasma model tables, such as Cloudy \citep{F98}, and then scaled according to the specified metallicity relative to the Solar value.
	
	In this work, the cooling parameter $\chi$ is estimated with Equation~\ref{eq:chi} but including an extra factor to account for a different metal content. 
	However, it still assumes that the radiative cooling rate is constant for the typical temperatures of shocked stellar winds ($10^6$--$10^7\rm\ K$). 
	Previously, \cite{C16} modified the expression in order to include the temperature dependence through the use an analytical expression for $\Lambda(T)$.
	In this study, this was not possible as we considered a non-analytic expression instead, which is a more realistic approach. 
	Let us bear in mind that the estimation of this parameter is to get an idea of the radiative nature of the system. 
	It has no impact on the simulation as the evolution of each model is self-consistent.
	
	In order to check the validity of the assumption of the medium being optically thin, we performed a posteriori estimations of the optical depth along the cartesian axes, as well as along the diagonal of the cubic domain in every simulated model. 
	The opacity for such calculations was taken from tabulated values of the OPAL Rosseland mean opacity \citep{I96}, according to the appropriate chemical abundances chosen (see Section~\ref{sec:ns}). 
	These tests confirmed the fact that the entire computational domain of every model was optically thin.

\section{Clump finder algorithm}
\label{app:algorithm}

	Here we describe the algorithm we used to identify clumps in our hydrodynamic models. 
	The method was inspired by the approach used in the \textsc{python} package \textsc{astrodendro}\footnote{\url{http://www.dendrograms.org/}}. 
	The input parameters, besides a \textsc{Ramses} output file, are $N_{\sigma}$ and $N_{\rm cell}$, where the former defines the density threshold $\rho_tr=\bar{\rho}+N_{\sigma}\times\sigma_{\rho}$ (being $\bar{\rho}$ the mean density and $\sigma_{\rho}$ the density dispersion), and the latter the minimum number of cells to identify a clump.

\label{sec:app1}

	\begin{algorithm}
		\caption{Clumpfinder}
		\begin{algorithmic}

		\Procedure{Find\_clumps}{snapshot,$N_{\sigma}$,$N_{\rm cell}$}\\
    			\State Read \textsc{Ramses} snapshot file
    			\State Consider only cells satisfying $\rho\geq\bar{\rho}+N_{\sigma}\times\sigma_{\rho}$
			\State Extract $x,y,z,\rho$ from remaining cells
			\State Find physically connected regions: structures\\
			
			\For{each structure}\\
			
				\State Define a clump per local density maximum
				\State Assign such cell to each clump
				\State Assign neighbour cells to corresponding clumps\\
				
				\While{$N$(unassigned cells)$>0$}\\
				
					\State Add neighbour cells recursively to each clump 
					\State as long as the density slope towards the 
					\State maximum is positive or zero
					
				\EndWhile\\
				
				\For{each clump in structure}\\
					\If{$N$(cells in clump) $>$ $N_{\rm cell}$}\\
						\State Calculate clump physical properties:
						\State $[m,x_{\rm cm},y_{\rm cm},z_{\rm cm},v_{x,\rm cm},v_{y,\rm cm},v_{z,\rm cm}]$\\
					
						\State Write clump properties into output file\\
					\EndIf
				\EndFor
				\State Flag structure as analysed
				
			\EndFor
	\EndProcedure

\end{algorithmic}
\end{algorithm}

\label{lastpage}

\begin{thebibliography}{99}
		\bibitem[\protect\citeauthoryear{Abdo et al.}{2010}]{A10} Abdo A.~A., et al., 2010, \apj, 723, 649
		\bibitem[\protect\citeauthoryear{Baganoff et al.}{2003}]{B03} Baganoff F.~K. et al., 2003, \apj, 591, 891
		\bibitem[\protect\citeauthoryear{Burkert et al.}{2012}]{B12} Burkert A., Schartmann M., Alig C., et al., 2012, \apj, 750, 58
		\bibitem[\protect\citeauthoryear{Burkert \& Lin}{2000}]{B00} Burkert A., Lin D.~N.~C., 2000, \apj, 537, 270 
		\bibitem[\protect\citeauthoryear{Calder\'on et al.}{2016}]{C16} Calder\'on D., Ballone A., Cuadra J., Schartmann M., Burkert A., Gillessen S., 2016, \mnras, 455, 4388
		\bibitem[\protect\citeauthoryear{Calder{\'o}n et al.}{2018}]{C18} Calder{\'o}n D., Cuadra J., Schartmann M., et al., 2018, \mnras, 478, 3494
		\bibitem[\protect\citeauthoryear{Calder{\'o}n, et al.}{2020}]{C19} Calder{\'o}n D., Cuadra J., Schartmann M., Burkert A., Russell C.~M.~P., 2020, \apjl, 888, L2
		\bibitem[\protect\citeauthoryear{Cuadra et al.}{2005}]{C05} Cuadra J., Nayakshin S., Springel V., Di Matteo T., 2005, \mnras, 360, L55
		\bibitem[\protect\citeauthoryear{Cuadra et al.}{2006}]{C06} Cuadra J., Nayakshin S., Springel V., Di Matteo T, 2006, \mnras, 366, 358
		\bibitem[\protect\citeauthoryear{Cuadra et al.}{2008}]{C08} Cuadra J., Nayakshin S., Martins F., 2008, \mnras, 383, 458
    		\bibitem[\protect\citeauthoryear{Cuadra et al.}{2015}]{C15} Cuadra J., Nayakshin S., Wang Q.~D., 2015, \mnras, 450, 277
		\bibitem[\protect\citeauthoryear{Crowther \& Willis}{1994}]{C94} Crowther P.~A., Willis A.~J., 1994, Sp. Sc. Rev., 66, 85
		\bibitem[\protect\citeauthoryear{Ferland et al.}{1998}]{F98} Ferland G.~J., Korista K.~T., Verner D.~A., Ferguson J.~W., Kingdon J.~B., Verner E.~M., 1998, PASP, 110, 761
		\bibitem[\protect\citeauthoryear{Fritz et al.}{2010}]{F10} Fritz T.~K., Gillessen S., Dodds-Eden K., et al., 2010, \apj, 721, 395 
		\bibitem[\protect\citeauthoryear{Genzel, Eisenhauer \& Gillessen}{2010}]{G10} Genzel R., Eisenhauer F., Gillessen S., 2010, Rev. Mod. Phys., 82, 3121
		\bibitem[\protect\citeauthoryear{Gillessen et al.}{2012}]{G12} Gillessen S., et al., 2012, Nature, 481, 51
		\bibitem[\protect\citeauthoryear{Gillessen et al.}{2017}]{G17} Gillessen S., Plewa P., Eisenhauer F., et al.\ 2017, \apj, 837, 30
		\bibitem[\protect\citeauthoryear{Hamaguchi et al.}{2016}]{Ha16}Hamaguchi K., Corcoran M.~F., Gull T.~R., et al., 2016, \apj, 817, 23 
		\bibitem[\protect\citeauthoryear{Hamaguchi et al.}{2018}]{H18} Hamaguchi K., Corcoran M.~F., Pittard J.~M., et al., 2018, Nature Astronomy, 2, 731
		\bibitem[\protect\citeauthoryear{Heitsch, et al.}{2007}]{H07} Heitsch F., Slyz A.~D., Devriendt J.~E.~G., Hartmann L.~W., Burkert A., 2007, \apj, 665, 445
		\bibitem[\protect\citeauthoryear{Hendrix et al.}{2016}]{H16} Hendrix T., Keppens R., van Marle A.~J., et al. 2016, \mnras, 460, 3975
		\bibitem[\protect\citeauthoryear{Hobbs et al.}{2013}]{H13} Hobbs A., Read J., Power C., Cole D., 2013, \mnras, 434, 1849
		\bibitem[\protect\citeauthoryear{Iglesias \& Rogers}{1996}]{I96} Iglesias C.~A., Rogers F.~J., 1996, \apj, 464, 943
		\bibitem[\protect\citeauthoryear{Kee et al.}{2014}]{K14} Kee N.~D., Owocki S., ud-Doula A. 2014, \mnras, 438, 3557 
		\bibitem[\protect\citeauthoryear{Lamberts et al.}{2011}]{L11} Lamberts A., Fromang S., Dubus G. 2011, \mnras, 418, 2618
		\bibitem[\protect\citeauthoryear{Lamberts et al.}{2012}]{L12} Lamberts A., Dubus G., Lesur G., Fromang S., 2012, \aap, 546, A60
		\bibitem[\protect\citeauthoryear{Lemaster  et al.}{2007}]{L07} Lemaster M.~N., Stone J.~M., Gardiner T.~A., 2007, \apj, 662, 582
		\bibitem[\protect\citeauthoryear{Lebedev \& Myasnikov}{1990}]{L90} Lebedev M.~G., Myasnikov A.~V., 1990, Fluid Dynamics, 25, 629
		\bibitem[\protect\citeauthoryear{L\"utzgendorf  et al.}{2016}]{L16} L\"utzgendorf N., Helm E. v.d., Pelupessy F.~I., Portegies Zwart S., 2016, \mnras, 456, 3645
		\bibitem[\protect\citeauthoryear{Martins et al.}{2007}]{M07} Martins F., Genzel R., Hillier D. J., Eisenhauer F., Paumard T., Gillessen S., Ott T., Trippe S., 2007, \aap, 468, 233
		\bibitem[\protect\citeauthoryear{McCourt, et al.}{2015}]{M15} McCourt M., O'Leary R.~M., Madigan A.-M., Quataert E., 2015, \mnras, 449, 2
		\bibitem[\protect\citeauthoryear{Muno et al.}{2007}]{MB07} Muno M.~P., Baganoff F.~K., Brandt W.~N., Park S., Morris M.~R., 2007, \apjl, 656, L69
		\bibitem[\protect\citeauthoryear{Panagiotou \& Walter}{2018}]{P18} Panagiotou C., Walter, R., 2018, \aap, 610, A37 
		\bibitem[\protect\citeauthoryear{Parkin \& Gosset}{2011}]{P11} Parkin E.~R., Gosset E., 2011, \aap, 530, A119
		\bibitem[\protect\citeauthoryear{Paumard et al.}{2006}]{P06} Paumard T. et al., 2006, \apj, 643, 1011
		\bibitem[\protect\citeauthoryear{Pittard}{2009}]{P09} Pittard J.~M., 2009, \mnras, 396, 1743
		\bibitem[\protect\citeauthoryear{Ponti et al.}{2010}]{P10} Ponti G., Terrier R., Goldwurm A., Belanger G., Trap G., 2010, \apj, 714,732
		\bibitem[\protect\citeauthoryear{Price et al.}{2011}]{P11} Price D.~J., Federrath C. \& Brunt, C.~M., 2011, \apjl, 727, L21 
		\bibitem[\protect\citeauthoryear{Puls, Vink \& Najarro}{2008}]{P08} Puls J., Vink J.~S., Najarro F., 2008, \aap~R, 16, 209
		\bibitem[\protect\citeauthoryear{Ressler et al}{2018}]{R18} Ressler S.~M., Quataert E. \& Stone J.~M., 2018, \mnras, 478, 3544
		\bibitem[\protect\citeauthoryear{Russell et al.}{2017}]{R17} Russell C.~M.~P., Wang Q.~D. \& Cuadra J., 2017, \mnras, 464, 4958
		\bibitem[\protect\citeauthoryear{Smith, Sigurdsson \& Abel}{2008}]{S08} Smith B., Sigurdsson S., Abel T., 2008, MNRAS, 385, 1443
		\bibitem[\protect\citeauthoryear{Stevens et al.}{1992}]{S92} Stevens I.~R., Blondin J.~M., Pollock A.~M., 1992, \apj, 386, 265
		\bibitem[\protect\citeauthoryear{Sunyaev et al.}{1993}]{S93} Sunyaev, R.~A. et al., 1993, \apj, 407, 606
		\bibitem[\protect\citeauthoryear{Sunyaev \& Churazov}{1998}]{S98} Sunyaev, R. \& Churazov, E., 1998, \mnras, 297, 1279 
		\bibitem[\protect\citeauthoryear{Teyssier}{2002}]{T02} Teyssier R., 2002, \aap, 385, 337
		\bibitem[\protect\citeauthoryear{Turk et al.}{2011}]{T11} Turk, M.~J., Smith, B.~D., Oishi, J.~S. et al., 2011, \apjs, 192, 9
		\bibitem[\protect\citeauthoryear{van Marle et al.}{2011}]{V11} van Marle A.~J., Keppens R., Meliani Z., 2011, \aap, 527, A3
		\bibitem[\protect\citeauthoryear{van Marle et al.}{2012}]{V12} van Marle A.~J., Keppens R., 2012, \aap, 547, A3
		\bibitem[\protect\citeauthoryear{Vink \& Harries}{2017}]{V17} Vink J.~S., Harries T.~J., 2017, \aap, 603, A120 
		\bibitem[\protect\citeauthoryear{Vishniac}{1983}]{V83} Vishniac E.~T., 1983, \apj, 274, 152
		\bibitem[\protect\citeauthoryear{Vishniac}{1994}]{V94} Vishniac E.~T., 1994, \apj, 428, 186
		\bibitem[\protect\citeauthoryear{Wang et al.}{2013}]{W13} Wang Q.~D. et al., 2013, Science, 341, 981
		\bibitem[\protect\citeauthoryear{Yelda et al.}{2014}]{Y14} Yelda S., Ghez A. M., Lu J. R., Do T., Meyer L., Morris M. R., Matthews K., 2014, \apj, 783, 131
	\end{thebibliography}
\end{document}